%% file: aapmsamp.tex
\documentclass[aps,prl,reprint,superscriptaddress]{revtex4-2}
\usepackage{graphicx}% Include figure files
\usepackage{dcolumn}% Align table columns on decimal point
\usepackage{bm}% bold math
\usepackage{siunitx}
\usepackage{booktabs}

\usepackage{amsmath,amssymb}

\begin{document}

\preprint{AAPM/123-QED}

\title{Bouncing microdroplets on hydrophobic surfaces}

\author{Jamie McLauchlan}
\affiliation{Department of Physics, University of Bath, 8 West Building, Claverton Down, Bath, BA2 7AY, United Kingdom}
\author{Jim S. Walker}
\affiliation{School of Chemistry, University of Bristol, Cantock's Close, Bristol, BS8 1TS, United Kingdom}
\author{Vatsal Sanjay}
\affiliation{Physics of Fluids Department, University of Twente, Horst, De Horst 2, 7522 NB Enschede, Netherlands}
\author{Maziyar Jalaal}
\affiliation{Van der Waals-Zeeman Institute, Institute of Physics, University of Amsterdam, Science Park 904, Amsterdam, 1098XH, The Netherlands}
\author{Jonathan P. Reid}
\affiliation{School of Chemistry, University of Bristol, Cantock's Close, Bristol, BS8 1TS, United Kingdom}
\author{Adam M. Squires}
\affiliation{Department of Chemistry, University of Bath, South Building, Soldier Down Ln, Claverton Down, Bath BA2 7AX}
\author{Anton Souslov}
\affiliation{TCM Group, Cavendish Laboratory, University of Cambridge, JJ Thomson Avenue, Cambridge CB3 0HE, United Kingdom}

\date{\today}

\begin{abstract}
Intuitively, slow droplets stick to a surface and faster droplets splash or bounce. However, recent work suggests that on non-wetting surfaces, whether microdroplets stick or bounce  depends only on their size and fluid properties, but not on the incoming velocity. Here, we show using theory and experiments that even poorly wetting surfaces have a velocity-dependent criterion for bouncing of aqueous droplets, which is as high as \textrm{6} \SI{}{\meter/\second} for diameters of \textrm{30--50}~\SI{}{\micro\meter} on hydrophobic surfaces such as Teflon. 
 We quantify this criterion by analyzing the interplay of dissipation, surface adhesion, and incoming kinetic energy, and describe a wealth of associated phenomena, including air bubbles and satellite droplets.
Our results on inertial microdroplets elucidate fundamental processes crucial to aerosol science and technology.
\end{abstract}


\maketitle
We breathe out small droplets and aerosols that splash, bounce, or stick onto surfaces, leading to contamination and disease transmission~\cite{Poon2020, Bourouiba2021, Katre2021,Onakpoya2021,Short2023,Shafaghi2020,Kumar2021}. Similar microdroplet phenomena are responsible for a variety of industrial~\cite{Lohse2021}, agricultural~\cite{Dorr2014,Massinon2017}, and environmental~\cite{Virtanen2011, Joung2015, Jain2015,Nietiadi2024,Tumminello2024} processes. 
Droplet impingement has been extensively explored in millimetric droplets and is often motivated by printing applications~\cite{Worthington1877, Worth1896, Worthington1896, Rioboo2001, Yarin06, Josserand16, Blanken2021,Lohse2021}, yet remains unexplored for fast microdroplets on poorly wetting substrates.

Here we ask a deceptively simple question: when does an aqueous microdroplet bounce? We contrast micron-size droplets with larger, millimeter-scale drops~\cite{Mao1997, Rioboo2008, Chen2016, Raman2016, Fink2018, Worth1896}. Large drops at small velocities stick to a substrate and spread to a flattened shape~\cite{Lohse2021, Aris1958, Scheller1995, Rioboo2002, Biance2004, Sikalo05, Attan07, Courbin2009, Snoeijer2013, Wang2023}, but fast millimeter-scale drops tend to splash~\cite{Palacios2012, Burzynski2020, Guo2022, Coppola2011, Kittel2018}, with the notable exception of bouncing on a thin air film~\cite{Sprittles2024, Lee2012, Kolinski2012, Langley2020} or bouncing off poorly wetting substrates~\cite{Jha2020, Sanjay2023}. In contrast, microdroplets with diameters of tens of microns occupy a different distinct regime due to their high surface-to-volume ratio, which creates a complex interplay between droplet dynamics, surface tension, and substrate adhesion. Splashing is rare, but sticking and bouncing are ubiquitous.

Substrate adhesion is a crucial factor dictating microdroplet bouncing, characterized by the contact angle $\theta$ via the Young--Dupr\'e equation. For wetting surfaces, the droplets always stick and never bounce~\cite{Visser2012, Visser2015, McCarthy2022,Mahato2024}. In contrast, for superhydrophobic (i.e., non-wetting) surfaces, substrate adhesion plays no role and the droplets bounce unless all inertia is damped out by viscous dissipation~\cite{Tai2021, Wang2023, Zhang2016}. The dimensionless ratio of (dissipative) viscosity to (non-dissipative) surface tension, the Ohnesorge number $\!\mathit{Oh}$, defines a simple criterion for microdroplets when surface adhesion is negligible: droplets bounce when $\!\mathit{Oh}$ is less than a constant of order 1, independent of incoming velocity~\cite{Jha2020, Sanjay2023}. For water microdroplets at room temperature, the Ohnesorge number has a relatively low value, $Oh \approx 0.02$, which shows that the droplets are underdamped and that both substrate adhesion and dissipation in the substrate-adjacent boundary layer must play a critical role. Together, these results hint towards a previously undiscovered transition on hydrophobic substrates that depends on contact angle and velocity. 

In this work, we discover that the transition from sticking to bouncing in microdroplets occurs when the incoming kinetic energy overcomes both dissipation and substrate adhesion. We focus on experiments with large microdroplets (around 30--\SI{50}{\micro\meter} in diameter) impinging on a hydrophobic Teflon surface with speeds of {1--10} \SI{}{m/s}. The mechanism of this transition can be understood using a simple under-damped ball-and-spring model, backed up by computational fluid dynamics that parallel our experimental results. These findings shed light on the previously unexplored interplay between inertia and adhesion crucial for applications from inkjet and 3D printing to sprays and aerosol resuspension.

\begin{figure*}[t]
\centering
\includegraphics[width=\textwidth]{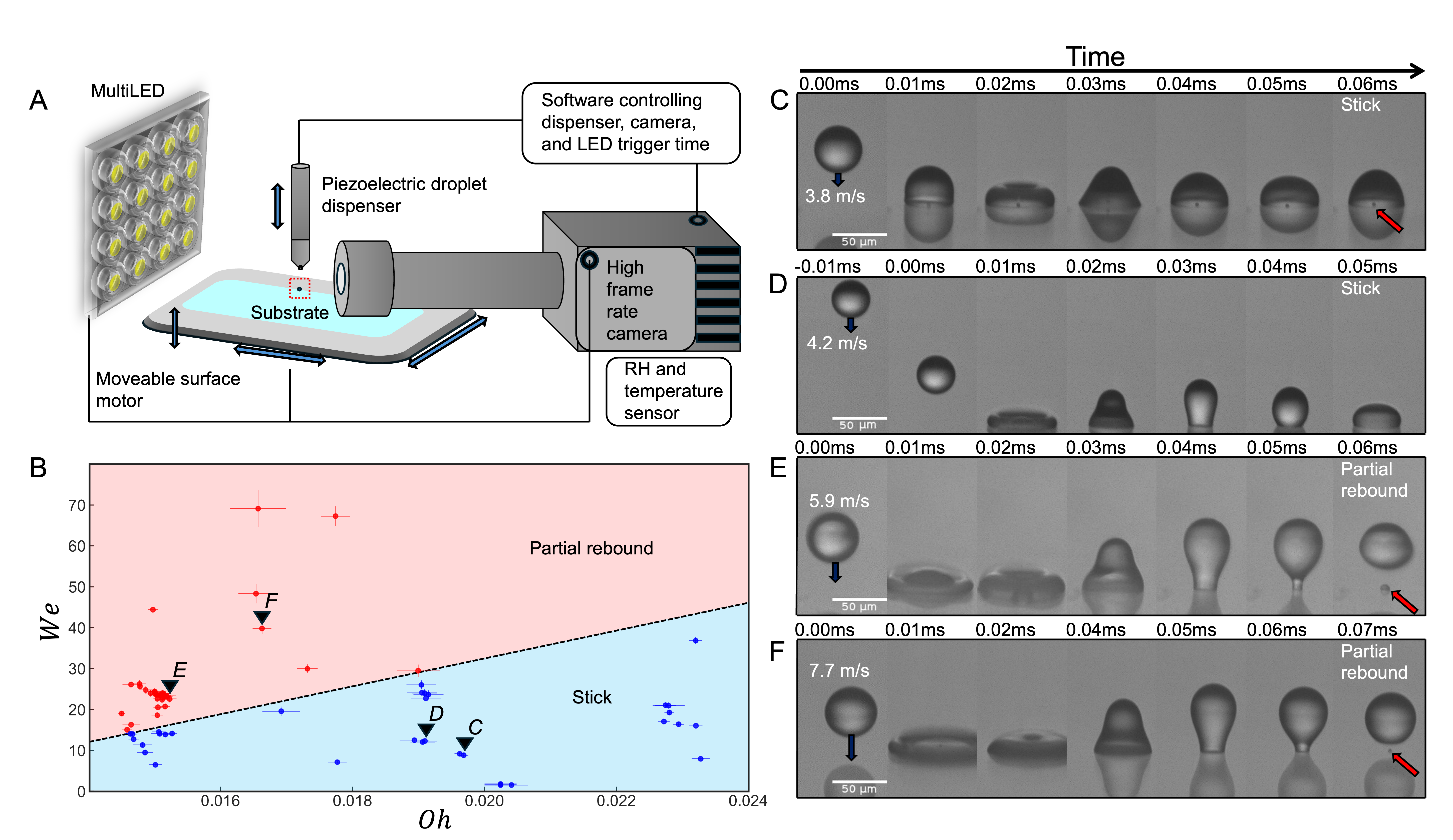}
\caption{Microdroplet experiments. (A) Schematic of experimental setup, in which an aqueous droplet is dispensed and imaged on a high-speed camera. (B) Experimental results plotted within the droplet phase space of Weber $\!\mathit{We}$ vs.~Ohnesorge $\!\mathit{Oh}$ numbers for a Teflon surface with a contact angle of 110$^\circ$. For each experimental point, the outcome was classified as a sticking event [e.g., points labelled as (C--D) and shown in the corresponding subfigure] or a partial rebound [e.g., subfigures (E--F)]. Error bars denote measurement uncertainties. The transition line is a guide to the eye separating the experimental outcomes. (C--F) High-speed imaging of droplet impact outcomes: The recorded outcomes are water droplets on a Teflon surface with a static contact angle of 110$^\circ$. (C) Sticking outcome, with an entrained air bubble shown by red arrow. Observed oscillations indicate that sticking occurs in the under-damped regime. $(\!\mathit{Oh}, \!\mathit{We}) = (0.0197, 8.80)$.  (D) Sticking outcome, without an air bubble but with oscillations. $(\!\mathit{Oh}, \!\mathit{We}) = (0.0191, 12.0)$. (E) Partial rebound, with a small sessile droplet shown by red arrow. $(\!\mathit{Oh}, \!\mathit{We}) = (0.0152, 23.7)$. (F) Partial rebound, with a smaller sessile droplet (red arrow). $(\!\mathit{Oh}, \!\mathit{We}) = (0.0166, 39.8)$.}
\label{fig:experiment1}
\end{figure*}

\section*{Experimental Stick-to-Bounce Transition}
%\subsection*{Set-up and Impact Variables}

In our experiments, we impact water-based microdroplets onto a poorly wetting Teflon substrate and use a high-speed camera to classify the outcomes as either sticking or partially rebounding, see Fig.~\ref{fig:experiment1}. In the experiments, each impact samples from a range of velocities $u$ (between 1 and \SI{10}{\meter/\second}) and droplet diameters $D$ (between $30$ and \SI{50}{\micro\meter}), and we vary the dynamic viscosity $\mu$ (between $0.88\times 10^{-3}$ and \SI{1.05e-3}{\pascal\cdot\second})  by varying temperature and glycerol concentration, while keeping surface tension $\gamma$ approximately constant, see Materials and Methods. Other relevant parameters, such as density $\rho$, and the surface-dependent static contact angle $\theta$ are also kept constant. Together, these parameters form a distinct region in the two-dimensional phase space spanned by the Weber number $\!\mathit{We}$, defined as
\begin{equation}
    \!\mathit{We} \equiv \frac{\rho D u^2}{\gamma},
\end{equation} 
and Ohnesorge number $\!\mathit{Oh}$, 
\begin{equation}
    \!\mathit{Oh} \equiv \frac{\mu}{\sqrt{\rho \gamma D}}, 
\end{equation}
where the denominator $\sqrt{\rho \gamma D}$ represents a combination of inertial and capillary effects, and is independent of the impact velocity. For our experimental parameters, $\!\mathit{We}$ varies between 1 and 70, and $\!\mathit{Oh}$ varies between 0.014 and 0.024, which is much lower than in inkjet printers~\cite{Lohse2021}. The prominence of inertia is captured in the Reynolds number $\textit{Re} \equiv \!\mathit{We}^{1/2} \!\mathit{Oh}$, which ranges from 50 to 500 in our experiments. Unlike millimetre-scale droplets~\cite{Jha2020, Sanjay2023} gravity does not play a role during impact, with a Bond number $\!\mathit{Bo} \equiv \rho g D^2/\gamma \approx O(10^{-4})$. We measure the static contact angle on our hydrophobic Teflon surface to be 110$^\circ$ with moderate hysteresis of 19$^\circ$ consistent with \SI{}{nm}-scale surface roughness as measured by atomic force microscopy (AFM), see Supporting Information (SI).

%\subsection*{Experimental Results and Discussion}
Slow microdroplets, with a relatively low $\!\mathit{We}$, adhere to the substrate after impact. Figs.~\ref{fig:experiment1}(C--D) illustrate how the droplet initially flattens and inertially spreads outwards. The contact line advances at a large angle until the radial spread reaches a maximum. Capillary forces then drive the fluid to retract and the droplet performs several underdamped oscillations (see SI Fig.~S3), in contrast to the previously explored overdamped regime~\cite{Sanjay2023}. In some cases, the droplet encloses an air bubble, see Fig.~\ref{fig:experiment1}(C), because of the collapse of the thin film of air formed underneath the droplet during impingement. Others have reported contactless bouncing on this air film~\cite{Sprittles2024, Lee2012, Kolinski2012, Langley2020}, but in our experiments the air cushion is always unstable and instead bouncing is controlled by surface adhesion. 

By contrast, fast microdroplets, with a relatively high $\!\mathit{We}$, perform a partial rebound while leaving behind a sessile remnant. In these cases, after the initial spreading and retraction, the droplet undergoes a necking instability and detaches from the surface, see Fig.~\ref{fig:experiment1}(E--F). Our combined data in Fig.~\ref{fig:experiment1}(B) shows the outcomes in $(\!\mathit{Oh},
\!\mathit{We})$ parameter space, with a line separating the sticking at  high $\!\mathit{Oh}$ and low $\!\mathit{We}$ from the partial rebounds at  low $\!\mathit{Oh}$ and high $\!\mathit{We}$. The positive slope of this line notably contrasts to the case of a superhydrophobic surface, Ref.~\cite{Sanjay2023}, for which the (purely vertical) transition line $\!\mathit{Oh} \approx 1$ is independent of $\!\mathit{We}$ and the incoming velocity. The effect of even a small surface adhesion is apparent: at small $\!\mathit{Oh}$, the droplet adheres to the hydrophobic surface and can stick even in the underdamped regime. However, at high incoming velocities (i.e., high $\!\mathit{We}$), the initial kinetic energy overcomes this surface adhesion and allows the droplet to escape. The small sessile droplet that stays on the surface provides additional evidence that, in this case, sticking results from surface adhesion.

\section*{Numerical Simulations}

\begin{figure*}[t]
\centering
\includegraphics[width=\textwidth]{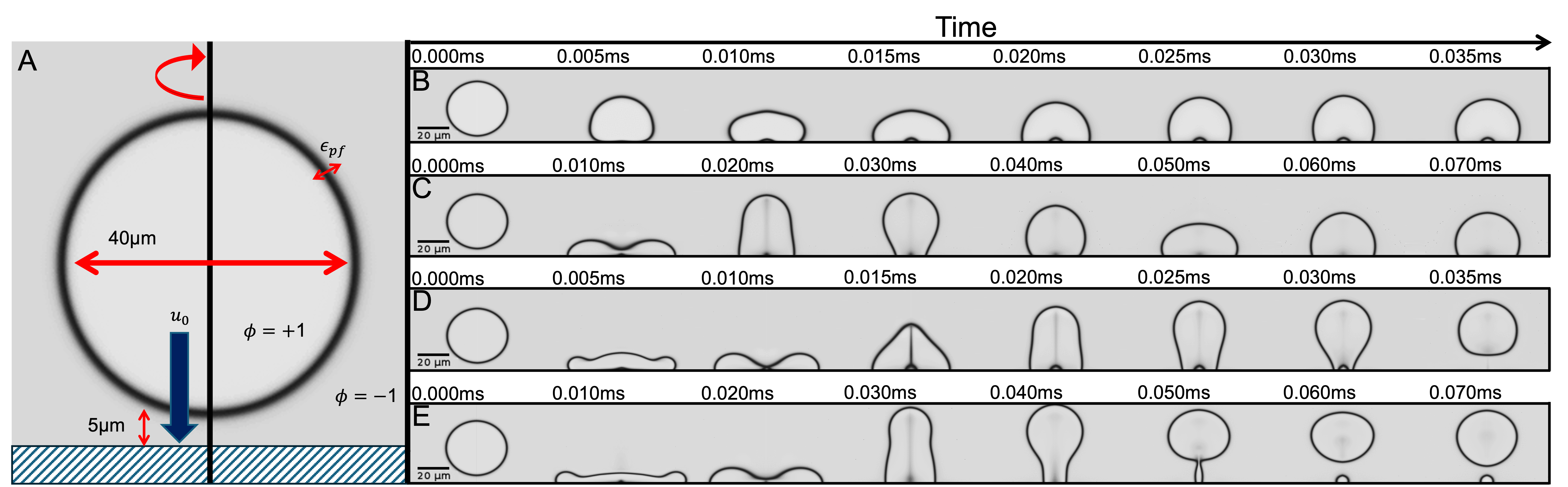}
\caption{(A) Schematic of finite-element phase-field simulations for microdroplet bouncing. (B--E) Numerical simulations reproduce the experimentally observed outcomes: (B) Sticking with an air bubble and small oscillations, $(\!\mathit{Oh},\!\mathit{We},\theta) = (0.019,2,110^\circ)$. (C) Sticking with a large maximum spread and large oscillations, $(\!\mathit{Oh},\!\mathit{We},\theta) = (0.028,56, 110^\circ)$. (D) Total rebound with a large maximum spread and an air bubble, $(\!\mathit{Oh},\!\mathit{We},\theta) = ( 0.019,27,110^\circ)$. (E) Partial rebound with a large initial spread and a necking instability, $(\!\mathit{Oh},\!\mathit{We},\theta) = (0.011,42, 100^\circ)$.}
\label{Fig2}
\end{figure*}

%\subsection*{Numerical Methods}
This experimental phenomenology can be quantitatively captured in finite-element numerical simulations based on a few simple ingredients. Once calibrated, we use these simulations to explore parameter regimes (in terms of $\!\mathit{Oh}$, $\!\mathit{We}$, and contact angle $\theta$) inaccessible in our experiments.

We perform computational fluid dynamics using finite-element software COMSOL~\cite{comsol} Multiphysics in an axially symmetric geometry (see Fig.~\ref{Fig2}A). We track the droplet-air interface using a phase-field variable $\phi$ which interpolates between $- 1$ and $+1$, and which follows Cahn-Hilliard dynamics~\cite{Phase1,Phase2,Phase3}, see Materials and Methods. Although we use a finite-element solver, we check that the droplet volume is conserved by computing a surface integral of $\phi$ across the entire domain. In the simulations, we integrate the incompressible Navier-Stokes equations:
\begin{equation}
\rho \partial_t \mathbf{u} + \rho (\mathbf{u} \cdot \nabla) \mathbf{u}  = \nabla \cdot \mathrm{\sigma} - \nabla p + \mathbf{F}_\mathrm{ST},
\end{equation}
where $\mathbf{u}(x,y)$ is the velocity field, $p$ is the pressure, $\mathbf{F}_\mathrm{ST}$ is the force due to surface tension, and $\mathbf{\sigma}$ is the viscous stress tensor. 
We implement incompressibility and a no-slip boundary condition for the substrate with a static contact angle at the contact line.

%\subsection*{Numerical Results and Discussion}

\begin{figure}[t]
    \centering
    \includegraphics[width=\linewidth]{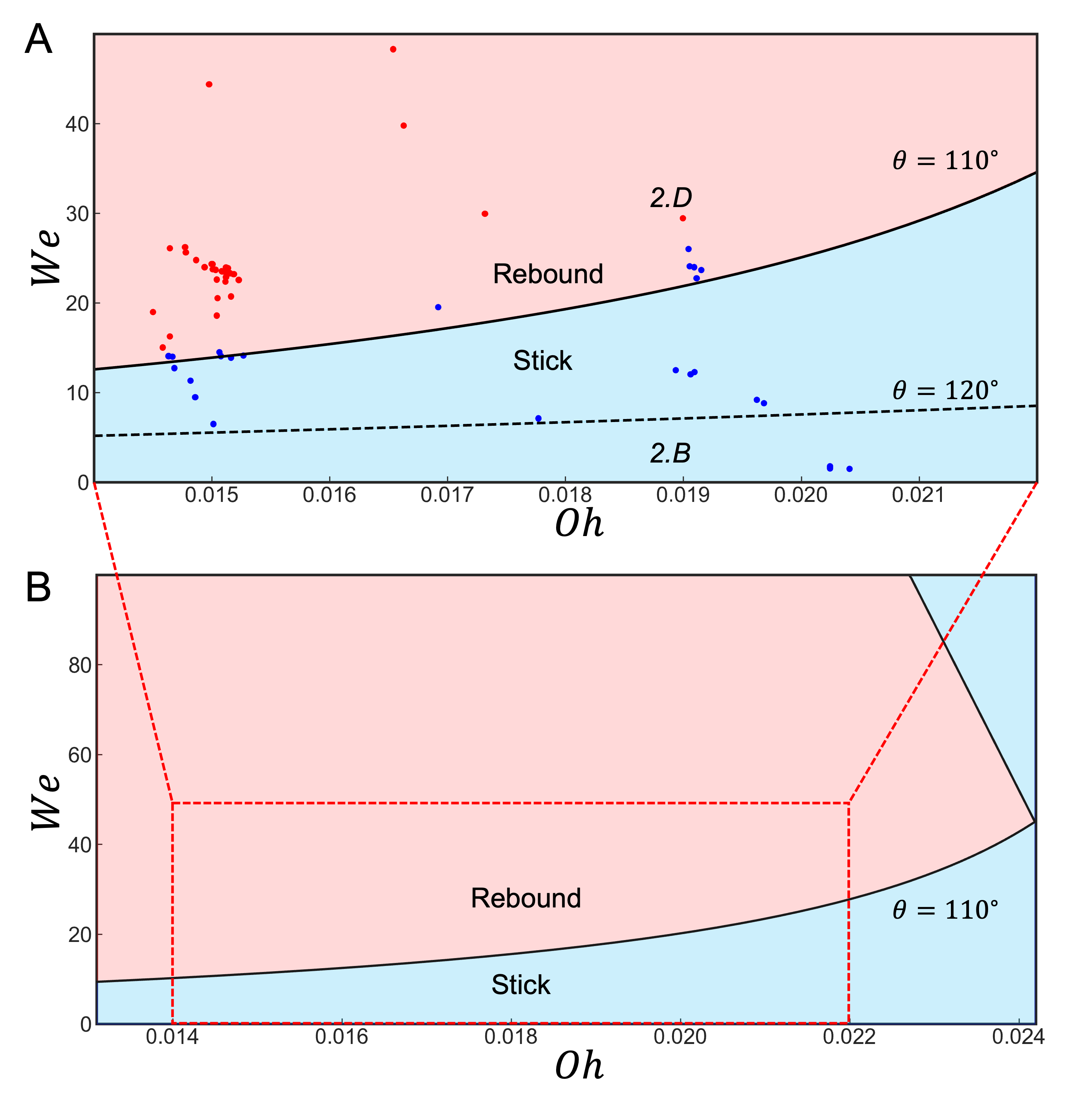}
    \caption{ (A) Droplet outcome comparing numerical and experimental results, where the black line indicates the numerical transition from sticking (blue) to bouncing (red) for $\theta = 110^\circ$. The dotted line shows that the same transition for the larger contact angle $\theta = 120^\circ$ occurs at lower $\!\textit{We}$ (See SI Fig.~S5 for numerical data points and Fig.~S6 for a rescaled plot in the Reynolds versus Ohnesorge numbers phase space). The data points are the experimental results from Fig.~\ref{fig:experiment1}. (B) A view of the $(\!\mathit{Oh}, \!\mathit{We})$ parameter space using simulations with a larger range of values and  $\theta = 110^\circ$. For higher impact velocities (i.e., higher $\!\mathit{We}$), the droplets again begin to stick, showing the reentrant nature of the transition. }
    \label{Fig3}
\end{figure}

We show sample simulation results in Fig.~\ref{Fig2}(B--E), which are broadly consistent with experiments. For low Weber number $\!\mathit{We}$ and $\theta = 110^\circ$  (consistent with experiments on Teflon), the droplets sticks to the substrate and can entrap an air bubble, see Figs.~\ref{Fig2}(B--C). At higher $\!\mathit{We}$, the air bubble remains, but the droplet rebounds, Fig.~\ref{Fig2}(D). Combining all of the numerical outcomes, Fig.~\ref{Fig3}(A) shows the transition line from sticking to bouncing for two values of the contact angle.

Increasing the contact angle in the simulations above $\theta =  110^\circ$ lowers the $\!\mathit{We}$ of the transition to sticking for a given $\!\mathit{Oh}$, with no sticking in this inertial regime when the contact angle approaches 180$^\circ$. This confirms our intuition that the inertial effects encoded in  $\!\mathit{We}$ interplay against the surface adhesion encoded in $\theta$: the larger the contact angle, the weaker the adhesion, and the less inertia is necessary to overcome it.

\begin{figure}[t]
    \centering
    \includegraphics[width=\linewidth]{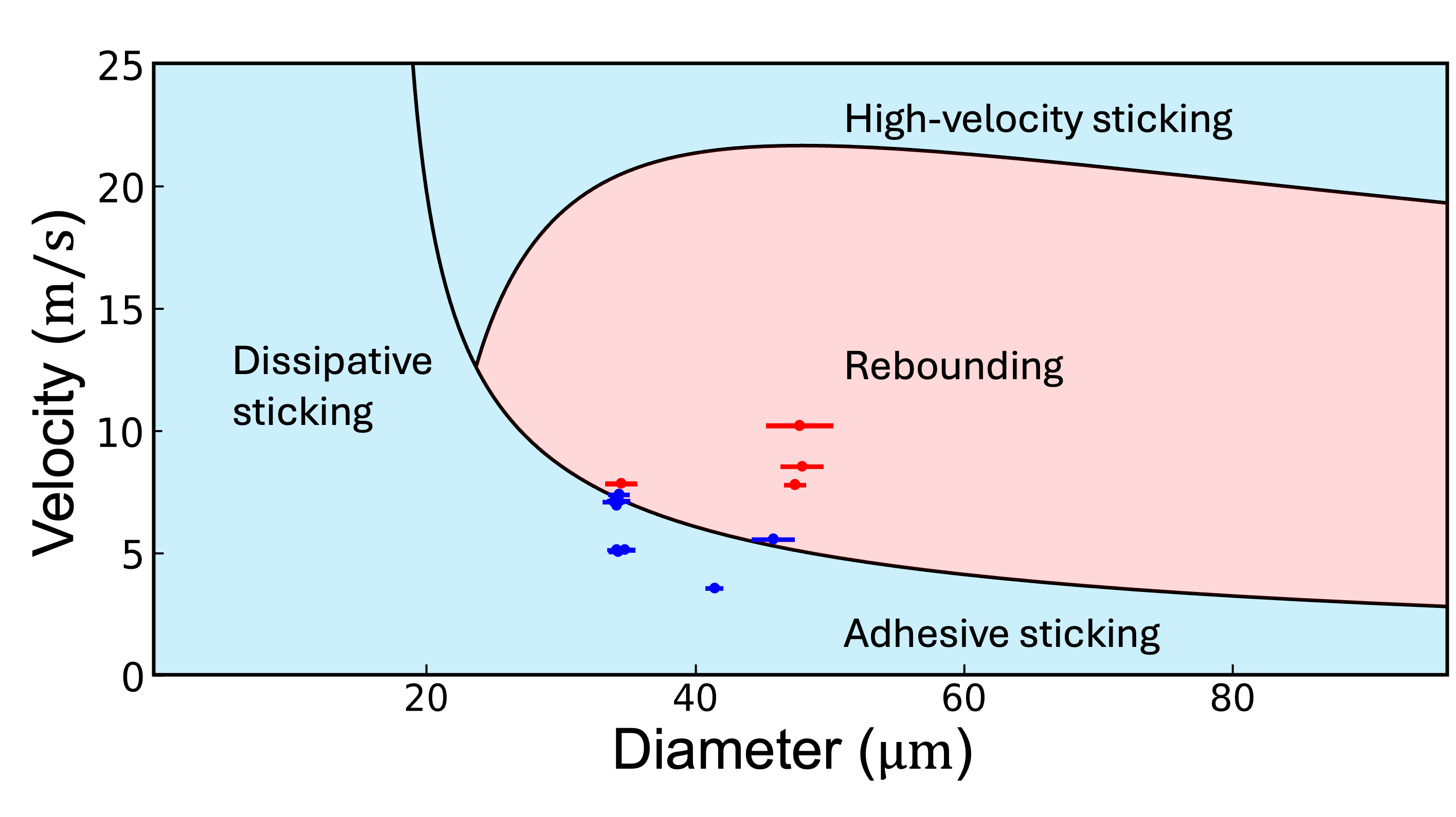}
    \caption{Rebounding and sticking in the phase space of Velocity versus droplet Diameter for poorly wetting surface, $\theta = 110^\circ$, corresponding to the re-dimensionalized $\!\mathit{We}-\!\mathit{Oh}$ axes in Fig.~\ref{Fig3}. We keep other parameters constant and plot experimental data for water on Teflon at room temperature as points (subset of Fig.~\ref{fig:experiment1} data recorded at the same viscosity and size) and finite-element simulations as lines. The graph indicates the different sticking mechanisms: for dissipative sticking, the droplets are small and their incoming velocity is fully dissipated by their viscosity; for adhesive sticking, the droplet is underdamped but cannot rebound due to surface adhesion; for high-velocity sticking, the incoming kinetic energy is dissipated due to the larger maximum spread diameter.}
    \label{Fig3E}
\end{figure}

The simulations allow us to access a counterintuitive regime for $\!\mathit{We}$ larger than in experiments, in which the droplets transition from bouncing back to sticking as their velocity is increased, Fig.~\ref{Fig3}(B). This upper transition occurs because  the droplet spreads more and therefore stores a larger amount of energy in the surface. In turn, the droplet then dissipates a larger fraction of its incoming kinetic energy~\cite{Jha2020} at larger incoming velocities. For larger values of the viscosity (i.e., larger $\!\mathit{Oh}$), this extra dissipation is sufficient to overcome the kinetic energy needed for a rebound (at sufficiently small contact angles $\theta$). The intersection of these two reentrant transitions leads to an overall maximum value of $\!\mathit{Oh} \approx 0.024$ above which only sticking outcomes are observed (for $\theta =  110^\circ$). Notably, this value is significantly smaller than $1$, indicating that sticking still occurs in the inertial regime and for any incoming droplet velocity accessible in our experiments (i.e., any $\!\mathit{We}$). From the criterion $\!\mathit{Oh} < 0.024$, we conclude that there is a universal size limit for aqueous droplets, below which bouncing does not occur for any velocity. We find that water droplets at room temperature cannot bounce from a surface with contact angle $\theta =  110^\circ$ if $D < \SI{25}{\micro\meter}$ as shown in Fig. ~\ref{Fig3E}, and this threshold decreases for more hydrophobic surfaces, for example $D < \SI{10}{\micro\meter}$ for $\theta =  120^\circ$. We conclude that aqueous aerosols with $\SI{10}{\nano\meter} \lesssim D \lesssim \SI{1}{\micro\meter}$ can only bounce off superhydrophobic surfaces with $\theta > 150^\circ$ and bouncing is suppressed for $\!\mathit{Oh} \gtrsim 1$.
Overall, bouncing for moderately hydrophobic surfaces occurs within a 'Goldilocks zone' of moderately large  $\!\mathit{We}$ and moderately small $\!\mathit{Oh}$, where the droplet has sufficient incoming kinetic energy to overcome adhesive effects without excessive dissipation during the spreading and receding process. 

The rebounds in Fig.~\ref{Fig2}(B--D) do not deposit a sessile droplet, which suggests that these simulations miss some of the experimental complexity, such as contact angle hysteresis. Nevertheless, partial rebounds can be simulated by changing the contact angle and other simulation parameters such as the fluid density, see Fig.~\ref{Fig2}(E). However, the bouncing mechanism appears to be the same for both partial and total rebounds, and we proceed to formulate quantitative models that capture both cases.

\section*{Energy balance criterion}
%\subsection*{Energy Considerations}
The complex dynamics that govern the outcome of the droplet-surface interaction can be heuristically understood through the lens of energy conservation. A droplet will only rebound if the kinetic energy after lift-off is positive, $E_{\mathrm{k,f}} > 0$. This criterion can be restated as $E_{\mathrm{k,f}} = E_{\mathrm{k,0}} - E_\mathrm{\gamma} - E_\mathrm{\mu} > 0$, where the initial kinetic energy $E_{\mathrm{k,0}}$ decreases by the energy of the newly created droplet-surface contact $E_\mathrm{\gamma}$ and the energy $E_\mathrm{\mu}$ associated with viscous dissipation during the impact process. Phenomenologically, almost all dissipation occurs during the retraction process, so $E_{\mathrm{k,0}}$ can be considered as the energy at maximum droplet spread.
This kinetic energy contains the velocity-dependence of the stick-to-bounce transition: $E_{\mathrm{k,0}} \equiv \frac{1}{2} m u_0^2$, where $u_0$ is the incoming velocity of the droplet center-of-mass before surface contact. A heuristic criterion for the velocity $u_0$ at this transition is then: 
\begin{equation}
\frac{1}{2} m u_0^2
= E_\mathrm{\gamma} + E_\mathrm{\mu}.
\end{equation}

The kinetic energy loss during a bouncing process, $E_\mathrm{\gamma} + E_\mathrm{\mu}$, can be quantified using simple scaling arguments. The surface energy scales as $E_\mathrm{\gamma} = \pi \gamma D^2 f(\theta)$, where $\pi \gamma D^2$ is the surface energy scale for a spherical surface, $\pi D^2$, and the contact angle $\theta$ encodes the difference between the air-droplet and the surface-droplet energies per unit area. A form for $f(\theta)$ is derived in the SI under the assumption that a small sessile droplet remains on the surface, but will be left arbitrary here.

The energy $E_\mathrm{\mu}$ dissipated by viscosity is more complex to quantify because it involves contributions from three different mechanisms: (i) $E_\mathrm{\mu,3D}$ in the three-dimensional flows in the droplet bulk, (ii) $E_\mathrm{\mu,2D}$ in the boundary layer associated with the contact area between the droplet and the surface~\cite{Bound2}, and (iii) $E_\mathrm{\mu,1D}$ along the air-droplet-surface contact line. Based on inertial dynamics, dimensional analysis requires that both bulk and surface dissipation scale as~\cite{Okumura2003}
\begin{equation}
E_\mathrm{\mu,3D} \sim E_\mathrm{\mu,2D} \sim \mu u_0^2 D t_c,
\end{equation}
in terms of the contact time $t_c$ (See SI Fig.S1), but with different scaling coefficients, so in combination $E_\mathrm{\mu,3D} + E_\mathrm{\mu,2D} = \kappa \mu u_0^2 D t_c$, where $\kappa$ is a dimensionless proportionality constant. Although $\kappa$ can itself depend on droplet parameters such as $\!\mathit{We}$ and boundary layer thickness, we assume it to be a constant for simplicity.
\begin{figure}[t]
    \centering
\includegraphics[width=\linewidth]{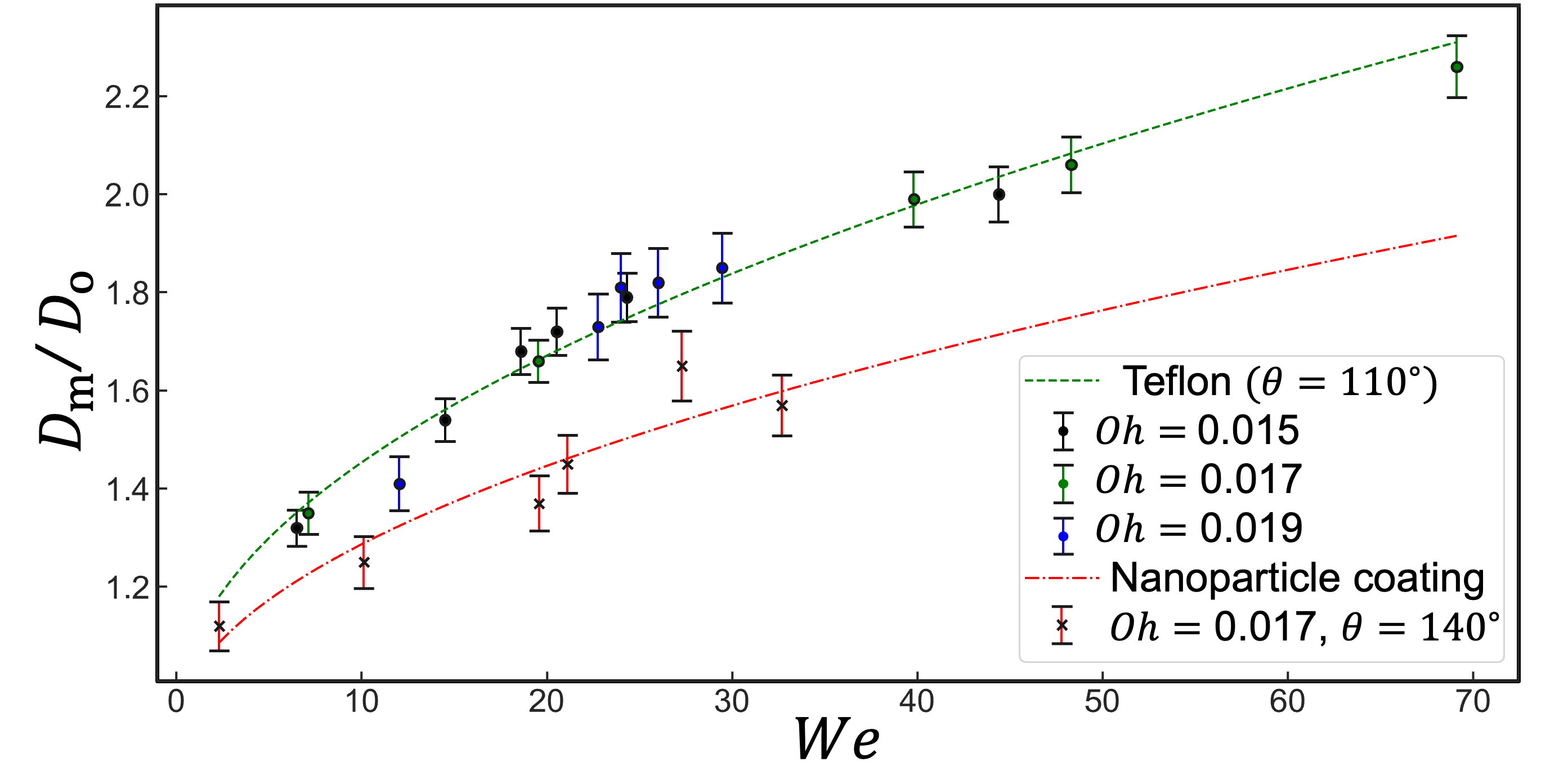}
    \caption{Experimental measurement of the diameter at maximum spread, $D_\textrm{m}$, normalized by droplet diameter $D$, for $(\!\mathit{Oh},\theta) = (0.015,110^\circ)$, $(0.017,110^\circ)$, $(0.019,110^\circ)$, and $(0.017,140^\circ)$, for two different surfaces: Teflon and  nanoparticle-coated glass. Both fits are consistent with  theoretical prediction in Eq.~[\ref{eq:dmax}] with different $\theta$-dependent prefactors. This data shows that the maximum spread is independent of $\!\mathit{Oh}$ (See SI Fig. S2).}
    \label{FigX}
\end{figure}

Contact line dissipation is a distinct process dominated by the friction of the contact line moving along the surface and depinning from surface imperfections. The energy $E_\mathrm{\mu,1D}$ depends instead on the maximum spread diameter $D_{\mathrm{m}}$, the surface tension, and the contact angle hysteresis $\Delta\!\cos \theta \equiv \cos\theta_\mathrm{a}-\cos\theta_\mathrm{r}$ via~\cite{Bound1, Quere}: 
\begin{equation}
E_\mathrm{\mu,1D} \approx D_{\mathrm{m}}^2 \gamma \, \Delta\!\cos \theta.
\label{eq:e1d}
\end{equation}
The maximum spread diameter $D_{\mathrm{m}}$ is an important variable in droplet systems  but its exact form has been widely debated in the literature ~\cite{Clanet2004,Eggers2010,BOSSA_2011,Bruin2014, Wildeman2016, sanjay2024, Mahato2024}. We fit to a widely accepted general form:
\begin{equation}
    \frac{D_{\mathrm{m}}}{D} = g(\theta)( 1 + C \!\mathit{We}^{1/2}),
\label{eq:dmax}
\end{equation}
where $C$ is a constant and $g(\theta)$ accounts for the wettability of the surface. This expression was previously derived in Ref.~\cite{BOSSA_2011} and has shown to hold true for a wide range of drop impacts~\cite{Bruin2014, sanjay2024}. In the narrow range of $ 10 < \!\mathit{We} < 100$, another phenomenological expression, $D_{\mathrm{m}}/D \approx \!\mathit{We}^{1/4}$, is also commonly used~\cite{Clanet2004,Mahato2024}. From our experimental data, Fig.~\ref{FigX}, we find that $C \approx 0.14$ and the spread is independent of $\!\mathit{Oh}$.
Substituting Eq.~[\ref{eq:dmax}] into Eq.~[\ref{eq:e1d}], we arrive at the scaling result for contact line dissipation, $E_\mathrm{\mu,1D} \approx D^2 \gamma h(\theta,\Delta \theta) ( 1 + C\!\mathit{We}^{1/2})^2$, where $h = g^2(\theta) \, \Delta\!\cos\theta$. 

Combining these scaling laws, we find that the droplet will have sufficient kinetic energy to bounce if 
\begin{equation}
E_{\mathrm{k,0}} > \pi \gamma D^2 f(\theta) + \kappa \mu u_0^2 D t_c + D^2 \gamma h(\theta,\Delta \theta)( 1 + C\!\mathit{We}^{1/2})^2.
\end{equation}
Rescaling by $D^2 \gamma$, we obtain the prediction for the stick-to-bounce transition (see SI for a complete derivation):
\begin{equation}
\label{Kap}
\mathit{We} = \frac{ f(\theta) + h(\theta, \Delta \theta)(1 + 2C\!\mathit{We}^{1/2})} {1 -  \kappa \mathit{Oh} - C^2h(\theta, \Delta \theta)}.
\end{equation}
This is the most general expression, however excluding the $E_\mathrm{\mu,1D}$ hysteresis contributions, it simplifies to a more intuitive form that provides a good approximation for systems with low hysteresis: 
\begin{equation}
\label{Kap3}
\mathit{We} = \frac{ f(\theta) }{1 -  \kappa \mathit{Oh} }.
\end{equation}

Equation~[\ref{Kap3}] interpolates between two extremes: (i) the regime of small $\!\mathit{Oh}$ (and small $\!\mathit{We}$) in this work, for which the stick-to-bounce transition is given by the linear relation $\!\mathit{We}(\!\mathit{Oh})$ and the slope of the line is determined by the contact angle, and (ii) the asymptotic limit of the transition for large $\!\mathit{We}$, $\mathit{Oh} \approx 1/\kappa$, which recovers the velocity-independent transition predicted for superhydrophobic surfaces~\cite{Sanjay2023}. For case (ii), the (constant) parameter $\kappa$ captures how the threshold $\mathit{Oh}$ is lowered by boundary-layer dissipation due to the interplay between wetting and droplet spreading. Overall, Eq.~[\ref{Kap3}] matches well the observed phenomenology and provides quantitative confirmation for the energy-balance mechanism underlying droplet sticking and bouncing.

\begin{figure}[t]
    \centering
    \includegraphics[width=\linewidth]{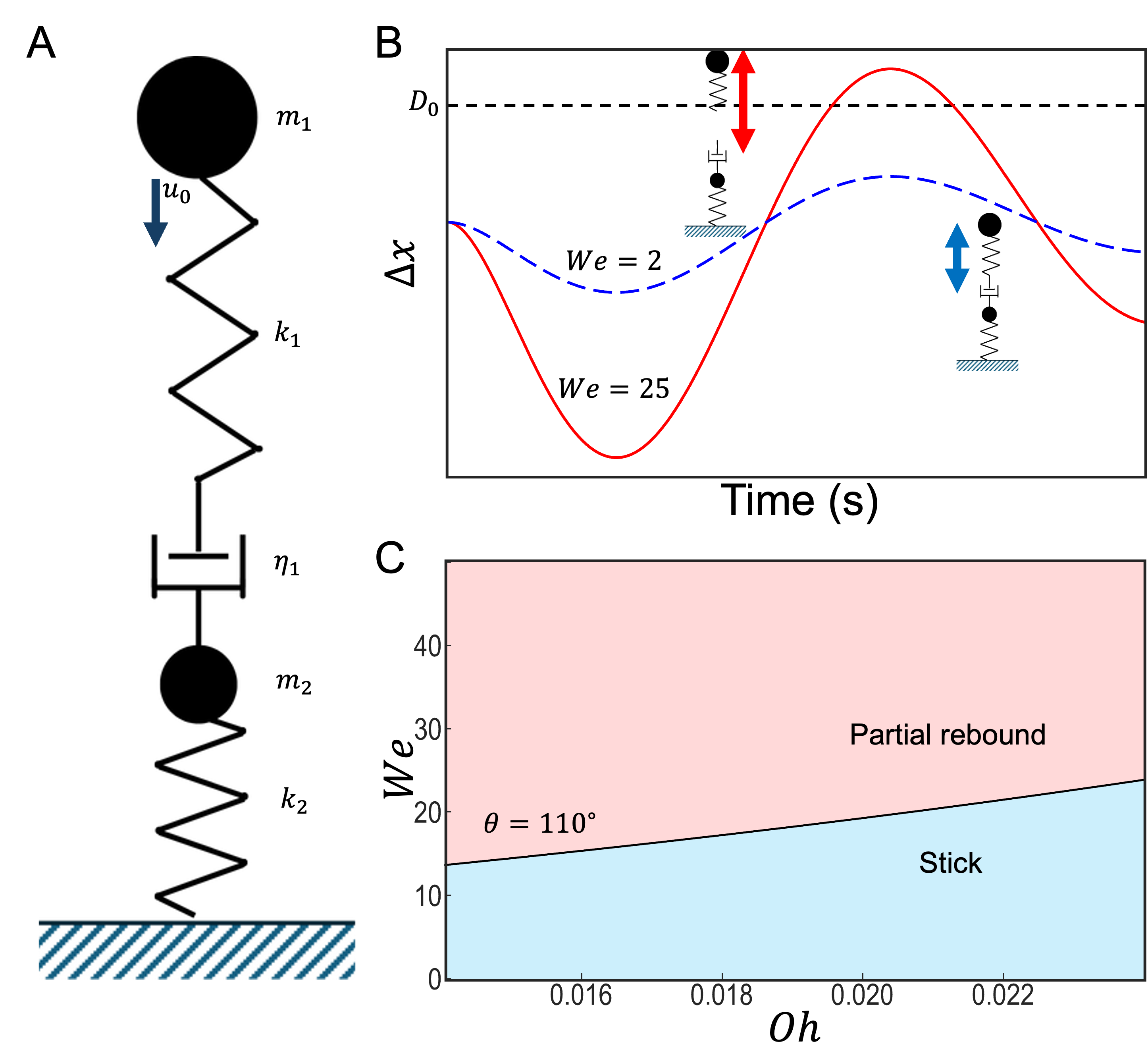}
    \caption{ (A) An illustration of the simple model of two masses, two springs, and a damper to represent a microdroplet impact. (B) The extension $\Delta x$ is plotted for two representative cases, one where $\Delta x$ exceeds the bouncing threshold (red) and one where it does not (blue, with lower initial velocity). (C) $\!\mathit{We}-\!\mathit{Oh}$ phase space for this model with ($\alpha, \zeta,\theta$)  = (50,1,110$^\circ$) . This minimal model reproduces the simulation and experimental results. (See SI Fig. S7 for a further phase space).}
    \label{Model}
\end{figure}

The phenomenological argument for energy balance leads us to construct a simple ball-and-spring model that captures both bouncing and sticking, Fig.~\ref{Model}(A).
A large upper mass $m_1$ (representing the bulk of the droplet) is connected to a smaller mass via a spring and dashpot in series, with the smaller mass $m_2$ (representing the part of the droplet that makes contact with the surface) attached to the surface through a second spring. A damper represents the total dissipation in the system, the upper spring represents droplet surface tension, and the lower spring represents the adhesion between the surface and the fluid. 
The particles first impact the surface until the springs are maximally compressed, corresponding to maximum droplet spreading. The energy stored in the springs corresponds to the energy in both the fluid-air and fluid-solid interfaces at maximum droplet spreading. As the springs relax from this compression, in the underdamped regime, some of the energy is dissipated by the dashpot, and the rest of the energy is converted into spring extension. Fig.~\ref{Model}(B) illustrates how we define a bouncing outcome in this model: the upper spring is considered to break above a threshold extension, during the first period of the spring oscillation. 

Using dimensional rescaling of the spring model, the dynamics can be re-expressed in terms of the $\!\mathit{We}$ and $\!\mathit{Oh}$ numbers,
\begin{align}
m_1\ddot{x}_1 &= -\frac{\zeta }{\mathit{We}} \Delta x - \frac{\alpha \mathit{Oh}}{\sqrt{We}} \Delta \dot{x} \\
m_2 \ddot{x}_2 &= \frac{\zeta }{\mathit{We}} \Delta x
+ \frac{\alpha \mathit{Oh}}{\sqrt{We}}\Delta \dot{x}  - [ 1 + \cos(\theta)] \zeta \mathit{We}^{-1} x_2, \nonumber
\end{align}
where  all of the quantities have been non-dimensionalized, $m_{1,2}$ are the mass fractions of the two droplet parts, $\mathit{\zeta}$ is the spring constant (of order 1), $\mathit{\alpha}$ is the damping (of order 50), $x_{1,2}$ are the coordinates of the upper and lower particles, respectively, and $\Delta x = x_1 - x_2$ is the extension of the upper spring (see SI for rescaling). Fig.~\ref{Model}(C) shows the stick-to-bounce transition using experimental values of $\!\mathit{We}$ and $\!\mathit{Oh}$ within this model, reproducing the transition line observed in simulations and experiments.

Overall, the ball-and-spring model translates the energy balance argument into the simplest dynamical model and shows the mechanism that relates surface adhesion to the velocity at which the transition is observed for a given $\!\mathit{Oh}$. 

\section*{Experiments on nanoparticle-coated surface}
\begin{figure*}[t]
\centering
\includegraphics[width=\textwidth]{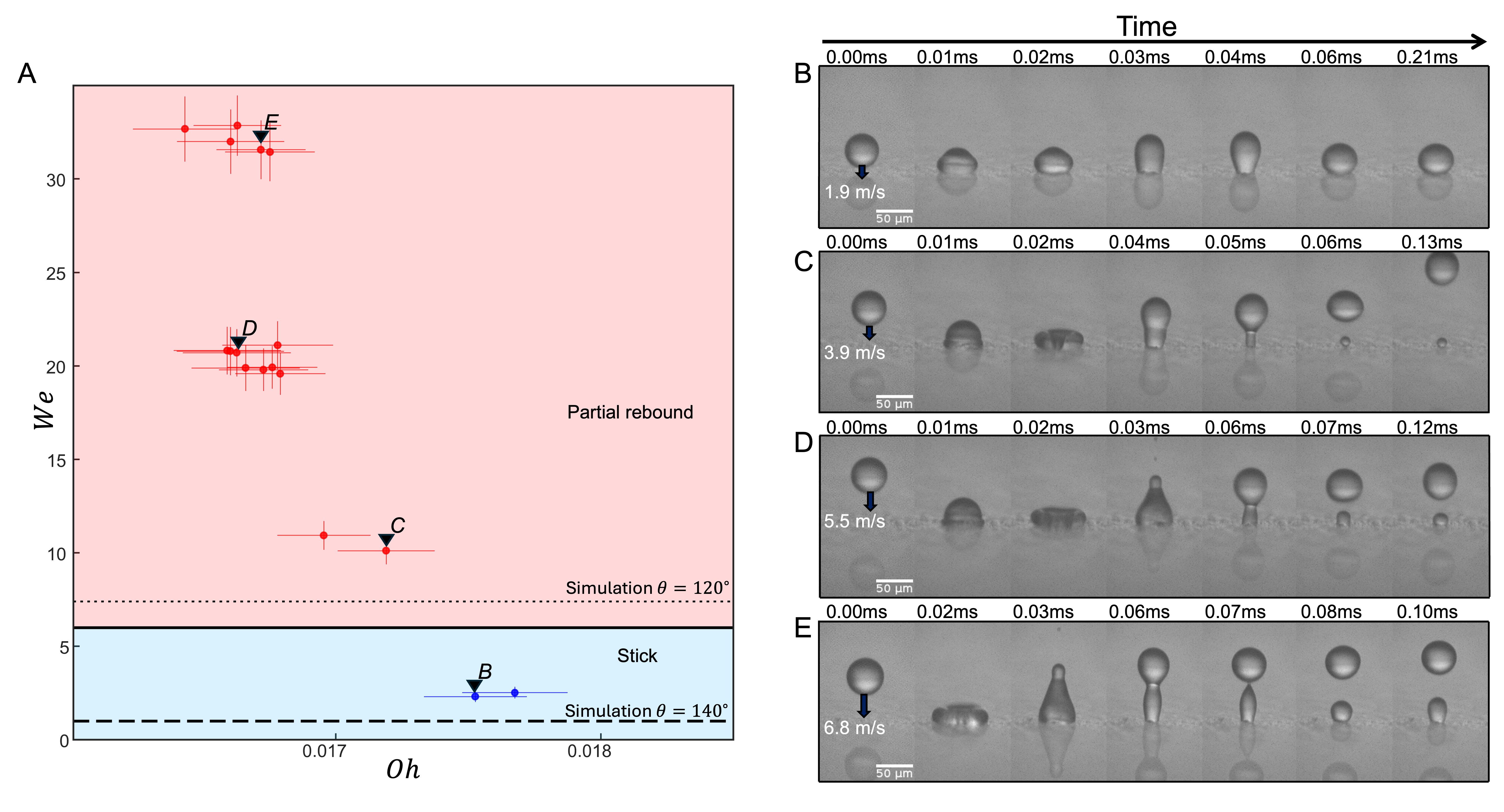}
\caption{(A) Phase space of microdroplet impact experiments on a silicon nanosphere coated glass surface of static contact angle $\theta = 140^\circ$. (C-F) High-speed imaging of droplet impact outcomes: (C) Sticking with large oscillations,  $(\!\mathit{We}, \!\mathit{Oh}) = (2.31, 0.0177)$. (D) Partial rebound with a small sessile droplet, $(\!\mathit{We}, \!\mathit{Oh}) = (10.1, 0.0172)$.   (E) Partial rebound with a medium sessile droplet,
$(\!\mathit{We}, \!\mathit{Oh}) = (20.7, 0.0166)$.
 (F) Partial rebound with a large sessile droplet,
 $(\!\mathit{We}, \!\mathit{Oh}) = (31.6, 0.0167)$.}
\label{final}
\end{figure*}

To further test model predictions, we performed experiments on a hydrophobic surface with a higher contact angle and experimentally verified the prediction that bouncing occurs for lower $\!\mathit{We}$. This surface, a silicon nanoparticle-coated glass, has a static contact angle of $140^\circ$. We plot the experimental results for sticking and partial rebounding in $(\!\mathit{Oh},\!\mathit{We})$ phase space in Fig.~\ref{final}(A). Consistent with our intuition that weaker adhesion makes bouncing easier, for this surface, bouncing occurs at a much lower $\!\mathit{We} \approx 10$. However, when we plot the simulation transition for $\theta = 140^\circ$, we find it significantly underestimates $\!\mathit{We}$ of the experimental transition. We posit the reason for this discrepancy is the greater roughness of the nanoparticle coated surfaces compared to Teflon. We measure the roughness using surface profilometry (see SI) and find a correspondingly larger contact angle hysteresis, with a receding contact angle $\theta_\textrm{r} = 120^\circ$ for mm-scale water droplets. When we use this value of the receding contact angle in the numerics we find the transition line moves to a higher value of $\!\mathit{We}$ which is consistent with experiment, Fig.~\ref{final}(A). We then modify the numerics to introduce contact angle hysteresis, and find that the droplet outcome only depends on $\theta_\textrm{r}$ and is largely independent of the advancing angle $\theta_\textrm{a}$ and the static angle $\theta$. Although these simulations do not fully account for the effects of surface roughness in increasing the contact line dissipation and pinning, we nevertheless obtain quantitative agreement for the location of the stick-to-bounce transition.
Fig.~\ref{final}(B) shows the underdamped sticking that we experimentally observe in the regime where the numerics without hysteresis and with a static $\theta = 140^\circ$ would incorrectly predict a rebound. The fluid-solid adhesion is significantly reduced from Teflon, but the droplet still lacks enough energy to escape the surface. However, increasing the incoming speed, Fig.~\ref{final}(C) shows a bouncing outcome on the nanoparticle-coated surface for droplet parameters that lead to sticking outcomes on Teflon. Notably, impacts at high $\!\mathit{We}$ result in an additional break up of the droplet with small satellite droplets as seen at \SI{0.03}{ms} in Fig.~\ref{final}(D--E). These additional effects are due to higher surface roughness, but despite this the fundamental bouncing criteria still hold. 
\section*{Conclusion}

In summary, we discovered that the weak surface adhesion of hydrophobic surfaces leads to a velocity-dependent stick-to-bounce transition in aqueous microdroplets. We have observed that the bouncing of microdroplets on a substrate occurs in a regime especially relevant for microdroplet processes governed by high inertia and small surface adhesion. In addition to the fundamental criterion for this transition, we have observed accompanying phenomenology such as the formation of sessile droplets during bouncing and the entrapment of air bubbles during deposition. 

Significantly, we predict a universal size limit for bouncing of water-based droplets, governed by the Ohnesorge number $\!\mathit{Oh}$. We find that for Teflon, bouncing is completely suppressed for droplets smaller than \SI{25}{\micro\meter}, and for less hydrophobic surfaces, we expect sticking for an even broader range of droplet sizes. In the context of aqueous bioaerosols, we expect that resuspension is suppressed and pathogen deposition is possible below this size limit. This observation may inform future studies on the relevant disease-transmission pathways for different droplet-size regimes.

More generally, these fundamental mechanisms have applications to natural and industrial processes where aerosols and microdroplet sprays interact with surfaces. For example, we predict that adding a hydrophilic coating will result in droplet deposition across a broad space of microdroplet parameters by suppressing the bouncing mechanism. We have identified a fundamental limit on the droplet velocity above which reliable inkjet printing on hydrophobic surfaces becomes impossible, a limit independent of the previously explored constraints imposed by the printing nozzle~\cite{Lohse2021}. Similarly, future 3D printing technologies that use Newtonian fluids will be affected by this speed limit, beyond which microdroplets bounce. Broadly, our research has implications across the vast application space that seeks maximum surface coverage and efficiency for industrial spray coating and crop spraying. 

%Bioaerosols do not resuspend on most surfaces.

\section{Materials and Methods}
\subsection{Experiments}
Microdroplet surface impact experiments were carried out using a MicroFab MJ-APB-01 \SI{30}{\micro\meter} piezoelectric droplet dispenser, which dispenses droplets onto a prepared substrate attached to a movable surface motor. A Photron FASTCAM NOVA S6, set at 100,000 frames per second (FPS), is used to image these droplets. This camera is attached to a Navitar Resolv4K zoom lens with a MoticPlan APO 20× objective and is backlit by a GSVitec MultiLED, see Ref.~\cite{McCarthy2022}.

For the initial substrates, glass microscope slides are cleaned with alcohol before a sheet of polytetrafluoroethylene (PTFE, i.e., Teflon) is placed on the glass. 
 The hydrophobic Teflon surface is measured to have a static contact angle of $108^\circ \pm 2^\circ$ with water, as verified by imaging sessile drops at the millimeter and micron scales. Full details of surface roughness and contact angles are provided in the SI.

Nanoparticle-coated surfaces were created by spraying SOFT99 Glaco Mirror Coat Zero onto cleaned glass slides for \SI{5}{\second}, creating a surface coated in silicon nanospheres. The average roughness of bare glass at \SI{40}{\micro\meter} is measured to be \SI{0.5}{\nano\meter}, and with the Teflon sheet, it is \SI{6}{\nano\meter}. The glass with the spray coat has a roughness of \SI{30}{\nano\meter} at \SI{30}{\micro\meter}.

In the impact experiments, we varied droplet impact velocity, viscosity, and size, affecting the $\!\mathit{We}$ and $\!\mathit{Oh}$ numbers. Impact velocity was adjusted by changing the height of the droplet generator relative to the substrate. Upon generation, pre-impact droplets oscillate and dampen within the first 0.5 milliseconds, reaching a stable spherical morphology. A minimum impact height is necessary to prevent these oscillations from affecting the impact dynamics; this was chosen as the distance of travel (DOT) during 0.5 milliseconds. Increasing the height beyond this minimum resulted in a decrease in impact velocity due to drag --an inverted height--velocity relationship compared to a freely falling droplet. The input voltage on the piezoelectric generator controlled the droplet size, resulting in two size modes of approximately 30 and \SI{50}{\micro\meter}, with the latter being the most common size for our impacts. Droplet velocities and sizes are analyzed using image analysis code, allowing measurement of the impact $\!\mathit{We}$ and $\!\mathit{Oh}$ numbers. Uncertainties in $\!\mathit{We}$ and $\!\mathit{Oh}$ are propagated from the spatial and temporal resolution limits.

Deionized water was initially used for impacts. As laboratory temperature varies daily, the fluid is left to reach equilibrium before use, after which its temperature is measured to ensure room temperature. The fluid and room temperature ranged from \SI{20}{\celsius} to \SI{26}{\celsius}, leading to variations in dynamic viscosity from \SI{1.00}{\milli\pascal\second} to \SI{0.89}{\milli\pascal\second}. For some runs, 5$\%$ v/v glycerol is added to increase the viscosity to \SI{1.11}{\milli\pascal\second}. These changes in viscosity alter the surface tension by less than 2$\%$, a variation deemed negligible in terms of contact angle changes. Relative humidity (RH) is measured and maintained within 10$\%$ across all runs.

Impacts at a DOT of \SI{1}{\milli\second} and \SI{20}{\celsius} result in droplets spreading, oscillating, and sticking to the surface. Experiments are then conducted down to \SI{0.5}{\milli\second} DOT and up to an environmental temperature of \SI{26}{\celsius}. The secondary surface experiments followed the same methodology, using only water at \SI{20}{\celsius}. Control impacts were also conducted on glass, which resulted in droplet deposition outcomes only.

\subsection{Simulations}

We make use of a phase-field method to model the two-phase droplet system. This approach introduces a phase-field variable, $\phi$, which distinguishes between the liquid ($\phi = 1$) and the gas ($\phi = -1$) phases. The phases are initialized as in Fig.~\ref{Fig2}. The Cahn-Hilliard equation governs the time evolution of the phase field $\phi$, which ensures a smooth evolution of the interfaces in the system:
\begin{equation}
\frac{\partial \phi}{\partial t} + \mathbf{u} \cdot \nabla \phi = \nabla \cdot  \chi  \lambda \nabla \psi. 
\end{equation}
Here $\mathbf{u}$ represents the velocity field and $\chi$ the interface mobility tuning parameter set as \SI{1}{\meter\per\second\per\kilogram}. The helper function, $\psi$, is defined as:
\begin{equation}
\psi = -\nabla \cdot \left( \epsilon_{\text{pf}}^2 \nabla \phi \right) +  \phi^2 - 1  \phi, 
\end{equation}
where $\epsilon_{\text{pf}}$ is the interface thickness, which we define by the max and min mesh elements $h_{\text{max}}$ and $h_{\text{max}}$: 
\begin{equation} 
 \epsilon_{\text{pf}} = \begin{cases} h_{\text{max}}, & \text{if } h_{\text{max}} > 1.3 \cdot h_{\text{min}} \\ 2 \cdot h_{\text{max}}, & \text{if } h_{\text{max}} \leq 1.3 \cdot h_{\text{min}} \end{cases} 
\end{equation} 
The parameter $\lambda$ is defined to be:
\begin{equation}
\lambda = \frac{3 \epsilon_{\text{pf}} \gamma}{\sqrt{8}},
\end{equation}
where $\gamma$ is the surface tension of the fluid-air system. This formulation accurately tracks the droplet-air interface and ensures a smooth transition between the liquid and gas phases.

To model the contact angle boundary condition, which describes the angle at which the liquid phase meets the solid substrate, we impose the following boundary condition:
\begin{equation}
    \mathbf{n} \cdot \chi \lambda \nabla\phi = 0,
\end{equation}
which equation ensures that the gradient of the phase field is consistent with the mobility at the interface, and 
\begin{equation}
  \mathbf{n} \cdot\epsilon_{\text{pf} }^2  \nabla\phi = \epsilon_{\text{pf} }^2 \cos (\theta) |{\nabla\phi}|,
\end{equation}
which enforces the correct relationship between the interface orientation and the contact angle. 
The effect of the phase field $\phi$ evolution is incorporated into the Navier-Stokes equations via the surface tension term $\textbf{F}_{\text{ST}}$, which couples to fluid flow:
\begin{equation}
    \textbf{F}_{\text{ST}} =  \frac{\lambda}{\epsilon_{\text{pf}}^2} \psi  \nabla \phi + \left( \frac{|\nabla \phi |^2}{2} + \frac{ (\phi^2 - 1)^2}{4\epsilon_{\text{pf}}^2}\right) \nabla \lambda - (\nabla \lambda \cdot \nabla \phi ) \nabla \phi
\end{equation}

\section{acknowledgments}
The authors gratefully acknowledge EPSRC (Grant No. EP/S023593/1) and the University of Bath Department of Chemistry for funding, Sara Dale for useful discussions and for help with AFM measurements of the nanoparticle-coated surface, as well as Lukesh Mahato, Lauren McCarthy, and Aditya Jha for fruitful discussions.

\clearpage
\onecolumngrid 
\appendix*
\section{\Large Supplementary Information}

\section{Experimental Results}
In our work we present experimental data on microdroplet impingements, demonstrating a stick-to-bounce transition dependent on the Weber ($\!\mathit{We}$) and Ohnesorge ($\!\mathit{Oh}$) numbers. In this supplementary information, we provide additional experimental results and analysis, including explicit surface characterization. We present surface roughness data and dynamic contact angle measurements for all surfaces involved. We also consider bounce contact time, showing that it scales with the inertiocapillary timescale. Furthermore, we examine spreading behavior and demonstrate that it is independent of $\!\mathit{Oh}$. Additionally, we analyze droplet oscillations following a sticking impact, showing that the system is underdamped. Finally, we derive a version of the coefficient of restitution relevant for partial rebounds and calculate its value across the stick-to-bounce transition. 

\subsection{Surface Roughness and Contact Angles}

This section analyzes all experimental surfaces using two key measurements: surface roughness and dynamic contact angle with water. Table ~\ref{Tab1} presents Atomic Force Microscopy (AFM) measurements of surface roughness parameters—root mean square roughness (Rq), arithmetic mean roughness (Ra), and peak-to-valley roughness (Rh)—for Teflon, untreated glass, and nanoparticle-coated glass at different scales. The data shows that nanoparticle-coated glass exhibits significantly higher roughness than untreated glass and Teflon, particularly at larger scales. The Teflon surface exhibits only a slightly higher roughness than the untreated glass. 
\begin{table}[ht]
    \centering
    \caption{Roughness Measurements Comparison by Surface}
    \begin{tabular}{lrrr}
        \toprule
        Substrate and Scale & Rq (\SI{}{nm}) & Ra (\SI{}{nm}) & Rh (\SI{}{nm}) \\
        \midrule
        Glass, \SI{1}{\micro\meter} & 0.491 & 0.322 & 5.61 \\
        Glass, \SI{40}{\micro\meter} & 1.53 & 0.567 & 97.9 \\
        Coated Glass, \SI{10}{\micro\meter} & 19.6 & 15.9 & 95.8 \\
        Coated Glass, \SI{30}{\micro\meter} & 34.5 & 27.8 & 198 \\
        Teflon, \SI{1}{\micro\meter} & 2.97 & 2.40 & 22.4 \\
        Teflon, \SI{40}{\micro\meter} & 6.31 & 5.01 & 87.5 \\ 
        \bottomrule
    \end{tabular}
    \label{Tab1}
\end{table}

The static contact angle between a sessile water droplet and the surface was measured both at the micron scale using the high frame rate camera and with a \SI{1}{\milli\meter} droplet using an Ossila Contact Angle Goniometer.
We were able to measure the advancing $\theta_\mathrm{a}$ and receding $\theta_\mathrm{r}$ contact angles for the mm-scale droplets, but not for the micron-scale droplets where we were limited by the camera frame rate. For the 1 mm droplet, deionised water was slowly dispensed from a needle to measure an advancing contact angle and then drawn back in to measure the receding contact angle. The results of these measurements are presented in Table~\ref{Tab2}. The Teflon surface showed consistent static contact angle values between the micron and millimeter scales. The advancing and receding contact angles for Teflon were in line with literature values with a hysteresis of approximately 19°~\cite{Kamuse85}.
\begin{table}[ht]
    \centering
    \caption{Contact Angle Measurements}
    \begin{tabular}{lcccc}
        \toprule
        Substrate & Droplet Size (\SI{}{\milli\meter}) & Static Angle (°) & Advancing (°) & Receding (°) \\
        \midrule
        Teflon & 0.05 & 108 ± 2 & N/A & N/A \\
        Teflon & 1.00 & 107 ± 2 & 111 ± 5 & 93 ± 3 \\
        Coated Glass & 0.05 & 139 ± 5 & N/A & N/A \\
        Coated Glass & 1.00 & 147 ± 2 & 163 ± 5 & 121 ± 4 \\
        \bottomrule
    \end{tabular}
    \label{Tab2}
\end{table}

The coated glass exhibited inconsistent static angles between the millimeter and micron scales, likely due to variations in roughness across different length scales, as shown in Table~\ref{Tab1}. A large hysteresis $\theta_\mathrm{a} - \theta_\mathrm{r}$ of 42° was measured at the millimeter scale, consistent with a rougher surface.  
\subsection{Contact Time}
\begin{figure}[t]
    \centering
    \includegraphics[width=\linewidth]{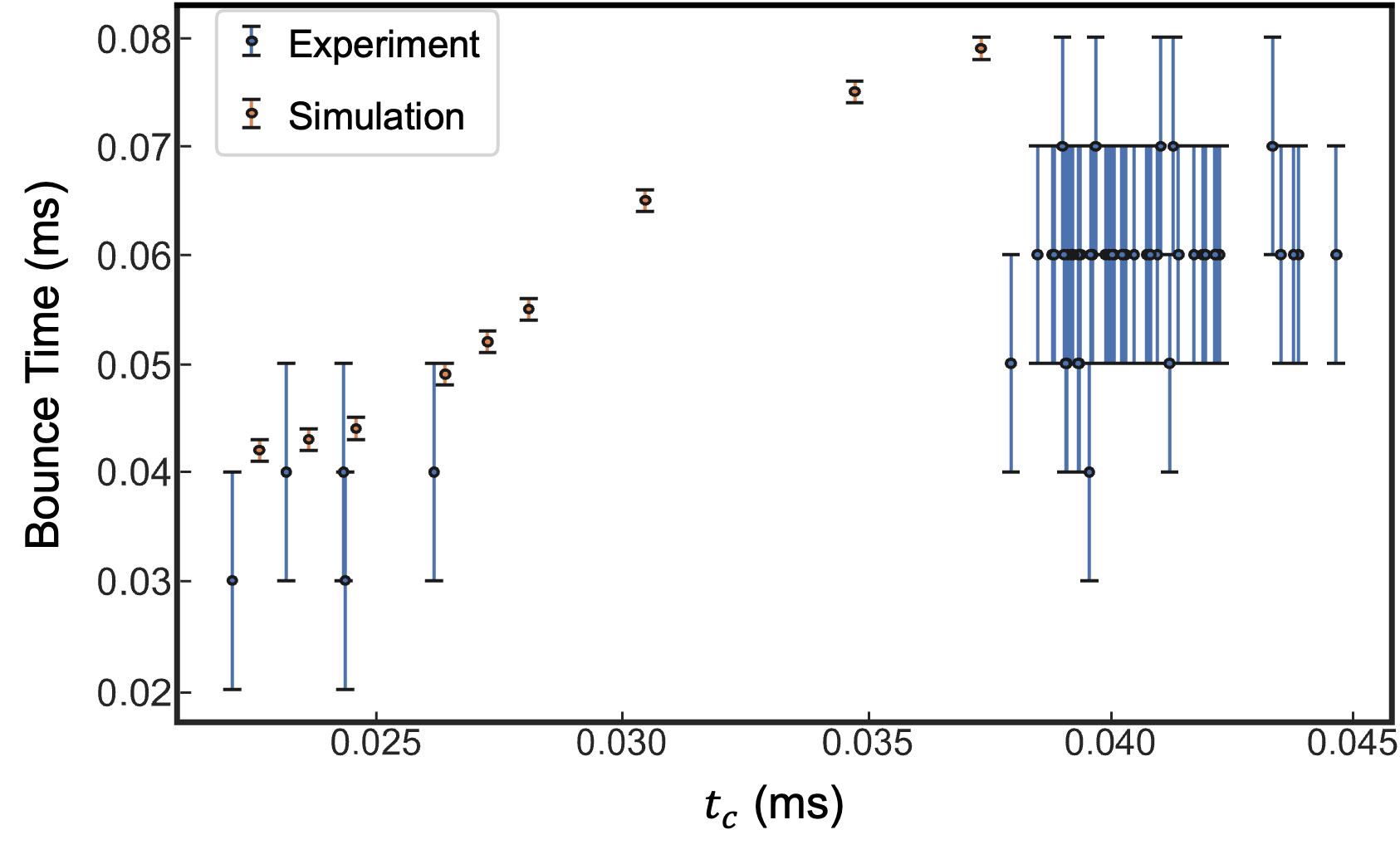}
    \caption{Measured time for a microdroplet to bounce versus the inertiocapillary time scale, $t_c$, for both experimental and numerical impacts. Measured bounce time is the time the majority of a droplet is in contact with the surface, this measurement is limited by the temporal resolution of \SI{0.01}{\milli \second} for experiments and \SI{0.001}{\milli \second} for the simulations, which define  the error bars. The graph shows that partial rebound events take place on approximately the same timescale as $t_c$.}
    \label{Fig3k}
\end{figure}
We measured the contact time during a bounce and compared with previous results. Ref.~\cite{Okumura2003} used a spring model to derive the bounce time of  a droplet to be the inertiocapillary time $t_c$: 
\begin{equation}
    t_c \approx \sqrt{\frac{\pi \rho D^3}{6 \gamma}}
\end{equation} 
 Bounce time vs $t_c$ is graphed for several experiments and numerical results in Fig. ~\ref{Fig3k}. Most experiments with bouncing occurred with droplets of \SI{50}{\micro\meter} in diameter, but \SI{30}{\micro\meter} did show a shorter contact time, as expected from the above expression for $t_c$. There was a factor of approximately two between the experimental bounce time and the inertiocapillary timescale. In simulations, we varied the fluid density to observe different impacts with different inertiocapillary times, which were all consistent with the predicted scaling. The deviations from linearity in Fig.~\ref{Fig3k} is attributed to the transition from total rebound at low densities to partial rebound at higher densities, where surface interactions increasingly influence the dynamics.

\subsection{Droplet Spreading}
\begin{figure}[t]
    \centering
    \includegraphics[width=\linewidth]{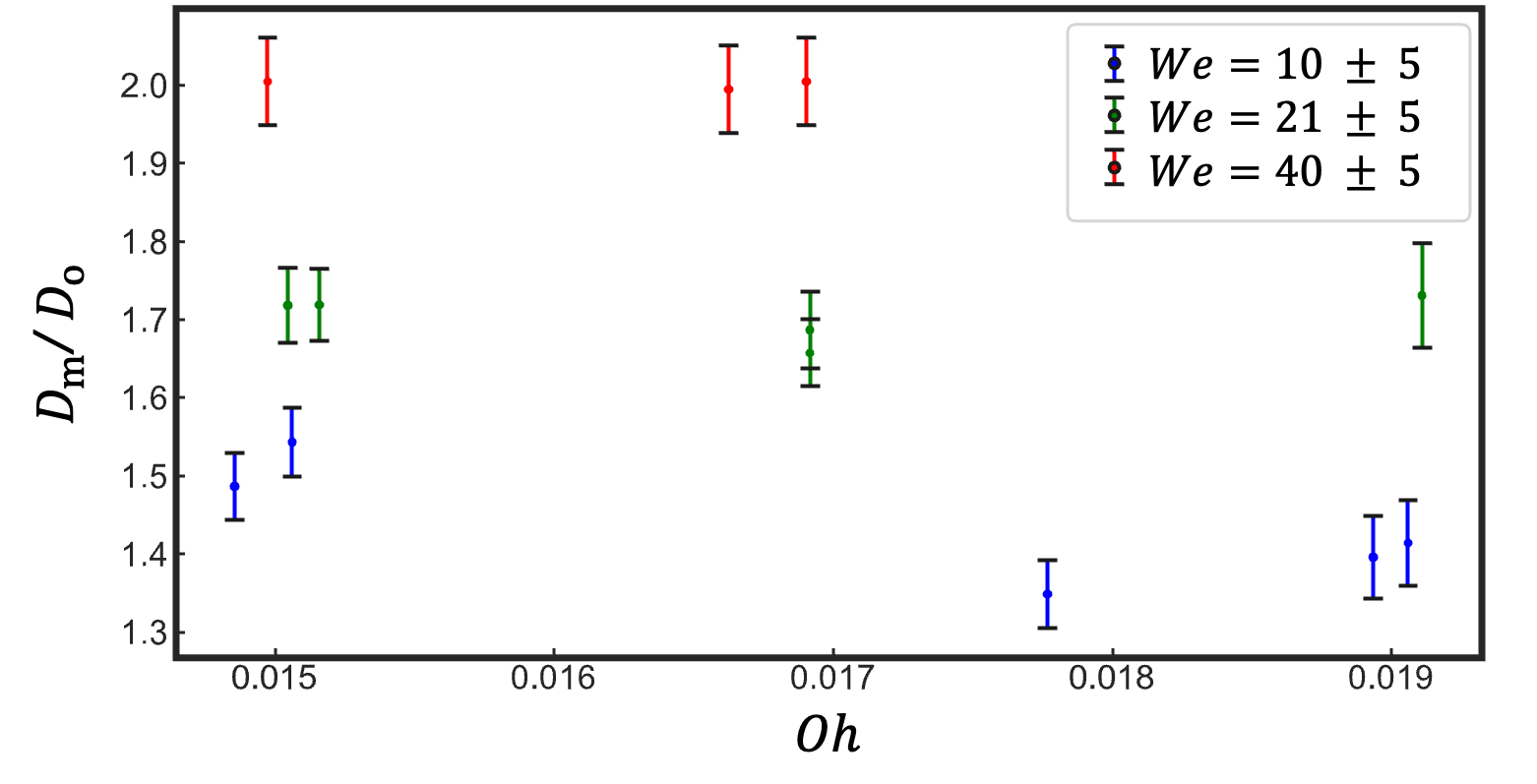}
    \caption{{
    Maximum spread of an impacting droplet (diameter $D_\textrm{m}$)  divided by the original droplet diameter $D_\textrm{o}$ versus $\!\mathit{Oh}$ for several fixed $\!\mathit{We}$ impacts on a Teflon surface. This graph shows that maximum spreading is independent of $\!\mathit{Oh}$. This graph is the companion to Fig.~5. in the main text, where we show spreading is $\!\mathit{We}$ dependant. The data points are from the impacts of both bouncing and non-bouncing events.} }
    \label{Spread2}
\end{figure}
We measure the maximum spread of a droplet during impact and compare it to past literature. We find good agreement with literature values~\cite{BOSSA_2011,Bruin2014,sanjay2024}, which provides a check that any surface contamination does not affect the droplet dynamics. We take the maximum spread as the maximum horizontal distance the droplet covers on the surface. 

\begin{equation}
\frac{D_{\mathrm{m}}}{D} = g(\theta)( 1 + C \!\mathit{We}^{1/2}).
\end{equation}
his expression was previously derived using superhydrophobic surfaces; in the hydrophobic case, we assume surface effects can be captured in the factor $g(\theta)$. This fit validates the spreading of the droplets and allows this expression of maximum spreading to be used for the energy criterion. 

Fig.~\ref{Spread2} shows a similar graph but for $\!\mathit{Oh}$ at fixed $\!\mathit{We}$. The graph shows that spreading is independent of $\!\mathit{Oh}$ in the low-viscosity microdroplet regime. These results confirm that the microdroplets' initial spreading behaviour is driven by inertia and counteracted by capillary forces. Hence we conclude that dissipation is minimal in the spreading phase.

\subsection{Droplet Oscillations}

\begin{figure}[t]
    \centering
    \includegraphics[width=\linewidth]{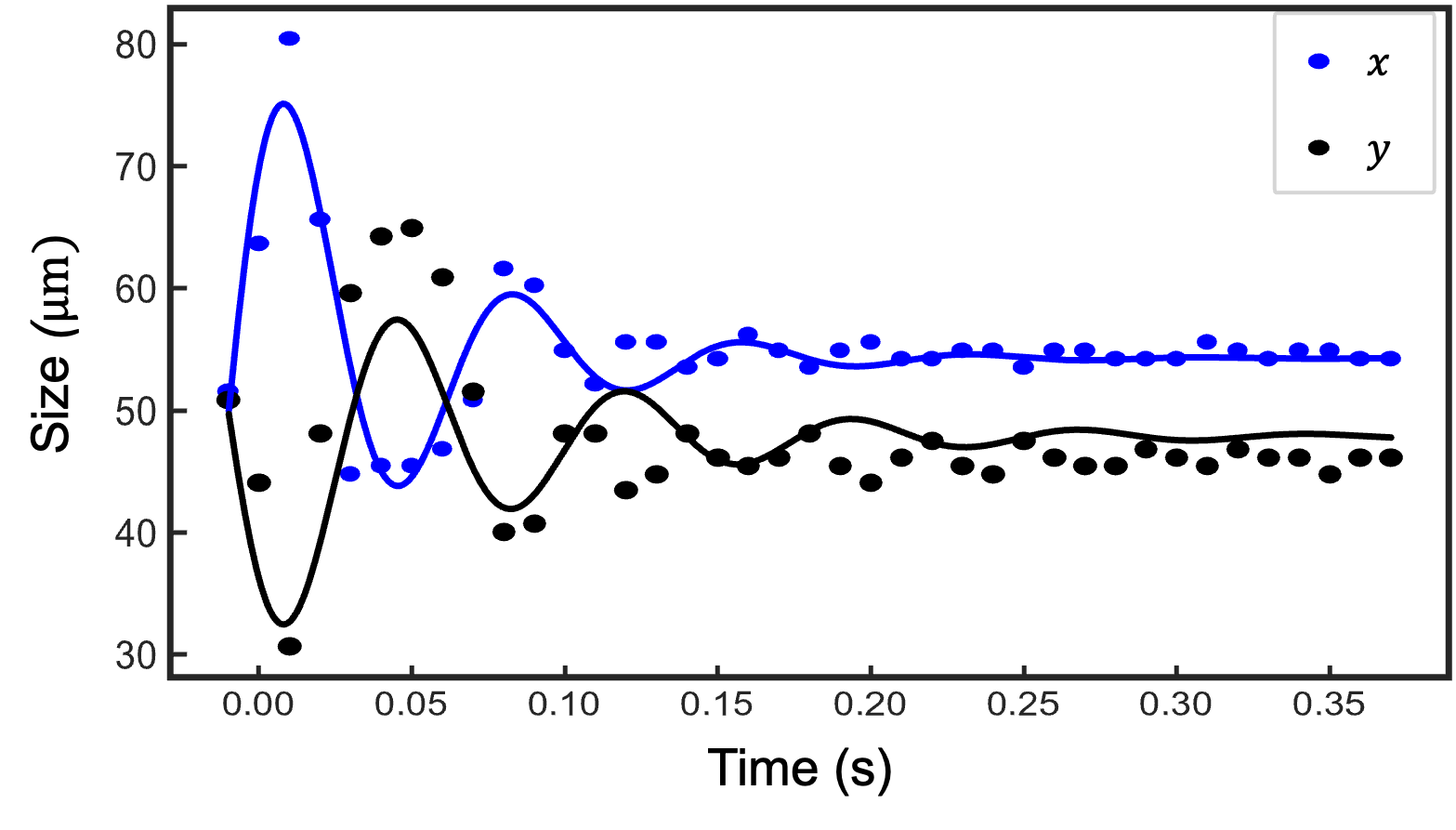}
    \caption{Droplet oscillations graphed for a near bounce incident of ($\!\mathit{We}$, $\!\mathit{Oh}$) = (12, 0.015) on a Teflon surface. The graph shows x (height) and y (width) lengths of the droplet graphed against time. Error bars are neglected to make trends clearer. The lines represents fits for a single-mode decaying sine wave. The graph shows the underdamped nature of the sticking events we consider.}
    \label{Fig5}
\end{figure}
We briefly analyze the oscillations in near-bouncing incidents to verify that the microdroplet system is underdamped. Upon impact in sticking events, the droplet oscillates over time. We measure this oscillation using the droplet's $x$ and $y$ length over time, see Fig.~\ref{Fig5}. Here, the droplet oscillations last an order of magnitude longer than the spreading process, indicating the system is underdamped. As $\!\mathit{Oh}$ increases, the oscillation time decreases. The fit in Fig.~\ref{Fig5} is a single-mode decaying sinusoidal function, which does not fully capture the multimodal oscillations which are induced by the surface. Further details of oscillations in microdroplet systems can be found in Ref.~\cite{McCarthy2022}, which analyzes post-impact oscillations on several surface types. This is in contract to the overdamped regime in which the droplets do not oscillate after impact. Oscillations are always seen experimentally along the stick-to-bounce boundary for microdroplets in this work, indicating this transition occurs in the underdamped regime, which highlights the role of surface adhesion in sticking.

\subsection{Coefficient of Restitution for a Partial Rebound Event}

\begin{figure}[t]
    \centering
    \includegraphics[width=\linewidth]{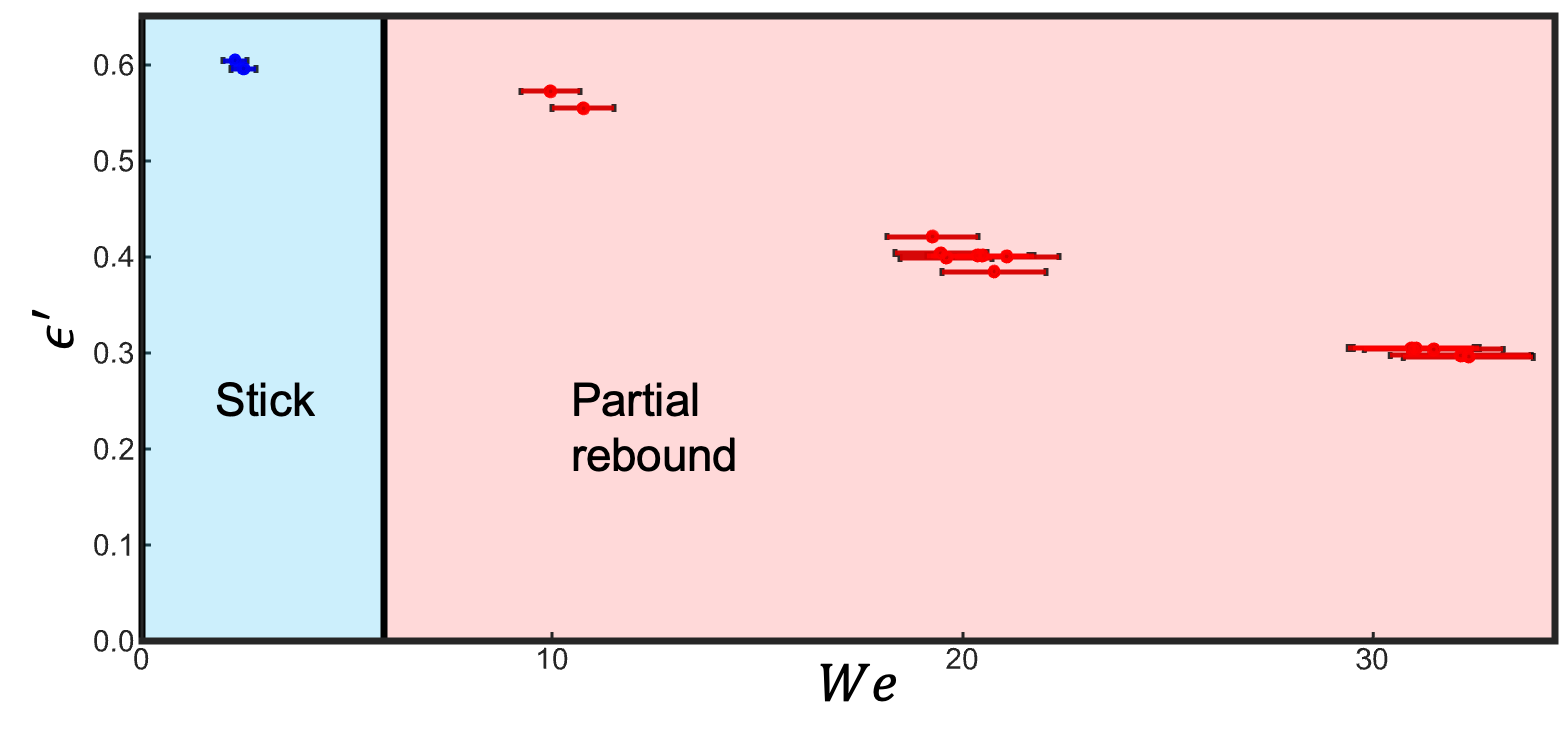}
    \caption{Graph of the altered coefficient of restitution $\epsilon'$ vs $\!\mathit{We}$ for all impacts on the nanoparticle coated surface. The coefficient $\epsilon'$ measures the energy remaining in the system after impact and decreases as $\!\mathit{We}$ increases across the stick-to-bounce transition, showing greater dissipation at higher $\!\mathit{We}$. Error bars are from the uncertainties in size resolution and velocity. }
    \label{COR}
\end{figure}

 A quantity used to measure energy dissipated in droplet bouncing is the coefficient of restitution. In this section, we discuss a form of the coefficient of restitution that also considers the change in surface energy during a partial rebound and can be used on both sides of the stick-to-bounce transition. We show an example of this for the nanoparticle coated surface.

 The coefficient of restitution is the ratio of a droplet's incoming to outgoing velocity, which quantifies how much momentum and kinetic energy is retained:
\begin{equation}
    \epsilon = \frac{u_{f}}{u_{o}}
\end{equation}

In a partial rebound, in addition to viscous loss, there is a change in the interfacial energy. To account for this, we define a ratio involving both kinetic and surface energies. This is the ratio of the final kinetic and surface energy of the upward primary droplet ($D_o$) and the surface energy of the sessile drop ($D_s$) to the initial droplet's kinetic and surface energies. 
\begin{equation}
    \epsilon' = \frac{ \frac{\pi}{12} \rho D_o^3 u_f^2  + \pi D_o^2 \gamma_{FS}  + \frac{\pi}{4} D_s^2 ( \gamma_{FS} - \gamma_{SA}) + \frac{\pi}{2} D_s^2 \gamma_{FA} (1 - \cos(\theta)) }{  \frac{\pi}{12} \rho D^3 u_o^2  + \pi D^2 \gamma_{FA}  }.
\end{equation}
Here $\frac{\pi}{12} \rho D_o^3 u_f^2$ is the upward kinetic energy of the rebounding droplet; $\pi D_o^2 \gamma_{FS}$ is the surface energy of the rebounding droplet; $\frac{\pi}{4} D_s^2 ( \gamma_{FS} - \gamma_{SA})$ is the surface energy of the new fluid-solid interface for the sessile droplet minus the energy of the previous solid-air interface; $\frac{\pi}{2} D_s^2 \gamma_{FA} (1 - \cos(\theta))$ is the fluid-air interface for the sessile drop; $\frac{\pi}{12} \rho D^3 u_o^2$ is the initial kinetic energy of the impacting drop; and $\pi D^2 \gamma_{FA}$ is the surface energy for the impacting droplet. Writing all interfacial energies in terms of surface tension from Young's equation and simplifying gives: 
\begin{equation}
    \epsilon' = \frac{ \frac{\pi}{12} \rho D_o^3 u_f^2  + \pi D_o^2 \sigma + \frac{\pi}{4} D_s^2 \sigma (2 - 3\cos(\theta)) }{  \frac{\pi}{12} \rho D^3 u_o^2  + \pi D^2 \sigma  }.
\end{equation}
Dividing through by $ \pi D^2 \sigma $.
\begin{equation}
    \epsilon' = \frac{ \frac{1}{12} \frac{\rho D_o^3 u_f^2}{D^2 \sigma} + \frac{D_o^2}{D^2} + \frac{1}{4} \frac{D_s^2}{D^2} (2 - 3\cos(\theta)) }{ \frac{1}{12} \frac{\rho D^3 u_o^2}{D^2 \sigma} + 1 }
\end{equation}
Here we define the ratio of droplet sizes 
\begin{equation}
    R_o = D_o^2 / D^2,
\end{equation}
\begin{equation}
    R_s = D_s^2 / D^2,
\end{equation}
and introduce the $\!\mathit{We}$ number: 
\begin{equation}
    \epsilon' = \frac{ \frac{1}{12} \mathit{We} R_o^{3/2} \frac{u_f^2}{u_o^2} +  R_o + \frac{1}{4} R_s (2 - 3\cos(\theta)) }{  \frac{1}{12} \mathit{We} + 1 }.
\end{equation}
Simplifying, we find
\begin{equation}
    \epsilon' = \frac{   R_o ( 12 + \mathit{We} R_o^{1/2} \epsilon^2 ) + 6R_s (1 - \frac{3}{2} \cos(\theta) )        }{\mathit{We} + 12}.
\end{equation}
This altered coefficient of restitution can be used to estimate the change in energy during a bounce event. In the case of a total rebound, $R_s = 0$ and $R_o = 1$, so that
\begin{equation}
    \epsilon' = \frac{      \epsilon^2 \mathit{We}  +12     }{\mathit{We} + 12}
\end{equation}
Which scales with usual coefficient of restitution $\epsilon$ but is not equivalent. For a sticking event, $R_o = 0$ and
\begin{equation}
    \epsilon' = \frac{ 6 R_s (1 - \frac{3}{2} \cos(\theta))}{\mathit{We} + 12}.
\end{equation}
We plot $\epsilon'$ for the nanoparticle coated surface impacts. Fig.~\ref{COR} shows that as $\!\mathit{We}$ increases, $\epsilon'$ decreases, even across the stick-to-bounce transition, highlighting a greater $\%$ of energy dissipated at higher $\!\mathit{We}$.

\section{Numerical Results}
\begin{figure}[t]
    \centering
    \includegraphics[width=\linewidth]{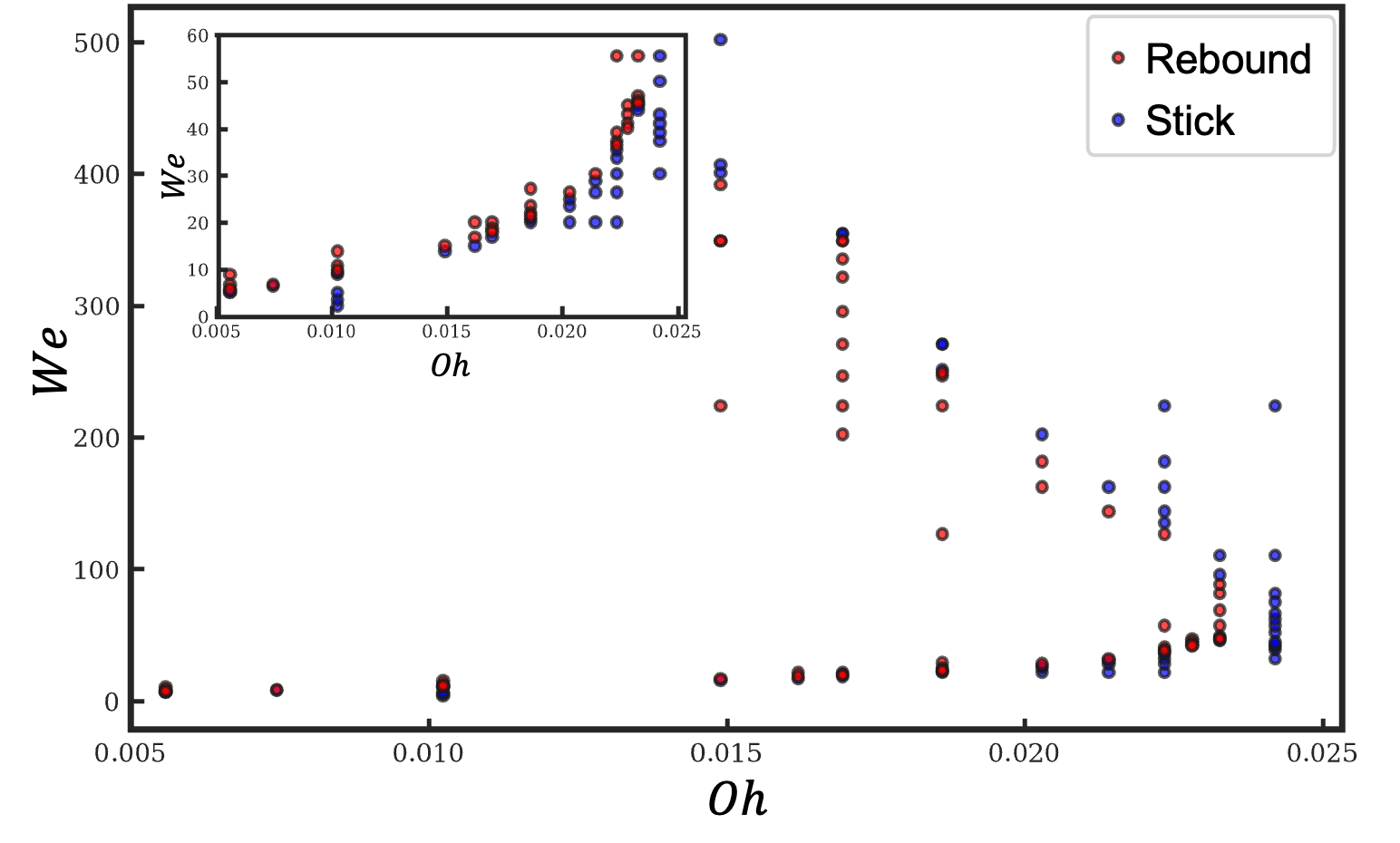}
    \caption{All droplet simulation outcomes used to plot Fig.~2(B) in the main text. Each point corresponds to a simulation result in ($\!\mathit{We}$, $\!\mathit{Oh}$) space for a surface with a contact angle of 110$^\circ$. A bounce event is defined as where most of the fluid is not in contact with the surface for at least one time-step. The simulation input values ($\!\mathit{We}$, $\!\mathit{Oh}$) are chosen to as fill the parameter space as much as possible.}
    \label{Sim}
\end{figure}
In this section, we outline further details of the simulation that allow us to make the transition plots in the main body of the paper; the numerical details are in the materials and methods section of the main text. 

We carry out 120 simulations for a contact angle of 110$^\circ$ for various combinations of the parameters $\!\mathit{We}$ and $\!\mathit{Oh}$. Although for almost all numerical simulations, we neglected hysteresis, we performed some simulations varying the advancing contact angle between 110$^\circ$ and 140$^\circ$ while keeping the receding contact angle fixed, and found that the advancing contact angle values did not alter the bouncing or sticking outcome. 

As shown in the main text, increasing the $\!\mathit{We}$ number for a set $\!\mathit{Oh}$ numbers led to a transition of microdroplets from sticking to bouncing at sufficiently low  $\!\mathit{Oh}$. To accurately pinpoint the transition, at fixed $\!\mathit{Oh}$ number, we vary the $\!\mathit{We}$ number in increasingly smaller intervals until both a sticking and a bouncing simulation took place with a difference of $\mathit{We}$ numbers less than 1. We then took the transition point as the midpoint of these two simulations. All simulations for a receding contact angle of 110$^\circ$ are seen in Fig.~\ref{Sim}. The lower bounds of this transition were connected by a smooth line that joins the points to give the transition line in Fig.~3(A) of the paper. The upper transition line was also included in Fig.~3(B) for completeness. 

The simulation was repeated for 100 values of a contact angle of 120$^\circ$, with the transition connected smoothly. Several additional simulations were performed at contact angles between 60$^\circ$ and 180$^\circ$. Bouncing did not occur at any hydrophilic angle, and bouncing remained $\!\mathit{We}$-independent in the superhydrophobic limit, consistent with previous results from literature.

\section{Reynolds Number Phase Diagrams}
\begin{figure}[t]
    \centering
    \includegraphics[width=\linewidth]{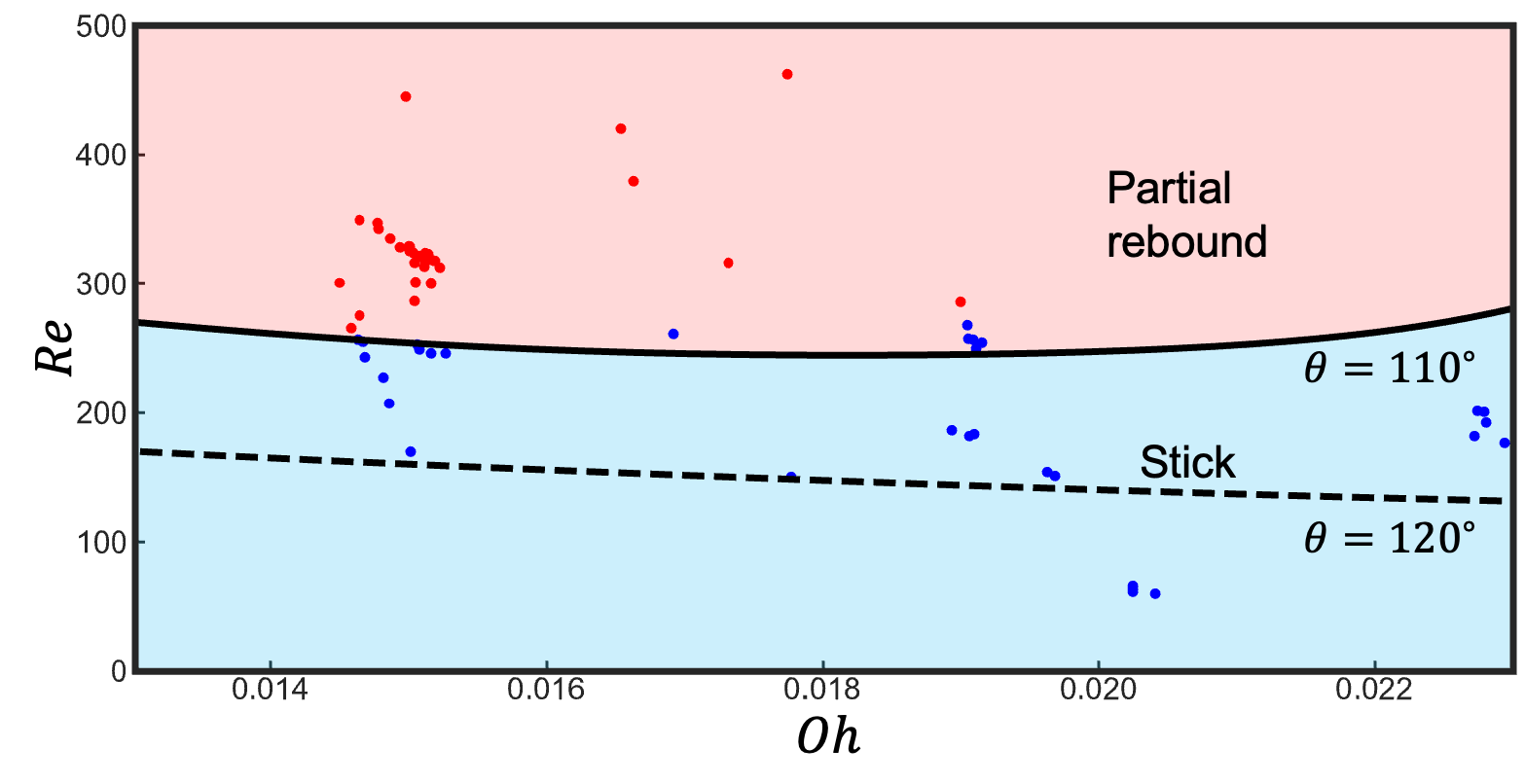}
    \caption{A rescaling of Fig.~3(A) from the main text in terms of the $\mathit{Re}$ vs $\!\mathit{Oh}$ parameter space with both experimental points and numerical data from~\ref{Sim}, and for a surface of contact angle of 120$^\circ$. Fig.~3(A) shows a transition from bouncing to sticking at an increasing $\mathit{We}$ as $\!\mathit{Oh}$ increases. However, this transition line is not monotonic in $\mathit{Re}$ vs $\!\mathit{Oh}$ space, and the minimum $\mathit{Re}$ for the transiition occurs at a non-zero $\!\mathit{Oh}$. This non-monotonicity suggests that the transition line is simpler to model in $\mathit{We}$ vs $\!\mathit{Oh}$ parameter space. Error estimates are the same as Fig.~3(A).}
    \label{figRe}
\end{figure}
In this section, we re-express Fig.~3(A) from the main text into a $\mathit{Re}$ vs $\!\mathit{Oh}$ parameter space instead of the $\mathit{We}$ vs $\!\mathit{Oh}$ space. This rescaling provides an alternative perspective on the system since $\mathit{Re}$ is a commonly used dimensionless number to describe the role of inertia. In the main text, we choose the $\!\mathit{We}$ vs $\!\mathit{Oh}$ space because it combines a velocity-independent variable ($\!\mathit{Oh}$) with one that is independent of viscosity ($\!\mathit{We}$), making it well suited to capture the transition dynamics.

The transition can be represented in a $\mathit{Re}$ vs $\!\mathit{Oh}$ plot, as shown in Fig.~~\ref{figRe}. The $\mathit{Re}$ number is a capillary-independent composite of $\!\mathit{We}$ and $\!\mathit{Oh}$, surface-tension effects are captured within $\!\mathit{Oh}$ in this representation. However, this form is intuitively more challenging to interpret. The lowest $\mathit{Re}$ at which bouncing occurs is approximately 280 for a receding angle of 110$^\circ$. This corresponds to a plateau in the graph, where, at lower $\!\mathit{Oh}$, a higher $\mathit{Re}$ indicates that for a lower viscosity, inertia dominates over viscous effects. The minimum $\mathit{Re}$ required for bouncing shifts to higher $\mathit{Re}$ as $\!\mathit{Oh}$ decreases. Then, at high $\!\mathit{Oh}$, an even higher $\mathit{Re}$ is needed for the same energy to remain in the system. The combination of viscous and inertial effects that enters $\mathit{Re}$ makes the phase-space results more difficult to interpret, which is why we choose the $(\!\mathit{Oh},\!\mathit{We})$ phase space for plotting results and for the energy-balance argument.

\section{Droplet Impact Energy Considerations}

In this section, we expand upon the discussion of the energy balance of a microdroplet that impacts a surface for the specific case of a partial rebound. We derive a condition to rebound based on the postulation that there must be remaining energy after an impact, expressed as
\begin{equation}
E_{\mathrm{k,f}} = E_{\mathrm{k,0}} - E_\mathrm{\gamma} - E_\mathrm{\mu}.
\end{equation}
To approach this derivation in more depth, we can discuss each term separately.
\subsection{Surface Energy Term}
The surface energy term in the main text $E_\mathrm{\gamma}$ is not explicitly computed. Here, we write a specific expression from the change in interfacial energy during a partial rebound. The initial droplet splits into a primary upward rebounding drop and a smaller sessile part deposited on the surface. From the conservation of volume, the sizes of these interfaces are linked:
\begin{equation}
     D^3 = D_\mathrm{o}^3 + \frac{1}{4} D_\mathrm{s}^3 ( 2 + \cos(\theta)) (1 - \cos(\theta))^2,
\end{equation}
where $D_\mathrm{o}$ is the rebounding droplet diameter and $D_\mathrm{s}$ is the spread of secondary droplet on the surface. We use the change in interfacial energy from pre- to post-impact as an approximation of $E_\mathrm{\gamma}$, the kinetic energy converted into extra surface energy. Then,
\begin{equation}
E_\mathrm{\gamma} = \pi D_o^2 \gamma  + \frac{\pi}{4}D_s^2 \gamma_{FS}  + \frac{\pi}{2}D_s^2 \gamma(1 - \cos(\theta)) - \pi D^2 \gamma - \frac{\pi}{4}D_s^2 \gamma_{SA}.
\end{equation}
This expression considers all interfaces between fluid, solid, and air and can be simplified using Young's equation linking interfacial energies to contact angle:
\begin{equation}
    \gamma_{SA} - \gamma_{FS} = \gamma_{FA} \cos(\theta) = \gamma \cos(\theta).
\end{equation}
Substituting, we find:
\begin{equation}
E_\mathrm{\gamma} = \pi \gamma ( D_o^2 - D^2 + \frac{\pi}{4}D_s^2 (2 - 3 \cos(\theta)), 
\end{equation}
which can be expressed in the general case as
\begin{equation}
E_\mathrm{\gamma} = \pi \gamma D^2 \left( \frac{D_o^2}{D^2} - 1 + \frac{\pi}{4} \frac{D_s^2}{D^2} (2 - 3 \cos(\theta))\right) \approx \pi \gamma D^2 f(\theta).
\end{equation}
This expression approximates the extra energy the interfaces now store at the point of rebounding where a fluid-solid interface is present. We use an experimental example to estimate this energy scale quantitatively. A bouncing droplet of diameter $D \approx $ \SI{50}{\micro\meter} is observed to have a sessile drop of diameter $D_s \approx $ \SI{8}{\micro\meter} on Teflon. From volume conservation this makes diameter $D_o \approx $ \SI{49.8}{\micro\meter}, and $E_\mathrm{\gamma} \approx$ \SI{10e-11}{\joule}, meaning around $5\%$ extra energy is stored in the interfaces. This approximation represents the energy stored in the interfaces during the point of the necking instability, where a fluid-solid interface remains.

\subsection{Dissipation Term}
\subsection{Dissipation Term}
In the main body of the paper, we discuss the form of the dissipated energy $E_\mathrm{\mu}$:\begin{equation}
 E_\mathrm{\mu} = E_\mathrm{\mu,3D} + E_\mathrm{\mu,2D} + E_\mathrm{\mu,1D}.
\end{equation}
The bulk dissipation and boundary layer dissipation both scale as 
\begin{equation}
E_\mathrm{\mu,3D} + E_\mathrm{\mu,2D} = \kappa \mu u_0^2 D t_c.
\end{equation}
The contact time we use here is the inertiocapillary time, which we show in the main text is approximately  the bounce time, that is, 
\begin{equation}
    t_c \approx \sqrt{\frac{ \rho D^3}{ \gamma}},
\end{equation} 
where any constant terms are absorbed into $\kappa$, so the dissipation term can be written in terms of a viscosity-dependant part as:
\begin{equation}
 E_\mathrm{\mu} =  \kappa \mu u_0^2 D^{5/2} \rho^{1/2} \gamma^{-1/2} + E_\mathrm{\mu,1D}
\end{equation}
The $E_\mathrm{\mu,1D}$ term is from friction at the contact line as the fluid pins and depins from the surface. We write the frictional force per unit length as 
\begin{equation}
    F_f = \gamma ( \cos (\theta_r) - \cos(\theta_a))
\end{equation}
and we write the full frictional energy loss as
\begin{equation}
    E_\mathrm{\mu,1D} = \pi \int_{0}^{D_{m}} r \gamma ( \cos (\theta_r) - \cos(\theta_a)) dr = \frac{\pi \gamma}{2} D_{m}^2 ( \cos (\theta_r) - \cos(\theta_a)).
\end{equation}
We show in the main text that in this regime, microdroplets maximum spread diameter $D_{\mathrm{m}}$ scales with the Weber number, 
\begin{equation}
   \frac{D_{\mathrm{m}}}{D} = g(\theta)( 1 + C \!\mathit{We}^{1/2}).
\label{eq:dmax1}
\end{equation}
Using this scaling, we write the contact line dissipation as
\begin{equation}
    E_\mathrm{\mu,1D} = D^2 \gamma \!\mathit g^2(\theta) \Delta \cos(\theta)( 1 + C \!\mathit{We}^{1/2})^{2} = h(\theta,\Delta \theta) D^2 \gamma ( 1 + C \!\mathit{We}^{1/2})^{2}. 
\end{equation}
Combining these contributions, we write the total dissipated energy as
\begin{equation}
 E_\mathrm{\mu} = \kappa \mu u_0^2 D^{5/2} \rho^{1/2} \gamma^{-1/2} + h(\theta,\Delta \theta) D^2 \gamma  ( 1 + C \!\mathit{We}^{1/2})^{2}
\end{equation}

\subsection{Non-dimensionalization of Energy Balance}
In this section, we rescale energy balance in terms of the Weber and Ohnesorge numbers. Using the results of the previous sections, the expression for the energy at the end of the sticking or bouncing process can be written as
\begin{equation}
E_{\mathrm{k,f}} = \frac{1}{2} m u_0^2 - \pi \gamma D^2 f(\theta) - \kappa \mu u_0^2 D^{5/2} \rho^{1/2} \gamma^{-1/2} - D^2 \gamma  h(\theta,\Delta \theta) ( 1 + C \!\mathit{We}^{1/2})^{2}.
\end{equation}
We can take the condition for bouncing to occur when the upward kinetic energy increases just above zero,
\begin{equation}
\frac{\pi}{12} D^3\rho u_0^2 =  \pi \gamma D^2 f(\theta) + \kappa \mu u_0^2 D^{5/2} \rho^{1/2} \gamma^{-1/2} + D^2 \gamma  h(\theta,\Delta \theta)( 1 + C \!\mathit{We}^{1/2})^{2}.
\end{equation}
Re-scaling by $\gamma D^2$ non-dimensionalises the system and we obtain
\begin{equation}
\frac{\pi}{12} D \rho u_0^2 \gamma^{-1} =  \pi f(\theta) + \kappa \mu u_0^2 D^{1/2} \rho^{1/2} \gamma^{-3/2} +   h(\theta,\Delta \theta)( 1 + C \!\mathit{We}^{1/2})^{2}.
\end{equation}
We now use the definition of the numbers $\!\mathit{We}$ and $\!\mathit{Oh}$ and absorb the constants in this expression to obtain:
\begin{equation}
 \mathit{We} =  f(\theta) + \kappa \mathit{We}\mathit{Oh} +  h(\theta,\Delta \theta) ( 1 + 2C\!\mathit{We}^{1/2} + C^2\!\mathit{We})
\end{equation}
or
\begin{equation}
\label{Kap2}
\mathit{We} = \frac{ f(\theta) + h(\theta, \Delta \theta)(1 + 2C\!\mathit{We}^{1/2})} {1 -  \kappa \mathit{Oh} - C^2h(\theta, \Delta \theta)}.
\end{equation}
From this expression, we obtain the limit $\kappa\mathit{Oh} + C^2h(\theta, \Delta \theta) = 1$, or assuming $C^2h(\theta, \Delta \theta)$ is small, $\kappa\mathit{Oh} = 1$. For higher values of $\mathit{Oh}$, bouncing does not occur due high dissipation, independent of $\!\mathit{We}$. This heuristic approach has allowed for a simplified expression the stick-to-bounce transition parameters. This expression fits well with our numerical data when the hysteresis terms are neglected and displays linear behaviour at small $\!\mathit{Oh}$, in agreement with experimental and numerical observation. 

Expression~\ref{Kap2} is written in terms of $\!\mathit{We}$ and $\!\mathit{Oh}$ as they directly encode the inertia and dissipation in the system. The expression can similarly be expressed using the Reynolds number instead of either $\!\mathit{We}$ or $\!\mathit{Oh}$, with $\mathit{Re} = \mathit{We}^{1/2} \mathit{Oh}^{-1}$. Using this substitution we find,
\begin{equation}
\mathit{Re}^2 = \frac{f(\theta) + h(\theta, \Delta \theta) (1 + 2C\mathit{Oh}\mathit{Re}) }{\mathit{Oh}^2(1 - \kappa \mathit{Oh} - C^2h(\theta, \Delta \theta))},
\end{equation}
see Fig.~\ref{figRe} for the corresponding plot of the numerical and simulation data.

\section{Analytical Ball-and-Spring Model}
In this section, we present details of the ball-and-spring model, including dimensionless rescaling to make contact with experiments and the estimation of the model parameters. 
\subsection{Model Set-up}
The ball-spring model shows the physical mechanism behind the transition from sticking to bouncing using a simple mechanical argument. Spring models have previously been used to understand droplet bouncing and contact time~\cite{Okumura2003, Jha2020} for superhydrophobic surfaces. We build on these previous results by introducing the droplet-surface contact energy into the model.

 The model consists of two masses, two springs, and a viscous damper. The two masses ($M_1$ and $M_2$) are connected to a spring ($k_1$) that represents surface tension and a damper ($\mu$) that represents the viscous dissipation in the system. The second mass is connected to a spring ($k_2$) that represents fluid-surface adhesion. In the model, spring 1 extends beyond a critical extension, the spring breaks, the system rebounds, and mass $M_1$ escapes. This is representative of a necking instability. This simplified model reproduces the fundamental behaviour of the system in the bouncing and sticking regimes. 

We use the following mapping between the ball-and-spring model parameters (on the left) and the fluid dynamic parameters (on the right):
\begin{equation}
k_1 = \zeta \gamma
\end{equation}
\begin{equation}
k_2  = f(\theta)  k_1 = (1 + \cos(\theta))  \zeta \gamma
\end{equation}
\begin{equation}
\mu = \alpha \eta D
\end{equation}
\begin{equation}
M_1 =  \frac{\pi}{6} \rho D_o^{3}
\end{equation}
\begin{equation}
M_2 =  \frac{\pi}{6} \rho D_{s}^{3}
\end{equation}
Here, we introduce dimensionless scale factors $\zeta$ and $\alpha$. The scale factor $\zeta$ is the scale factor between the spring constant and the surface tension. This factor can be computed from the ratio of the spring and droplet oscillation frequencies. The scale factor $\alpha$ is the damping scaling between the spring system and the droplet. Although a linear damper is a simplified model of viscous dissipation in a drop, it can be a valid approximation in a small range of parameter values. We also expect $\alpha$ to be surface dependent, with a larger spread and hysteresis corresponding to a larger dissipation. In the parameter estimation section, we estimate $\alpha$ and $\zeta$ for a Teflon surface in the microdroplet regime. 

The equations of motion of the two masses can then be expressed as:
\begin{equation}
\label{1}
M_1 \ddot{x}_1 = -k_1( x_1 - x_2 ) - \mu (\dot{x}_1 - \dot{x}_2),
\end{equation}
and
\begin{equation}
\label{2}
M_2 \ddot{x}_2 = k_1( x_1 - x_2 ) + \mu (\dot{x}_1 - \dot{x}_2) - k_2 x_2.
\end{equation}
We then re-express these equations using droplet parameters:
\begin{equation}
 \frac{\pi}{6} \rho D_o^{3} \ddot{x}_1 = -\zeta \gamma( x_1 - x_2 ) - \alpha \eta D (\dot{x}_1 - \dot{x}_2)
\end{equation}
and
\begin{equation}
 \frac{\pi}{6} \rho D_{s}^{3} \ddot{x}_2 = \zeta \gamma( x_1 - x_2 ) + \alpha \eta D (\dot{x}_1 - \dot{x}_2) - \zeta  \gamma x_2.
\end{equation}
We then non-dimensionalize this system of equations, first by introducing the dimensionless ratios of diameters:
\begin{equation}
m_1 = \frac{\pi}{6} \frac{D_o^3}{D^3},
\end{equation}
\begin{equation}
m_2 = \frac{\pi}{6} \frac{D_{s}^{3}}{D^3}
\end{equation}
Then, we rescale:
\begin{equation}
\tilde{x}_{1,2} = \frac{x_{1,2}}{D}
\end{equation}
\begin{equation}
\tilde{t} = t \frac{u}{D}
\end{equation}
\begin{equation}
\dot{\tilde{x}}_{1,2} = \frac{\dot{x}_{1,2}}{u}
\end{equation}
\begin{equation}
\ddot{\tilde{x}}_{1,2} = \frac{\ddot{x}_{1,2} D}{u^2}
\end{equation}
The equations of motion are then written using dimensionless variables:
\begin{equation}
 m_1 \rho D^2 u^2 \ddot{\tilde{x}}_1 = -\zeta \gamma D( \tilde{x}_1 - \tilde{x}_2 ) - \alpha \eta D u (\dot{\tilde{x}}_1 - \dot{\tilde{x}}_2)
\end{equation}
and
\begin{equation}
\label{6}
 m_2 \rho D^2 u^2 \ddot{\tilde{x}}_2 = \zeta \gamma D( \tilde{x}_1 - \tilde{x}_2 ) + \alpha \eta D u (\dot{\tilde{x}}_1 - \dot{\tilde{x}}_2) - \zeta  \gamma D \tilde{x}_2.
\end{equation}
We simplify the equations by dividing by $\rho D^2 u^2$:
\begin{equation}
 m_1 \ddot{\tilde{x}}_1 = -\zeta \frac{\gamma}{\rho D u^2}( \tilde{x}_1 - \tilde{x}_2 ) - \alpha \frac{\eta}{\rho D v} (\dot{\tilde{x}}_1 - \dot{\tilde{x}}_2)
\end{equation}
and
\begin{equation}
 m_2 \ddot{\tilde{x}}_2 = \zeta \frac{\gamma}{\rho D u^2}( \tilde{x}_1 - \tilde{x}_2 ) + \alpha \frac{\eta}{\rho D v} (\dot{\tilde{x}}_1 - \dot{\tilde{x}}_2) - (1 - \cos(\theta)) \zeta  \frac{\gamma}{\rho D u^2} \tilde{x}_2.
\end{equation}
Here, we introduce the Weber and Ohnesorge numbers to allow a direct link with the droplet system.
\begin{equation}
m_1 \ddot{\tilde{x}}_1 = - \zeta \frac{1}{\mathit{We}}( \tilde{x}_1 - \tilde{x}_2 ) - \alpha \frac{Oh}{\sqrt{\mathit{We}}} (\dot{\tilde{x}}_1 - \dot{\tilde{x}}_2)
\end{equation}
and
\begin{equation}
m_2 \ddot{\tilde{x}}_2 = \zeta \frac{1}{\mathit{We}}( \tilde{x}_1 - \tilde{x}_2 ) + \alpha \frac{Oh}{\sqrt{\mathit{We}}} (\dot{\tilde{x}}_1 - \dot{\tilde{x}}_2) - (1 - \cos(\theta)) \zeta \frac{1}{\mathit{We}} \tilde{x}_2.
\end{equation}
For convenience, we introduce the extension of spring 1: $\Delta x = \tilde{x}_1  - \tilde{x}_2$ and arrive at the form presented in the main text:
\begin{equation}
 m_1\ddot{\tilde{x}}_1 = - \zeta \frac{1}{\mathit{We}} \Delta x  - \alpha \frac{Oh}{\sqrt{\mathit{We}}} \Delta\dot{x}
\end{equation}
and
\begin{equation}
m_2\ddot{\tilde{x}}_2 = \zeta \frac{1}{\mathit{We}}\Delta x  + \alpha \frac{Oh}{\sqrt{\mathit{We}}} \Delta\dot{x} - (1 - \cos(\theta)) \zeta \frac{1}{\mathit{We}} \tilde{x}_2
\end{equation}
The system's initial condition is that it starts at its equilibrium and moves downwards, equivalent to the droplet spreading and energy being stored in the interfaces. This downward velocity is encoded in the system's $\!\mathit{We}$ number, such that the dimensionless velocity is unity:
\begin{equation}
\tilde{x}_1 = \tilde{x}_2 = 0
\end{equation}
and
\begin{equation}
\dot{\tilde{x}}_1 = \dot{\tilde{x}}_2 = -1
\end{equation}
We solve this system for combinations of $\!\mathit{We}$, $\!\mathit{Oh}$, and $\theta$ for given parameter values. We state a bounce occurs when for spring 1, $\Delta x$ exceeds a critical length scale. We this lengthscale to be $\Delta x  = 1$ in the non-dimensionalized units. 

\subsection{Parameter estimation}
This section estimates the dimensionless scaling factors between the ball-and-spring model and the microdroplet system. 

The value of $\zeta = 1$ matches the experimentally observed oscillation frequencies. To estimate $\alpha$, we use two independent methods. First, we compare the damping of oscillation in a microdroplet impact (i.e. Fig.~\ref{Fig5}) to the damping of the spring system and the coefficient of restitution to the damping for a bouncing microdroplet. We observe that for an impact with a Teflon surface, microdroplets lose between 30 and 50$\%$ of their initial energy, depending on the numbers $\!\mathit{We}$ and $\!\mathit{Oh}$. This compares to 1$\%$ in the model with the same $\!\mathit{Oh}$ and an $\alpha$ of 1. Setting $\alpha$ to 50 matches the energy dissipation approximately for the range of $\!\mathit{Oh}$  0.015 to 0.025, which corresponds to our experiments. Note that this estimate is only valid for the specific surface used, which is Teflon. Indicating additional dissipative mechanisms makes dissipation approximately 50 times greater than the linear bulk.

\begin{figure}[t]
    \centering
    \includegraphics[width=\linewidth]{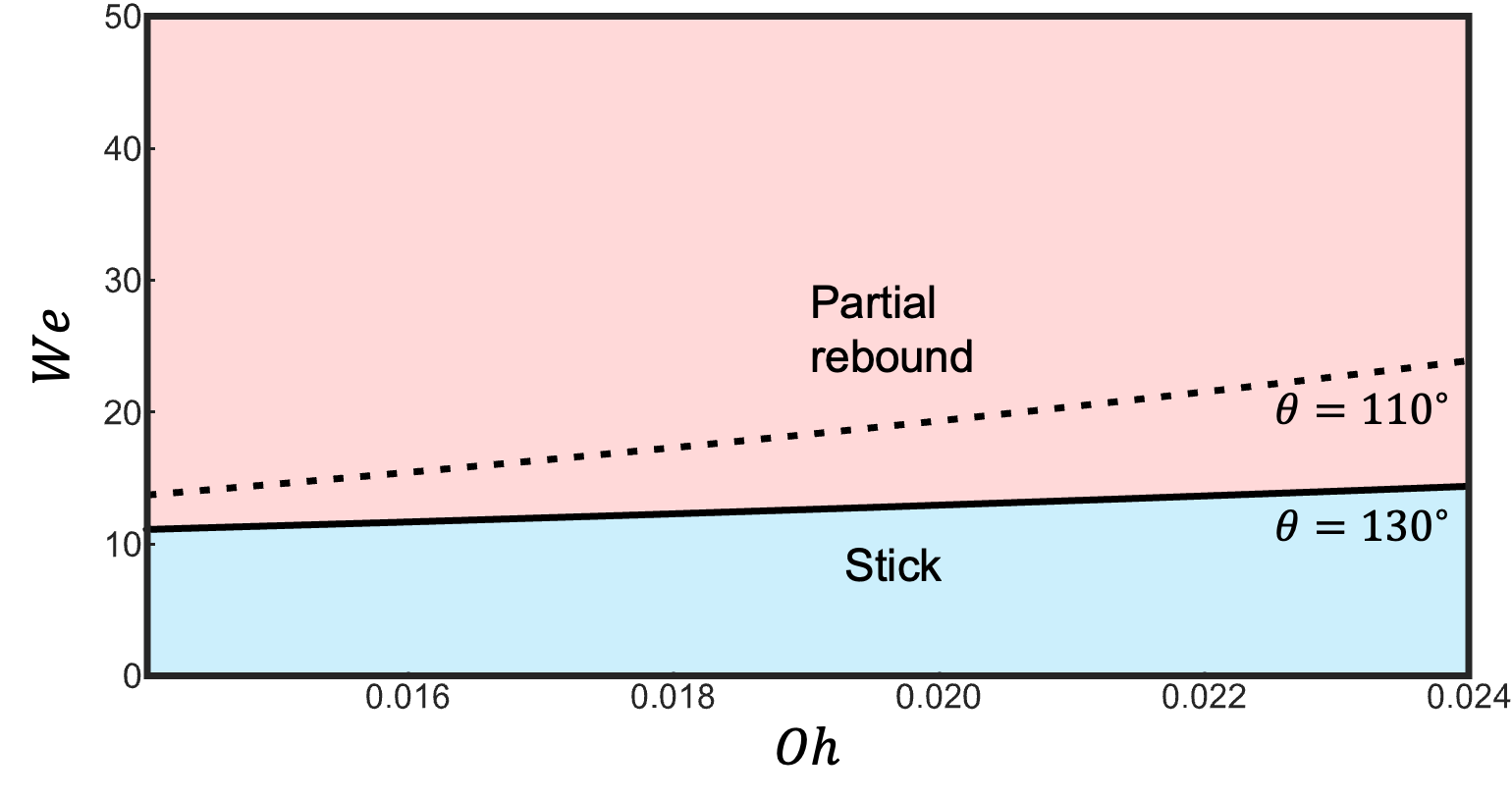}
    \caption{Phase space of sticking vs bouncing based on the ball-and-spring model for different sets of parameters. Here the two lines correspond to $(\kappa,\theta) = (30,130^\circ)$ (solid) and $(\kappa,\theta) = (50,110^\circ)$ (dashed), with this second line corresponding to the line plotts in Fig.~6(c) of the main text. }
    \label{Modela}
\end{figure}

The second estimate of $\alpha$ uses the energy dissipated in a microdroplet. In the main text, we say the dissipation is given by:
\begin{equation}
 E_\mathrm{\mu} = E_\mathrm{\mu,3D} + E_\mathrm{\mu,2D} + E_\mathrm{\mu,1D}
\end{equation}
where $E_\mathrm{\mu,3D}$ for microdroplets impacting between 1 and \SI{10}{m/s}  is between \SI{1e-12}{\joule} to \SI{1e-10}{\joule}. The value $E_\mathrm{\mu,2D}$ is approximately an order of magnitude larger, scaling as $\mathit{Oh}^{1/2}$. We also estimate the contact line dissipation using Ref.~\cite{Quere}:
\begin{equation}
    E_\mathrm{\mu,1D} = \frac{\pi \gamma}{2} D_{\textrm{m
    }}^2 ( \cos (\theta_r) - \cos(\theta_a)).
\end{equation}
Contact angle hysteresis for Teflon is measured to be approximately $19^\circ$, and the maximum spread varies from 40 to \SI{90}{\micro\meter} for the range of $\!\mathit{We}$ numbers we consider. This gives a value of $E_\mathrm{\mu,1D}$ approximately 3 to 6 $\times$ greater than $E_\mathrm{\mu,3D}$. 

Combining these approximations suggests a scaling of between  30 and 60 for $\alpha$. The value $\alpha = 50$ aligns well with both of these estimation methods. The $\alpha$ value increases with hysteresis and hydrophilicity, so estimates must be adjusted for a relevant surface. 

This model predicts the stick-to-bounce transition for the parameters we considered. Fig.~\ref{Modela} shows that the ball-and-spring model replicates some of the physics of varying the contact angle in experiments. For example, a droplet bounces on a more hydrophobic surface at a lower value of $\!\mathit{We}$.

\section{Movies}
Movie 1 (A) Water microdroplet impacting and sticking to a Teflon substrate. $\!\mathit{We} = 12$ and $\!\mathit{Oh} = 0.015$. (B) 5\% V/V Water/glycerol microdroplet impacting and sticking to a Teflon substrate with a bubble. $\!\mathit{We} = 2.0$ and $\!\mathit{Oh} = 0.020$. (C) Water microdroplet impacting and partially rebounding off a Teflon substrate. $\!\mathit{We} = 24$ and $\!\mathit{Oh} = 0.015$. (D) Water microdroplet impacting and partially rebounding off a Teflon substrate. $\!\mathit{We} = 40$ and $\!\mathit{Oh} = 0.017$. The videos correspond to Fig 1.~(C-F) in the main text. All videos were recorded at 100,000 FPS and played at 6 FPS.

Movie 2 (A) Water microdroplet impacting and sticking to a nano particle-coated substrate. $\!\mathit{We} = 2.5$ and $\!\mathit{Oh} = 0.017$. (B) Water microdroplet impacting and bouncing off nanoparticle-coated substrate. $\!\mathit{We} = 10$ and $\!\mathit{Oh} = 0.017$. The videos used compose Fig 7. (B-C) in the main text. All videos were recorded at 100,000 FPS and played at 6 FPS. 

\twocolumngrid
\input{refs.bbl}  
%\bibliography{refs-new}

\end{document}

%% file: refs.bbl

%% file: aapmsamp.bbl
\begin{thebibliography}{68}%
\makeatletter
\providecommand \@ifxundefined [1]{%
 \@ifx{#1\undefined}
}%
\providecommand \@ifnum [1]{%
 \ifnum #1\expandafter \@firstoftwo
 \else \expandafter \@secondoftwo
 \fi
}%
\providecommand \@ifx [1]{%
 \ifx #1\expandafter \@firstoftwo
 \else \expandafter \@secondoftwo
 \fi
}%
\providecommand \natexlab [1]{#1}%
\providecommand \enquote  [1]{``#1''}%
\providecommand \bibnamefont  [1]{#1}%
\providecommand \bibfnamefont [1]{#1}%
\providecommand \citenamefont [1]{#1}%
\providecommand \href@noop [0]{\@secondoftwo}%
\providecommand \href [0]{\begingroup \@sanitize@url \@href}%
\providecommand \@href[1]{\@@startlink{#1}\@@href}%
\providecommand \@@href[1]{\endgroup#1\@@endlink}%
\providecommand \@sanitize@url [0]{\catcode `\\12\catcode `\$12\catcode `\&12\catcode `\#12\catcode `\^12\catcode `\_12\catcode `\%12\relax}%
\providecommand \@@startlink[1]{}%
\providecommand \@@endlink[0]{}%
\providecommand \url  [0]{\begingroup\@sanitize@url \@url }%
\providecommand \@url [1]{\endgroup\@href {#1}{\urlprefix }}%
\providecommand \urlprefix  [0]{URL }%
\providecommand \Eprint [0]{\href }%
\providecommand \doibase [0]{https://doi.org/}%
\providecommand \selectlanguage [0]{\@gobble}%
\providecommand \bibinfo  [0]{\@secondoftwo}%
\providecommand \bibfield  [0]{\@secondoftwo}%
\providecommand \translation [1]{[#1]}%
\providecommand \BibitemOpen [0]{}%
\providecommand \bibitemStop [0]{}%
\providecommand \bibitemNoStop [0]{.\EOS\space}%
\providecommand \EOS [0]{\spacefactor3000\relax}%
\providecommand \BibitemShut  [1]{\csname bibitem#1\endcsname}%
\let\auto@bib@innerbib\@empty
%</preamble>
\bibitem [{\citenamefont {Poon}\ \emph {et~al.}(2020)\citenamefont {Poon}, \citenamefont {Brown}, \citenamefont {Direito}, \citenamefont {Hodgson}, \citenamefont {Nagard}, \citenamefont {Lips}, \citenamefont {MacPhee}, \citenamefont {Marenduzzo}, \citenamefont {Royer}, \citenamefont {Silva}, \citenamefont {Thijssen},\ and\ \citenamefont {Titmuss}}]{Poon2020}%
  \BibitemOpen
  \bibfield  {author} {\bibinfo {author} {\bibfnamefont {W.~C.}\ \bibnamefont {Poon}}, \bibinfo {author} {\bibfnamefont {A.~T.}\ \bibnamefont {Brown}}, \bibinfo {author} {\bibfnamefont {S.~O.}\ \bibnamefont {Direito}}, \bibinfo {author} {\bibfnamefont {D.~J.}\ \bibnamefont {Hodgson}}, \bibinfo {author} {\bibfnamefont {L.~L.}\ \bibnamefont {Nagard}}, \bibinfo {author} {\bibfnamefont {A.}~\bibnamefont {Lips}}, \bibinfo {author} {\bibfnamefont {C.~E.}\ \bibnamefont {MacPhee}}, \bibinfo {author} {\bibfnamefont {D.}~\bibnamefont {Marenduzzo}}, \bibinfo {author} {\bibfnamefont {J.~R.}\ \bibnamefont {Royer}}, \bibinfo {author} {\bibfnamefont {A.~F.}\ \bibnamefont {Silva}}, \bibinfo {author} {\bibfnamefont {J.~H.}\ \bibnamefont {Thijssen}},\ and\ \bibinfo {author} {\bibfnamefont {S.}~\bibnamefont {Titmuss}},\ }\bibfield  {title} {\bibinfo {title} {Soft matter science and the covid-19 pandemic},\ }\href@noop {} {\bibfield  {journal} {\bibinfo  {journal} {Soft Matter}\ }\textbf {\bibinfo {volume} {16}},\ \bibinfo
  {pages} {8310} (\bibinfo {year} {2020})}\BibitemShut {NoStop}%
\bibitem [{\citenamefont {Bourouiba}(2021)}]{Bourouiba2021}%
  \BibitemOpen
  \bibfield  {author} {\bibinfo {author} {\bibfnamefont {L.}~\bibnamefont {Bourouiba}},\ }\bibfield  {title} {\bibinfo {title} {The fluid dynamics of disease transmission},\ }\href@noop {} {\bibfield  {journal} {\bibinfo  {journal} {Annual Review of Fluid Mechanics}\ }\textbf {\bibinfo {volume} {53}},\ \bibinfo {pages} {473} (\bibinfo {year} {2021})}\BibitemShut {NoStop}%
\bibitem [{\citenamefont {Katre}\ \emph {et~al.}(2021)\citenamefont {Katre}, \citenamefont {Banerjee}, \citenamefont {Balusamy},\ and\ \citenamefont {Sahu}}]{Katre2021}%
  \BibitemOpen
  \bibfield  {author} {\bibinfo {author} {\bibfnamefont {P.}~\bibnamefont {Katre}}, \bibinfo {author} {\bibfnamefont {S.}~\bibnamefont {Banerjee}}, \bibinfo {author} {\bibfnamefont {S.}~\bibnamefont {Balusamy}},\ and\ \bibinfo {author} {\bibfnamefont {K.~C.}\ \bibnamefont {Sahu}},\ }\bibfield  {title} {\bibinfo {title} {Fluid dynamics of respiratory droplets in the context of covid-19: Airborne and surfaceborne transmissions},\ }\href@noop {} {\bibfield  {journal} {\bibinfo  {journal} {Physics of Fluids}\ }\textbf {\bibinfo {volume} {33}},\ \bibinfo {pages} {081302} (\bibinfo {year} {2021})}\BibitemShut {NoStop}%
\bibitem [{\citenamefont {Onakpoya}\ \emph {et~al.}(2021)\citenamefont {Onakpoya}, \citenamefont {Heneghan}, \citenamefont {Spencer}, \citenamefont {Brassey}, \citenamefont {Plüddemann}, \citenamefont {Evans}, \citenamefont {Conly},\ and\ \citenamefont {Jefferson}}]{Onakpoya2021}%
  \BibitemOpen
  \bibfield  {author} {\bibinfo {author} {\bibfnamefont {I.~J.}\ \bibnamefont {Onakpoya}}, \bibinfo {author} {\bibfnamefont {C.~J.}\ \bibnamefont {Heneghan}}, \bibinfo {author} {\bibfnamefont {E.~A.}\ \bibnamefont {Spencer}}, \bibinfo {author} {\bibfnamefont {J.}~\bibnamefont {Brassey}}, \bibinfo {author} {\bibfnamefont {A.}~\bibnamefont {Plüddemann}}, \bibinfo {author} {\bibfnamefont {D.~H.}\ \bibnamefont {Evans}}, \bibinfo {author} {\bibfnamefont {J.~M.}\ \bibnamefont {Conly}},\ and\ \bibinfo {author} {\bibfnamefont {T.}~\bibnamefont {Jefferson}},\ }\bibfield  {title} {\bibinfo {title} {Sars-cov-2 and the role of fomite transmission: a systematic review},\ }\href@noop {} {\bibfield  {journal} {\bibinfo  {journal} {F1000Research}\ }\textbf {\bibinfo {volume} {10}},\ \bibinfo {pages} {233} (\bibinfo {year} {2021})}\BibitemShut {NoStop}%
\bibitem [{\citenamefont {Short}\ and\ \citenamefont {Cowling}(2023)}]{Short2023}%
  \BibitemOpen
  \bibfield  {author} {\bibinfo {author} {\bibfnamefont {K.~R.}\ \bibnamefont {Short}}\ and\ \bibinfo {author} {\bibfnamefont {B.~J.}\ \bibnamefont {Cowling}},\ }\bibfield  {title} {\bibinfo {title} {Assessing the potential for fomite transmission of sars-cov-2},\ }\href@noop {} {\bibfield  {journal} {\bibinfo  {journal} {The Lancet Microbe}\ }\textbf {\bibinfo {volume} {4}},\ \bibinfo {pages} {e380} (\bibinfo {year} {2023})}\BibitemShut {NoStop}%
\bibitem [{\citenamefont {Shafaghi}\ \emph {et~al.}(2020)\citenamefont {Shafaghi}, \citenamefont {Talabazar}, \citenamefont {Koşar},\ and\ \citenamefont {Ghorbani}}]{Shafaghi2020}%
  \BibitemOpen
  \bibfield  {author} {\bibinfo {author} {\bibfnamefont {A.~H.}\ \bibnamefont {Shafaghi}}, \bibinfo {author} {\bibfnamefont {F.~R.}\ \bibnamefont {Talabazar}}, \bibinfo {author} {\bibfnamefont {A.}~\bibnamefont {Koşar}},\ and\ \bibinfo {author} {\bibfnamefont {M.}~\bibnamefont {Ghorbani}},\ }\bibfield  {title} {\bibinfo {title} {on the effect of the respiratory droplet generation condition on covid-19 transmission},\ }\href@noop {} {\bibfield  {journal} {\bibinfo  {journal} {Fluids}\ }\textbf {\bibinfo {volume} {5}},\ \bibinfo {pages} {113} (\bibinfo {year} {2020})}\BibitemShut {NoStop}%
\bibitem [{\citenamefont {Kumar}\ \emph {et~al.}(2021)\citenamefont {Kumar}, \citenamefont {Chatterjee}, \citenamefont {Agrawal},\ and\ \citenamefont {Bhardwaj}}]{Kumar2021}%
  \BibitemOpen
  \bibfield  {author} {\bibinfo {author} {\bibfnamefont {B.}~\bibnamefont {Kumar}}, \bibinfo {author} {\bibfnamefont {S.}~\bibnamefont {Chatterjee}}, \bibinfo {author} {\bibfnamefont {A.}~\bibnamefont {Agrawal}},\ and\ \bibinfo {author} {\bibfnamefont {R.}~\bibnamefont {Bhardwaj}},\ }\bibfield  {title} {\bibinfo {title} {Evaluating a transparent coating on a face shield for repelling airborne respiratory droplets},\ }\href@noop {} {\bibfield  {journal} {\bibinfo  {journal} {Physics of Fluids}\ }\textbf {\bibinfo {volume} {33}},\ \bibinfo {pages} {111705} (\bibinfo {year} {2021})}\BibitemShut {NoStop}%
\bibitem [{\citenamefont {Lohse}(2022)}]{Lohse2021}%
  \BibitemOpen
  \bibfield  {author} {\bibinfo {author} {\bibfnamefont {D.}~\bibnamefont {Lohse}},\ }\bibfield  {title} {\bibinfo {title} {Fundamental fluid dynamics challenges in inkjet printing},\ }\href@noop {} {\bibfield  {journal} {\bibinfo  {journal} {Annual Review of Fluid Mechanics}\ }\textbf {\bibinfo {volume} {54}},\ \bibinfo {pages} {349} (\bibinfo {year} {2022})}\BibitemShut {NoStop}%
\bibitem [{\citenamefont {Dorr}\ \emph {et~al.}(2014)\citenamefont {Dorr}, \citenamefont {Kempthorne}, \citenamefont {Mayo}, \citenamefont {Forster}, \citenamefont {Zabkiewicz}, \citenamefont {McCue}, \citenamefont {Belward}, \citenamefont {Turner},\ and\ \citenamefont {Hanan}}]{Dorr2014}%
  \BibitemOpen
  \bibfield  {author} {\bibinfo {author} {\bibfnamefont {G.~J.}\ \bibnamefont {Dorr}}, \bibinfo {author} {\bibfnamefont {D.~M.}\ \bibnamefont {Kempthorne}}, \bibinfo {author} {\bibfnamefont {L.~C.}\ \bibnamefont {Mayo}}, \bibinfo {author} {\bibfnamefont {W.~A.}\ \bibnamefont {Forster}}, \bibinfo {author} {\bibfnamefont {J.~A.}\ \bibnamefont {Zabkiewicz}}, \bibinfo {author} {\bibfnamefont {S.~W.}\ \bibnamefont {McCue}}, \bibinfo {author} {\bibfnamefont {J.~A.}\ \bibnamefont {Belward}}, \bibinfo {author} {\bibfnamefont {I.~W.}\ \bibnamefont {Turner}},\ and\ \bibinfo {author} {\bibfnamefont {J.}~\bibnamefont {Hanan}},\ }\bibfield  {title} {\bibinfo {title} {Towards a model of spray-canopy interactions: Interception, shatter, bounce and retention of droplets on horizontal leaves},\ }\href@noop {} {\bibfield  {journal} {\bibinfo  {journal} {Ecological Modelling}\ }\textbf {\bibinfo {volume} {290}},\ \bibinfo {pages} {94} (\bibinfo {year} {2014})}\BibitemShut {NoStop}%
\bibitem [{\citenamefont {Massinon}\ \emph {et~al.}(2017)\citenamefont {Massinon}, \citenamefont {Cock}, \citenamefont {Forster}, \citenamefont {Nairn}, \citenamefont {McCue}, \citenamefont {Zabkiewicz},\ and\ \citenamefont {Lebeau}}]{Massinon2017}%
  \BibitemOpen
  \bibfield  {author} {\bibinfo {author} {\bibfnamefont {M.}~\bibnamefont {Massinon}}, \bibinfo {author} {\bibfnamefont {N.~D.}\ \bibnamefont {Cock}}, \bibinfo {author} {\bibfnamefont {W.~A.}\ \bibnamefont {Forster}}, \bibinfo {author} {\bibfnamefont {J.~J.}\ \bibnamefont {Nairn}}, \bibinfo {author} {\bibfnamefont {S.~W.}\ \bibnamefont {McCue}}, \bibinfo {author} {\bibfnamefont {J.~A.}\ \bibnamefont {Zabkiewicz}},\ and\ \bibinfo {author} {\bibfnamefont {F.}~\bibnamefont {Lebeau}},\ }\bibfield  {title} {\bibinfo {title} {Spray droplet impaction outcomes for different plant species and spray formulations},\ }\href@noop {} {\bibfield  {journal} {\bibinfo  {journal} {Crop Protection}\ }\textbf {\bibinfo {volume} {99}},\ \bibinfo {pages} {65} (\bibinfo {year} {2017})}\BibitemShut {NoStop}%
\bibitem [{\citenamefont {Virtanen}\ \emph {et~al.}(2011)\citenamefont {Virtanen}, \citenamefont {Kannosto}, \citenamefont {Kuuluvainen}, \citenamefont {Arffman}, \citenamefont {Joutsensaari}, \citenamefont {Saukko}, \citenamefont {Hao}, \citenamefont {Yli-Pirilä}, \citenamefont {Tiitta}, \citenamefont {Holopainen}, \citenamefont {Keskinen}, \citenamefont {Worsnop}, \citenamefont {Smith},\ and\ \citenamefont {Laaksonen}}]{Virtanen2011}%
  \BibitemOpen
  \bibfield  {author} {\bibinfo {author} {\bibfnamefont {A.}~\bibnamefont {Virtanen}}, \bibinfo {author} {\bibfnamefont {J.}~\bibnamefont {Kannosto}}, \bibinfo {author} {\bibfnamefont {H.}~\bibnamefont {Kuuluvainen}}, \bibinfo {author} {\bibfnamefont {A.}~\bibnamefont {Arffman}}, \bibinfo {author} {\bibfnamefont {J.}~\bibnamefont {Joutsensaari}}, \bibinfo {author} {\bibfnamefont {E.}~\bibnamefont {Saukko}}, \bibinfo {author} {\bibfnamefont {L.}~\bibnamefont {Hao}}, \bibinfo {author} {\bibfnamefont {P.}~\bibnamefont {Yli-Pirilä}}, \bibinfo {author} {\bibfnamefont {P.}~\bibnamefont {Tiitta}}, \bibinfo {author} {\bibfnamefont {J.~K.}\ \bibnamefont {Holopainen}}, \bibinfo {author} {\bibfnamefont {J.}~\bibnamefont {Keskinen}}, \bibinfo {author} {\bibfnamefont {D.~R.}\ \bibnamefont {Worsnop}}, \bibinfo {author} {\bibfnamefont {J.~N.}\ \bibnamefont {Smith}},\ and\ \bibinfo {author} {\bibfnamefont {A.}~\bibnamefont {Laaksonen}},\ }\bibfield  {title} {\bibinfo {title} {Bounce behavior of freshly nucleated biogenic
  secondary organic aerosol particles},\ }\href@noop {} {\bibfield  {journal} {\bibinfo  {journal} {Atmospheric Chemistry and Physics}\ }\textbf {\bibinfo {volume} {11}},\ \bibinfo {pages} {8759} (\bibinfo {year} {2011})}\BibitemShut {NoStop}%
\bibitem [{\citenamefont {Joung}\ and\ \citenamefont {Buie}(2015)}]{Joung2015}%
  \BibitemOpen
  \bibfield  {author} {\bibinfo {author} {\bibfnamefont {Y.~S.}\ \bibnamefont {Joung}}\ and\ \bibinfo {author} {\bibfnamefont {C.~R.}\ \bibnamefont {Buie}},\ }\bibfield  {title} {\bibinfo {title} {Aerosol generation by raindrop impact on soil},\ }\href@noop {} {\bibfield  {journal} {\bibinfo  {journal} {Nature Communications}\ }\textbf {\bibinfo {volume} {6}},\ \bibinfo {pages} {6083} (\bibinfo {year} {2015})}\BibitemShut {NoStop}%
\bibitem [{\citenamefont {Jain}\ and\ \citenamefont {Petrucci}(2015)}]{Jain2015}%
  \BibitemOpen
  \bibfield  {author} {\bibinfo {author} {\bibfnamefont {S.}~\bibnamefont {Jain}}\ and\ \bibinfo {author} {\bibfnamefont {G.~A.}\ \bibnamefont {Petrucci}},\ }\bibfield  {title} {\bibinfo {title} {A new method to measure aerosol particle bounce using a cascade electrical low pressure impactor},\ }\href@noop {} {\bibfield  {journal} {\bibinfo  {journal} {Aerosol Science and Technology}\ }\textbf {\bibinfo {volume} {49}},\ \bibinfo {pages} {390} (\bibinfo {year} {2015})}\BibitemShut {NoStop}%
\bibitem [{\citenamefont {Nietiadi}\ \emph {et~al.}(2024)\citenamefont {Nietiadi}, \citenamefont {Urbassek},\ and\ \citenamefont {Rosandi}}]{Nietiadi2024}%
  \BibitemOpen
  \bibfield  {author} {\bibinfo {author} {\bibfnamefont {M.~L.}\ \bibnamefont {Nietiadi}}, \bibinfo {author} {\bibfnamefont {H.~M.}\ \bibnamefont {Urbassek}},\ and\ \bibinfo {author} {\bibfnamefont {Y.}~\bibnamefont {Rosandi}},\ }\bibfield  {title} {\bibinfo {title} {An atomistic study of sticking, bouncing, and aggregate destruction in collisions of grains with small aggregates},\ }\href@noop {} {\bibfield  {journal} {\bibinfo  {journal} {Scientific Reports}\ }\textbf {\bibinfo {volume} {14}} (\bibinfo {year} {2024})}\BibitemShut {NoStop}%
\bibitem [{\citenamefont {Tumminello}\ \emph {et~al.}(2024)\citenamefont {Tumminello}, \citenamefont {Niles}, \citenamefont {Valdez}, \citenamefont {Madawala}, \citenamefont {Gamage}, \citenamefont {Kimble}, \citenamefont {Leibensperger}, \citenamefont {Huang}, \citenamefont {Kaluarachchi}, \citenamefont {Dinasquet}, \citenamefont {Malfatti}, \citenamefont {Lee}, \citenamefont {Deane}, \citenamefont {Stokes}, \citenamefont {Stone}, \citenamefont {Tivanski}, \citenamefont {Prather}, \citenamefont {Boor},\ and\ \citenamefont {Slade}}]{Tumminello2024}%
  \BibitemOpen
  \bibfield  {author} {\bibinfo {author} {\bibfnamefont {P.~R.}\ \bibnamefont {Tumminello}}, \bibinfo {author} {\bibfnamefont {R.}~\bibnamefont {Niles}}, \bibinfo {author} {\bibfnamefont {V.}~\bibnamefont {Valdez}}, \bibinfo {author} {\bibfnamefont {C.~K.}\ \bibnamefont {Madawala}}, \bibinfo {author} {\bibfnamefont {D.~K.}\ \bibnamefont {Gamage}}, \bibinfo {author} {\bibfnamefont {K.~A.}\ \bibnamefont {Kimble}}, \bibinfo {author} {\bibfnamefont {R.~J.~I.}\ \bibnamefont {Leibensperger}}, \bibinfo {author} {\bibfnamefont {C.}~\bibnamefont {Huang}}, \bibinfo {author} {\bibfnamefont {C.}~\bibnamefont {Kaluarachchi}}, \bibinfo {author} {\bibfnamefont {J.}~\bibnamefont {Dinasquet}}, \bibinfo {author} {\bibfnamefont {F.}~\bibnamefont {Malfatti}}, \bibinfo {author} {\bibfnamefont {C.}~\bibnamefont {Lee}}, \bibinfo {author} {\bibfnamefont {G.~B.}\ \bibnamefont {Deane}}, \bibinfo {author} {\bibfnamefont {M.~D.}\ \bibnamefont {Stokes}}, \bibinfo {author} {\bibfnamefont {E.}~\bibnamefont {Stone}}, \bibinfo {author}
  {\bibfnamefont {A.}~\bibnamefont {Tivanski}}, \bibinfo {author} {\bibfnamefont {K.~A.}\ \bibnamefont {Prather}}, \bibinfo {author} {\bibfnamefont {B.~E.}\ \bibnamefont {Boor}},\ and\ \bibinfo {author} {\bibfnamefont {J.~H.}\ \bibnamefont {Slade}},\ }\bibfield  {title} {\bibinfo {title} {Size-dependent nascent sea spray aerosol bounce fractions and estimated viscosity: The role of divalent cation enrichment, surface tension, and the kelvin effect},\ }\href@noop {} {\bibfield  {journal} {\bibinfo  {journal} {Environmental Science \& Technology}\ }\textbf {\bibinfo {volume} {58}},\ \bibinfo {pages} {19666} (\bibinfo {year} {2024})}\BibitemShut {NoStop}%
\bibitem [{\citenamefont {Worthington}(1877)}]{Worthington1877}%
  \BibitemOpen
  \bibfield  {author} {\bibinfo {author} {\bibfnamefont {A.~M.}\ \bibnamefont {Worthington}},\ }\bibfield  {title} {\bibinfo {title} {On the forms assumed by drops of liquids falling vertically on a horizontal plate},\ }\href@noop {} {\bibfield  {journal} {\bibinfo  {journal} {Proceedings of the Royal Society of London}\ }\textbf {\bibinfo {volume} {25}},\ \bibinfo {pages} {261} (\bibinfo {year} {1877})}\BibitemShut {NoStop}%
\bibitem [{\citenamefont {Worthington}(1895)}]{Worth1896}%
  \BibitemOpen
  \bibfield  {author} {\bibinfo {author} {\bibfnamefont {A.~M.}\ \bibnamefont {Worthington}},\ }\href@noop {} {\emph {\bibinfo {title} {The Splash of a Drop}}}\ (\bibinfo  {publisher} {Society for Promoting Christian Knowledge (S.P.C.K.)},\ \bibinfo {address} {London},\ \bibinfo {year} {1895})\BibitemShut {NoStop}%
\bibitem [{\citenamefont {Worthington}\ and\ \citenamefont {Cole}(1896)}]{Worthington1896}%
  \BibitemOpen
  \bibfield  {author} {\bibinfo {author} {\bibfnamefont {A.~M.}\ \bibnamefont {Worthington}}\ and\ \bibinfo {author} {\bibfnamefont {R.~S.}\ \bibnamefont {Cole}},\ }\bibfield  {title} {\bibinfo {title} {Impact with a liquid surface, studied by means of instantaneous photography},\ }\href@noop {} {\bibfield  {journal} {\bibinfo  {journal} {Proceedings of the Royal Society of London}\ }\textbf {\bibinfo {volume} {59}},\ \bibinfo {pages} {250} (\bibinfo {year} {1896})}\BibitemShut {NoStop}%
\bibitem [{\citenamefont {Rioboo}\ \emph {et~al.}(2001)\citenamefont {Rioboo}, \citenamefont {Tropea},\ and\ \citenamefont {Marengo}}]{Rioboo2001}%
  \BibitemOpen
  \bibfield  {author} {\bibinfo {author} {\bibfnamefont {R.}~\bibnamefont {Rioboo}}, \bibinfo {author} {\bibfnamefont {C.}~\bibnamefont {Tropea}},\ and\ \bibinfo {author} {\bibfnamefont {M.}~\bibnamefont {Marengo}},\ }\bibfield  {title} {\bibinfo {title} {Outcomes from a drop impact on solid surfaces},\ }\href@noop {} {\bibfield  {journal} {\bibinfo  {journal} {Atomization and Sprays}\ }\textbf {\bibinfo {volume} {11}},\ \bibinfo {pages} {155} (\bibinfo {year} {2001})}\BibitemShut {NoStop}%
\bibitem [{\citenamefont {Yarin}(2006)}]{Yarin06}%
  \BibitemOpen
  \bibfield  {author} {\bibinfo {author} {\bibfnamefont {A.}~\bibnamefont {Yarin}},\ }\bibfield  {title} {\bibinfo {title} {Drop impact dynamics: Splashing, spreading, receding, bouncing…},\ }\href@noop {} {\bibfield  {journal} {\bibinfo  {journal} {Annual Review of Fluid Mechanics}\ }\textbf {\bibinfo {volume} {38}},\ \bibinfo {pages} {159} (\bibinfo {year} {2006})}\BibitemShut {NoStop}%
\bibitem [{\citenamefont {Josserand}\ and\ \citenamefont {Thoroddsen}(2016)}]{Josserand16}%
  \BibitemOpen
  \bibfield  {author} {\bibinfo {author} {\bibfnamefont {C.}~\bibnamefont {Josserand}}\ and\ \bibinfo {author} {\bibfnamefont {S.~T.}\ \bibnamefont {Thoroddsen}},\ }\bibfield  {title} {\bibinfo {title} {Drop impact on a solid surface},\ }\href@noop {} {\bibfield  {journal} {\bibinfo  {journal} {Annual Review of Fluid Mechanics}\ }\textbf {\bibinfo {volume} {48}},\ \bibinfo {pages} {365} (\bibinfo {year} {2016})}\BibitemShut {NoStop}%
\bibitem [{\citenamefont {Blanken}\ \emph {et~al.}(2021)\citenamefont {Blanken}, \citenamefont {Saleem}, \citenamefont {Thoraval},\ and\ \citenamefont {Antonini}}]{Blanken2021}%
  \BibitemOpen
  \bibfield  {author} {\bibinfo {author} {\bibfnamefont {N.}~\bibnamefont {Blanken}}, \bibinfo {author} {\bibfnamefont {M.~S.}\ \bibnamefont {Saleem}}, \bibinfo {author} {\bibfnamefont {M.-J.}\ \bibnamefont {Thoraval}},\ and\ \bibinfo {author} {\bibfnamefont {C.}~\bibnamefont {Antonini}},\ }\bibfield  {title} {\bibinfo {title} {Impact of compound drops: a perspective},\ }\href@noop {} {\bibfield  {journal} {\bibinfo  {journal} {Current Opinion in Colloid \& Interface Science}\ }\textbf {\bibinfo {volume} {51}},\ \bibinfo {pages} {101389} (\bibinfo {year} {2021})}\BibitemShut {NoStop}%
\bibitem [{\citenamefont {Mao}\ \emph {et~al.}(1997)\citenamefont {Mao}, \citenamefont {Kuhn},\ and\ \citenamefont {Tran}}]{Mao1997}%
  \BibitemOpen
  \bibfield  {author} {\bibinfo {author} {\bibfnamefont {T.}~\bibnamefont {Mao}}, \bibinfo {author} {\bibfnamefont {D.~C.}\ \bibnamefont {Kuhn}},\ and\ \bibinfo {author} {\bibfnamefont {H.}~\bibnamefont {Tran}},\ }\bibfield  {title} {\bibinfo {title} {Spread and rebound of liquid droplets upon impact on flat surfaces},\ }\href@noop {} {\bibfield  {journal} {\bibinfo  {journal} {AIChE Journal}\ }\textbf {\bibinfo {volume} {43}},\ \bibinfo {pages} {2169} (\bibinfo {year} {1997})}\BibitemShut {NoStop}%
\bibitem [{\citenamefont {Rioboo}\ \emph {et~al.}(2008)\citenamefont {Rioboo}, \citenamefont {Voué}, \citenamefont {Vaillant},\ and\ \citenamefont {Coninck}}]{Rioboo2008}%
  \BibitemOpen
  \bibfield  {author} {\bibinfo {author} {\bibfnamefont {R.}~\bibnamefont {Rioboo}}, \bibinfo {author} {\bibfnamefont {M.}~\bibnamefont {Voué}}, \bibinfo {author} {\bibfnamefont {A.}~\bibnamefont {Vaillant}},\ and\ \bibinfo {author} {\bibfnamefont {J.~D.}\ \bibnamefont {Coninck}},\ }\bibfield  {title} {\bibinfo {title} {Drop impact on porous superhydrophobic polymer surfaces},\ }\href@noop {} {\bibfield  {journal} {\bibinfo  {journal} {Langmuir}\ }\textbf {\bibinfo {volume} {24}},\ \bibinfo {pages} {14074} (\bibinfo {year} {2008})}\BibitemShut {NoStop}%
\bibitem [{\citenamefont {Chen}\ \emph {et~al.}(2016)\citenamefont {Chen}, \citenamefont {Bonaccurso}, \citenamefont {Deng},\ and\ \citenamefont {Zhang}}]{Chen2016}%
  \BibitemOpen
  \bibfield  {author} {\bibinfo {author} {\bibfnamefont {L.}~\bibnamefont {Chen}}, \bibinfo {author} {\bibfnamefont {E.}~\bibnamefont {Bonaccurso}}, \bibinfo {author} {\bibfnamefont {P.}~\bibnamefont {Deng}},\ and\ \bibinfo {author} {\bibfnamefont {H.}~\bibnamefont {Zhang}},\ }\bibfield  {title} {\bibinfo {title} {Droplet impact on soft viscoelastic surfaces},\ }\href@noop {} {\bibfield  {journal} {\bibinfo  {journal} {Phys. Rev. E}\ }\textbf {\bibinfo {volume} {94}},\ \bibinfo {pages} {063117} (\bibinfo {year} {2016})}\BibitemShut {NoStop}%
\bibitem [{\citenamefont {Raman}\ \emph {et~al.}(2016)\citenamefont {Raman}, \citenamefont {Jaiman}, \citenamefont {Lee},\ and\ \citenamefont {Low}}]{Raman2016}%
  \BibitemOpen
  \bibfield  {author} {\bibinfo {author} {\bibfnamefont {K.~A.}\ \bibnamefont {Raman}}, \bibinfo {author} {\bibfnamefont {R.~K.}\ \bibnamefont {Jaiman}}, \bibinfo {author} {\bibfnamefont {T.~S.}\ \bibnamefont {Lee}},\ and\ \bibinfo {author} {\bibfnamefont {H.~T.}\ \bibnamefont {Low}},\ }\bibfield  {title} {\bibinfo {title} {Lattice boltzmann simulations of droplet impact onto surfaces with varying wettabilities},\ }\href@noop {} {\bibfield  {journal} {\bibinfo  {journal} {International Journal of Heat and Mass Transfer}\ }\textbf {\bibinfo {volume} {95}},\ \bibinfo {pages} {336} (\bibinfo {year} {2016})}\BibitemShut {NoStop}%
\bibitem [{\citenamefont {Fink}\ \emph {et~al.}(2018)\citenamefont {Fink}, \citenamefont {Cai}, \citenamefont {Stroh}, \citenamefont {Bernard}, \citenamefont {Kriegseis}, \citenamefont {Frohnapfel}, \citenamefont {Marschall},\ and\ \citenamefont {Wörner}}]{Fink2018}%
  \BibitemOpen
  \bibfield  {author} {\bibinfo {author} {\bibfnamefont {V.}~\bibnamefont {Fink}}, \bibinfo {author} {\bibfnamefont {X.}~\bibnamefont {Cai}}, \bibinfo {author} {\bibfnamefont {A.}~\bibnamefont {Stroh}}, \bibinfo {author} {\bibfnamefont {R.}~\bibnamefont {Bernard}}, \bibinfo {author} {\bibfnamefont {J.}~\bibnamefont {Kriegseis}}, \bibinfo {author} {\bibfnamefont {B.}~\bibnamefont {Frohnapfel}}, \bibinfo {author} {\bibfnamefont {H.}~\bibnamefont {Marschall}},\ and\ \bibinfo {author} {\bibfnamefont {M.}~\bibnamefont {Wörner}},\ }\bibfield  {title} {\bibinfo {title} {Drop bouncing by micro-grooves},\ }\href@noop {} {\bibfield  {journal} {\bibinfo  {journal} {International Journal of Heat and Fluid Flow}\ }\textbf {\bibinfo {volume} {70}},\ \bibinfo {pages} {271} (\bibinfo {year} {2018})}\BibitemShut {NoStop}%
\bibitem [{\citenamefont {Aris}\ \emph {et~al.}(1958)\citenamefont {Aris}, \citenamefont {Amudson}, \citenamefont {Beek}, \citenamefont {Alche}, \citenamefont {Berger}, \citenamefont {Lapidus}, \citenamefont {Berger}, \citenamefont {Perlmutter}, \citenamefont {Amudson}, \citenamefont {Douglas}, \citenamefont {Rippin}, \citenamefont {Gaitonde}, \citenamefont {Douglas}, \citenamefont {Alche}, \citenamefont {Lasalle}, \citenamefont {Lefschetz}, \citenamefont {Leucke}, \citenamefont {Mcgurie}, \citenamefont {Luus}, \citenamefont {Lapidus}, \citenamefont {Takahasi}, \citenamefont {Rabins}, \citenamefont {Auslander}, \citenamefont {Takahasi},\ and\ \citenamefont {Auslander}}]{Aris1958}%
  \BibitemOpen
  \bibfield  {author} {\bibinfo {author} {\bibfnamefont {R.}~\bibnamefont {Aris}}, \bibinfo {author} {\bibfnamefont {N.~R.}\ \bibnamefont {Amudson}}, \bibinfo {author} {\bibfnamefont {J.}~\bibnamefont {Beek}}, \bibinfo {author} {\bibfnamefont {J.}~\bibnamefont {Alche}}, \bibinfo {author} {\bibfnamefont {A.~J.}\ \bibnamefont {Berger}}, \bibinfo {author} {\bibfnamefont {L.}~\bibnamefont {Lapidus}}, \bibinfo {author} {\bibfnamefont {A.~J.}\ \bibnamefont {Berger}}, \bibinfo {author} {\bibfnamefont {J.~S.}\ \bibnamefont {Perlmutter}}, \bibinfo {author} {\bibfnamefont {D.~D.}\ \bibnamefont {Amudson}}, \bibinfo {author} {\bibfnamefont {N.~R.}\ \bibnamefont {Douglas}}, \bibinfo {author} {\bibfnamefont {J.~M.}\ \bibnamefont {Rippin}}, \bibinfo {author} {\bibfnamefont {D.~W.~T.}\ \bibnamefont {Gaitonde}}, \bibinfo {author} {\bibfnamefont {N.}~\bibnamefont {Douglas}}, \bibinfo {author} {\bibfnamefont {J.~M.}\ \bibnamefont {Alche}}, \bibinfo {author} {\bibfnamefont {J.}~\bibnamefont {Lasalle}}, \bibinfo {author}
  {\bibfnamefont {J.}~\bibnamefont {Lefschetz}}, \bibinfo {author} {\bibfnamefont {S.}~\bibnamefont {Leucke}}, \bibinfo {author} {\bibfnamefont {R.~H.}\ \bibnamefont {Mcgurie}}, \bibinfo {author} {\bibfnamefont {M.~L.}\ \bibnamefont {Luus}}, \bibinfo {author} {\bibfnamefont {R.}~\bibnamefont {Lapidus}}, \bibinfo {author} {\bibfnamefont {L.}~\bibnamefont {Takahasi}}, \bibinfo {author} {\bibfnamefont {Y.}~\bibnamefont {Rabins}}, \bibinfo {author} {\bibfnamefont {M.~J.}\ \bibnamefont {Auslander}}, \bibinfo {author} {\bibfnamefont {D.}~\bibnamefont {Takahasi}},\ and\ \bibinfo {author} {\bibfnamefont {J.}~\bibnamefont {Auslander}},\ }\bibfield  {title} {\bibinfo {title} {Stability by liapunov's direct method with applica-tions},\ }\href@noop {} {\bibfield  {journal} {\bibinfo  {journal} {Ind. Eng. Chem., Fundam}\ }\textbf {\bibinfo {volume} {7}},\ \bibinfo {pages} {161} (\bibinfo {year} {1958})}\BibitemShut {NoStop}%
\bibitem [{\citenamefont {Scheller}\ and\ \citenamefont {Bousfield}(1995)}]{Scheller1995}%
  \BibitemOpen
  \bibfield  {author} {\bibinfo {author} {\bibfnamefont {B.~L.}\ \bibnamefont {Scheller}}\ and\ \bibinfo {author} {\bibfnamefont {D.~W.}\ \bibnamefont {Bousfield}},\ }\bibfield  {title} {\bibinfo {title} {Newtonian drop impact with a solid surface},\ }\href@noop {} {\bibfield  {journal} {\bibinfo  {journal} {AIChE Journal}\ }\textbf {\bibinfo {volume} {41}},\ \bibinfo {pages} {1357} (\bibinfo {year} {1995})}\BibitemShut {NoStop}%
\bibitem [{\citenamefont {Rioboo}\ \emph {et~al.}(2002)\citenamefont {Rioboo}, \citenamefont {Marengo},\ and\ \citenamefont {Tropea}}]{Rioboo2002}%
  \BibitemOpen
  \bibfield  {author} {\bibinfo {author} {\bibfnamefont {R.}~\bibnamefont {Rioboo}}, \bibinfo {author} {\bibfnamefont {M.}~\bibnamefont {Marengo}},\ and\ \bibinfo {author} {\bibfnamefont {C.}~\bibnamefont {Tropea}},\ }\bibfield  {title} {\bibinfo {title} {Time evolution of liquid drop impact onto solid, dry surfaces},\ }\href@noop {} {\bibfield  {journal} {\bibinfo  {journal} {Experiments in Fluids}\ }\textbf {\bibinfo {volume} {33}},\ \bibinfo {pages} {112} (\bibinfo {year} {2002})}\BibitemShut {NoStop}%
\bibitem [{\citenamefont {Biance}\ \emph {et~al.}(2004)\citenamefont {Biance}, \citenamefont {Clanet},\ and\ \citenamefont {Quéré}}]{Biance2004}%
  \BibitemOpen
  \bibfield  {author} {\bibinfo {author} {\bibfnamefont {A.~L.}\ \bibnamefont {Biance}}, \bibinfo {author} {\bibfnamefont {C.}~\bibnamefont {Clanet}},\ and\ \bibinfo {author} {\bibfnamefont {D.}~\bibnamefont {Quéré}},\ }\bibfield  {title} {\bibinfo {title} {First steps in the spreading of a liquid droplet},\ }\href@noop {} {\bibfield  {journal} {\bibinfo  {journal} {Physical Review E - Statistical Physics, Plasmas, Fluids, and Related Interdisciplinary Topics}\ }\textbf {\bibinfo {volume} {69}},\ \bibinfo {pages} {4} (\bibinfo {year} {2004})}\BibitemShut {NoStop}%
\bibitem [{\citenamefont {Sikalo}\ \emph {et~al.}(2005)\citenamefont {Sikalo}, \citenamefont {Wilhelm}, \citenamefont {Roisman}, \citenamefont {Jakirlić},\ and\ \citenamefont {Tropea}}]{Sikalo05}%
  \BibitemOpen
  \bibfield  {author} {\bibinfo {author} {\bibfnamefont {S.}~\bibnamefont {Sikalo}}, \bibinfo {author} {\bibfnamefont {H.~D.}\ \bibnamefont {Wilhelm}}, \bibinfo {author} {\bibfnamefont {I.~V.}\ \bibnamefont {Roisman}}, \bibinfo {author} {\bibfnamefont {S.}~\bibnamefont {Jakirlić}},\ and\ \bibinfo {author} {\bibfnamefont {C.}~\bibnamefont {Tropea}},\ }\bibfield  {title} {\bibinfo {title} {Dynamic contact angle of spreading droplets: Experiments and simulations},\ }\href@noop {} {\bibfield  {journal} {\bibinfo  {journal} {Physics of Fluids}\ }\textbf {\bibinfo {volume} {17}},\ \bibinfo {pages} {1} (\bibinfo {year} {2005})}\BibitemShut {NoStop}%
\bibitem [{\citenamefont {Attané}\ \emph {et~al.}(2007)\citenamefont {Attané}, \citenamefont {Girard},\ and\ \citenamefont {Morin}}]{Attan07}%
  \BibitemOpen
  \bibfield  {author} {\bibinfo {author} {\bibfnamefont {P.}~\bibnamefont {Attané}}, \bibinfo {author} {\bibfnamefont {F.}~\bibnamefont {Girard}},\ and\ \bibinfo {author} {\bibfnamefont {V.}~\bibnamefont {Morin}},\ }\bibfield  {title} {\bibinfo {title} {An energy balance approach of the dynamics of drop impact on a solid surface},\ }\href@noop {} {\bibfield  {journal} {\bibinfo  {journal} {Physics of Fluids}\ }\textbf {\bibinfo {volume} {19}},\ \bibinfo {pages} {012101} (\bibinfo {year} {2007})}\BibitemShut {NoStop}%
\bibitem [{\citenamefont {Courbin}\ \emph {et~al.}(2009)\citenamefont {Courbin}, \citenamefont {Bird}, \citenamefont {Reyssat},\ and\ \citenamefont {Stone}}]{Courbin2009}%
  \BibitemOpen
  \bibfield  {author} {\bibinfo {author} {\bibfnamefont {L.}~\bibnamefont {Courbin}}, \bibinfo {author} {\bibfnamefont {J.~C.}\ \bibnamefont {Bird}}, \bibinfo {author} {\bibfnamefont {M.}~\bibnamefont {Reyssat}},\ and\ \bibinfo {author} {\bibfnamefont {H.~A.}\ \bibnamefont {Stone}},\ }\bibfield  {title} {\bibinfo {title} {Dynamics of wetting: from inertial spreading to viscous imbibition},\ }\href@noop {} {\bibfield  {journal} {\bibinfo  {journal} {Journal of Physics: Condensed Matter}\ }\textbf {\bibinfo {volume} {21}},\ \bibinfo {pages} {464127} (\bibinfo {year} {2009})}\BibitemShut {NoStop}%
\bibitem [{\citenamefont {Snoeijer}\ and\ \citenamefont {Andreotti}(2013)}]{Snoeijer2013}%
  \BibitemOpen
  \bibfield  {author} {\bibinfo {author} {\bibfnamefont {J.~H.}\ \bibnamefont {Snoeijer}}\ and\ \bibinfo {author} {\bibfnamefont {B.}~\bibnamefont {Andreotti}},\ }\bibfield  {title} {\bibinfo {title} {Moving contact lines: Scales, regimes, and dynamical transitions},\ }\href@noop {} {\bibfield  {journal} {\bibinfo  {journal} {Annual Review of Fluid Mechanics}\ }\textbf {\bibinfo {volume} {45}},\ \bibinfo {pages} {269} (\bibinfo {year} {2013})}\BibitemShut {NoStop}%
\bibitem [{\citenamefont {Wang}\ \emph {et~al.}(2023)\citenamefont {Wang}, \citenamefont {Yan}, \citenamefont {Du}, \citenamefont {Ji}, \citenamefont {Inanlu}, \citenamefont {Min},\ and\ \citenamefont {Miljkovic}}]{Wang2023}%
  \BibitemOpen
  \bibfield  {author} {\bibinfo {author} {\bibfnamefont {X.}~\bibnamefont {Wang}}, \bibinfo {author} {\bibfnamefont {X.}~\bibnamefont {Yan}}, \bibinfo {author} {\bibfnamefont {J.}~\bibnamefont {Du}}, \bibinfo {author} {\bibfnamefont {B.}~\bibnamefont {Ji}}, \bibinfo {author} {\bibfnamefont {M.~J.}\ \bibnamefont {Inanlu}}, \bibinfo {author} {\bibfnamefont {Q.}~\bibnamefont {Min}},\ and\ \bibinfo {author} {\bibfnamefont {N.}~\bibnamefont {Miljkovic}},\ }\bibfield  {title} {\bibinfo {title} {Spreading dynamics of microdroplets on nanostructured surfaces},\ }\href@noop {} {\bibfield  {journal} {\bibinfo  {journal} {Journal of Colloid and Interface Science}\ }\textbf {\bibinfo {volume} {635}},\ \bibinfo {pages} {221} (\bibinfo {year} {2023})}\BibitemShut {NoStop}%
\bibitem [{\citenamefont {Palacios}\ \emph {et~al.}(2012)\citenamefont {Palacios}, \citenamefont {Hernández}, \citenamefont {Gómez}, \citenamefont {Zanzi},\ and\ \citenamefont {López}}]{Palacios2012}%
  \BibitemOpen
  \bibfield  {author} {\bibinfo {author} {\bibfnamefont {J.}~\bibnamefont {Palacios}}, \bibinfo {author} {\bibfnamefont {J.}~\bibnamefont {Hernández}}, \bibinfo {author} {\bibfnamefont {P.}~\bibnamefont {Gómez}}, \bibinfo {author} {\bibfnamefont {C.}~\bibnamefont {Zanzi}},\ and\ \bibinfo {author} {\bibfnamefont {J.}~\bibnamefont {López}},\ }\bibfield  {title} {\bibinfo {title} {On the impact of viscous drops onto dry smooth surfaces},\ }\href@noop {} {\bibfield  {journal} {\bibinfo  {journal} {Experiments in Fluids}\ }\textbf {\bibinfo {volume} {52}},\ \bibinfo {pages} {1449} (\bibinfo {year} {2012})}\BibitemShut {NoStop}%
\bibitem [{\citenamefont {Burzynski}\ \emph {et~al.}(2020)\citenamefont {Burzynski}, \citenamefont {Roisman},\ and\ \citenamefont {Bansmer}}]{Burzynski2020}%
  \BibitemOpen
  \bibfield  {author} {\bibinfo {author} {\bibfnamefont {D.~A.}\ \bibnamefont {Burzynski}}, \bibinfo {author} {\bibfnamefont {I.~V.}\ \bibnamefont {Roisman}},\ and\ \bibinfo {author} {\bibfnamefont {S.~E.}\ \bibnamefont {Bansmer}},\ }\bibfield  {title} {\bibinfo {title} {On the splashing of high-speed drops impacting a dry surface},\ }\href@noop {} {\bibfield  {journal} {\bibinfo  {journal} {Journal of Fluid Mechanics}\ }\textbf {\bibinfo {volume} {892}},\ \bibinfo {pages} {A2} (\bibinfo {year} {2020})}\BibitemShut {NoStop}%
\bibitem [{\citenamefont {Guo}\ \emph {et~al.}(2022)\citenamefont {Guo}, \citenamefont {Liu}, \citenamefont {Sun}, \citenamefont {Liu},\ and\ \citenamefont {Liu}}]{Guo2022}%
  \BibitemOpen
  \bibfield  {author} {\bibinfo {author} {\bibfnamefont {C.}~\bibnamefont {Guo}}, \bibinfo {author} {\bibfnamefont {L.}~\bibnamefont {Liu}}, \bibinfo {author} {\bibfnamefont {J.}~\bibnamefont {Sun}}, \bibinfo {author} {\bibfnamefont {C.}~\bibnamefont {Liu}},\ and\ \bibinfo {author} {\bibfnamefont {S.}~\bibnamefont {Liu}},\ }\bibfield  {title} {\bibinfo {title} {Splashing behavior of impacting droplets on grooved superhydrophobic surfaces},\ }\href@noop {} {\bibfield  {journal} {\bibinfo  {journal} {Physics of Fluids}\ }\textbf {\bibinfo {volume} {34}},\ \bibinfo {pages} {052105} (\bibinfo {year} {2022})}\BibitemShut {NoStop}%
\bibitem [{\citenamefont {Coppola}\ \emph {et~al.}(2011)\citenamefont {Coppola}, \citenamefont {Rocco},\ and\ \citenamefont {de~Luca}}]{Coppola2011}%
  \BibitemOpen
  \bibfield  {author} {\bibinfo {author} {\bibfnamefont {G.}~\bibnamefont {Coppola}}, \bibinfo {author} {\bibfnamefont {G.}~\bibnamefont {Rocco}},\ and\ \bibinfo {author} {\bibfnamefont {L.}~\bibnamefont {de~Luca}},\ }\bibfield  {title} {\bibinfo {title} {Insights on the impact of a plane drop on a thin liquid film},\ }\href@noop {} {\bibfield  {journal} {\bibinfo  {journal} {Physics of Fluids}\ }\textbf {\bibinfo {volume} {23}},\ \bibinfo {pages} {022105} (\bibinfo {year} {2011})}\BibitemShut {NoStop}%
\bibitem [{\citenamefont {Kittel}\ \emph {et~al.}(2018)\citenamefont {Kittel}, \citenamefont {Roisman},\ and\ \citenamefont {Tropea}}]{Kittel2018}%
  \BibitemOpen
  \bibfield  {author} {\bibinfo {author} {\bibfnamefont {H.~M.}\ \bibnamefont {Kittel}}, \bibinfo {author} {\bibfnamefont {I.~V.}\ \bibnamefont {Roisman}},\ and\ \bibinfo {author} {\bibfnamefont {C.}~\bibnamefont {Tropea}},\ }\bibfield  {title} {\bibinfo {title} {Splash of a drop impacting onto a solid substrate wetted by a thin film of another liquid},\ }\href@noop {} {\bibfield  {journal} {\bibinfo  {journal} {Phys. Rev. Fluids}\ }\textbf {\bibinfo {volume} {3}},\ \bibinfo {pages} {073601} (\bibinfo {year} {2018})}\BibitemShut {NoStop}%
\bibitem [{\citenamefont {Sprittles}(2024)}]{Sprittles2024}%
  \BibitemOpen
  \bibfield  {author} {\bibinfo {author} {\bibfnamefont {J.~E.}\ \bibnamefont {Sprittles}},\ }\bibfield  {title} {\bibinfo {title} {Gas microfilms in droplet dynamics: When do drops bounce?},\ }\href@noop {} {\bibfield  {journal} {\bibinfo  {journal} {Annual Review of Fluid Mechanics}\ }\textbf {\bibinfo {volume} {47}},\ \bibinfo {pages} {17} (\bibinfo {year} {2024})}\BibitemShut {NoStop}%
\bibitem [{\citenamefont {Lee}\ \emph {et~al.}(2012)\citenamefont {Lee}, \citenamefont {Weon}, \citenamefont {Je},\ and\ \citenamefont {Fezzaa}}]{Lee2012}%
  \BibitemOpen
  \bibfield  {author} {\bibinfo {author} {\bibfnamefont {J.~S.}\ \bibnamefont {Lee}}, \bibinfo {author} {\bibfnamefont {B.~M.}\ \bibnamefont {Weon}}, \bibinfo {author} {\bibfnamefont {J.~H.}\ \bibnamefont {Je}},\ and\ \bibinfo {author} {\bibfnamefont {K.}~\bibnamefont {Fezzaa}},\ }\bibfield  {title} {\bibinfo {title} {How does an air film evolve into a bubble during drop impact?},\ }\href@noop {} {\bibfield  {journal} {\bibinfo  {journal} {Phys. Rev. Lett.}\ }\textbf {\bibinfo {volume} {109}},\ \bibinfo {pages} {204501} (\bibinfo {year} {2012})}\BibitemShut {NoStop}%
\bibitem [{\citenamefont {Kolinski}\ \emph {et~al.}(2012)\citenamefont {Kolinski}, \citenamefont {Rubinstein}, \citenamefont {Mandre}, \citenamefont {Brenner}, \citenamefont {Weitz},\ and\ \citenamefont {Mahadevan}}]{Kolinski2012}%
  \BibitemOpen
  \bibfield  {author} {\bibinfo {author} {\bibfnamefont {J.~M.}\ \bibnamefont {Kolinski}}, \bibinfo {author} {\bibfnamefont {S.~M.}\ \bibnamefont {Rubinstein}}, \bibinfo {author} {\bibfnamefont {S.}~\bibnamefont {Mandre}}, \bibinfo {author} {\bibfnamefont {M.~P.}\ \bibnamefont {Brenner}}, \bibinfo {author} {\bibfnamefont {D.~A.}\ \bibnamefont {Weitz}},\ and\ \bibinfo {author} {\bibfnamefont {L.}~\bibnamefont {Mahadevan}},\ }\bibfield  {title} {\bibinfo {title} {Skating on a film of air: Drops impacting on a surface},\ }\href@noop {} {\bibfield  {journal} {\bibinfo  {journal} {Phys. Rev. Lett.}\ }\textbf {\bibinfo {volume} {108}},\ \bibinfo {pages} {074503} (\bibinfo {year} {2012})}\BibitemShut {NoStop}%
\bibitem [{\citenamefont {Langley}\ \emph {et~al.}(2020)\citenamefont {Langley}, \citenamefont {Castrejón-Pita},\ and\ \citenamefont {Thoroddsen}}]{Langley2020}%
  \BibitemOpen
  \bibfield  {author} {\bibinfo {author} {\bibfnamefont {K.~R.}\ \bibnamefont {Langley}}, \bibinfo {author} {\bibfnamefont {A.~A.}\ \bibnamefont {Castrejón-Pita}},\ and\ \bibinfo {author} {\bibfnamefont {S.~T.}\ \bibnamefont {Thoroddsen}},\ }\bibfield  {title} {\bibinfo {title} {Droplet impacts onto soft solids entrap more air},\ }\href@noop {} {\bibfield  {journal} {\bibinfo  {journal} {Soft Matter}\ }\textbf {\bibinfo {volume} {16}},\ \bibinfo {pages} {5702} (\bibinfo {year} {2020})}\BibitemShut {NoStop}%
\bibitem [{\citenamefont {Jha}\ \emph {et~al.}(2020)\citenamefont {Jha}, \citenamefont {Chantelot}, \citenamefont {Clanet},\ and\ \citenamefont {Quéré}}]{Jha2020}%
  \BibitemOpen
  \bibfield  {author} {\bibinfo {author} {\bibfnamefont {A.}~\bibnamefont {Jha}}, \bibinfo {author} {\bibfnamefont {P.}~\bibnamefont {Chantelot}}, \bibinfo {author} {\bibfnamefont {C.}~\bibnamefont {Clanet}},\ and\ \bibinfo {author} {\bibfnamefont {D.}~\bibnamefont {Quéré}},\ }\bibfield  {title} {\bibinfo {title} {Viscous bouncing},\ }\href@noop {} {\bibfield  {journal} {\bibinfo  {journal} {Soft Matter}\ }\textbf {\bibinfo {volume} {16}},\ \bibinfo {pages} {7270} (\bibinfo {year} {2020})}\BibitemShut {NoStop}%
\bibitem [{\citenamefont {Sanjay}\ \emph {et~al.}(2023)\citenamefont {Sanjay}, \citenamefont {Chantelot},\ and\ \citenamefont {Lohse}}]{Sanjay2023}%
  \BibitemOpen
  \bibfield  {author} {\bibinfo {author} {\bibfnamefont {V.}~\bibnamefont {Sanjay}}, \bibinfo {author} {\bibfnamefont {P.}~\bibnamefont {Chantelot}},\ and\ \bibinfo {author} {\bibfnamefont {D.}~\bibnamefont {Lohse}},\ }\bibfield  {title} {\bibinfo {title} {When does an impacting drop stop bouncing?},\ }\href@noop {} {\bibfield  {journal} {\bibinfo  {journal} {Journal of Fluid Mechanics}\ }\textbf {\bibinfo {volume} {958}},\ \bibinfo {pages} {A26} (\bibinfo {year} {2023})}\BibitemShut {NoStop}%
\bibitem [{\citenamefont {Visser}\ \emph {et~al.}(2012)\citenamefont {Visser}, \citenamefont {Tagawa}, \citenamefont {Sun},\ and\ \citenamefont {Lohse}}]{Visser2012}%
  \BibitemOpen
  \bibfield  {author} {\bibinfo {author} {\bibfnamefont {C.~W.}\ \bibnamefont {Visser}}, \bibinfo {author} {\bibfnamefont {Y.}~\bibnamefont {Tagawa}}, \bibinfo {author} {\bibfnamefont {C.}~\bibnamefont {Sun}},\ and\ \bibinfo {author} {\bibfnamefont {D.}~\bibnamefont {Lohse}},\ }\bibfield  {title} {\bibinfo {title} {Microdroplet impact at very high velocity},\ }\href@noop {} {\bibfield  {journal} {\bibinfo  {journal} {Soft Matter}\ }\textbf {\bibinfo {volume} {8}},\ \bibinfo {pages} {10732} (\bibinfo {year} {2012})}\BibitemShut {NoStop}%
\bibitem [{\citenamefont {Visser}\ \emph {et~al.}(2015)\citenamefont {Visser}, \citenamefont {Frommhold}, \citenamefont {Wildeman}, \citenamefont {Mettin}, \citenamefont {Lohse},\ and\ \citenamefont {Sun}}]{Visser2015}%
  \BibitemOpen
  \bibfield  {author} {\bibinfo {author} {\bibfnamefont {C.~W.}\ \bibnamefont {Visser}}, \bibinfo {author} {\bibfnamefont {P.~E.}\ \bibnamefont {Frommhold}}, \bibinfo {author} {\bibfnamefont {S.}~\bibnamefont {Wildeman}}, \bibinfo {author} {\bibfnamefont {R.}~\bibnamefont {Mettin}}, \bibinfo {author} {\bibfnamefont {D.}~\bibnamefont {Lohse}},\ and\ \bibinfo {author} {\bibfnamefont {C.}~\bibnamefont {Sun}},\ }\bibfield  {title} {\bibinfo {title} {Dynamics of high-speed micro-drop impact: Numerical simulations and experiments at frame-to-frame times below 100 ns},\ }\href@noop {} {\bibfield  {journal} {\bibinfo  {journal} {Soft Matter}\ }\textbf {\bibinfo {volume} {11}},\ \bibinfo {pages} {1708} (\bibinfo {year} {2015})}\BibitemShut {NoStop}%
\bibitem [{\citenamefont {McCarthy}\ \emph {et~al.}(2023)\citenamefont {McCarthy}, \citenamefont {Reid},\ and\ \citenamefont {Walker}}]{McCarthy2022}%
  \BibitemOpen
  \bibfield  {author} {\bibinfo {author} {\bibfnamefont {L.~P.}\ \bibnamefont {McCarthy}}, \bibinfo {author} {\bibfnamefont {J.~P.}\ \bibnamefont {Reid}},\ and\ \bibinfo {author} {\bibfnamefont {J.~S.}\ \bibnamefont {Walker}},\ }\bibfield  {title} {\bibinfo {title} {High frame-rate imaging of the shape oscillations and spreading dynamics of picolitre droplets impacting on a surface},\ }\href@noop {} {\bibfield  {journal} {\bibinfo  {journal} {Physics of Fluids}\ }\textbf {\bibinfo {volume} {35}},\ \bibinfo {pages} {122010} (\bibinfo {year} {2023})}\BibitemShut {NoStop}%
\bibitem [{\citenamefont {Mahato}\ \emph {et~al.}(2025)\citenamefont {Mahato}, \citenamefont {Bracher}, \citenamefont {McLauchlan}, \citenamefont {Harniman},\ and\ \citenamefont {Walker}}]{Mahato2024}%
  \BibitemOpen
  \bibfield  {author} {\bibinfo {author} {\bibfnamefont {L.~K.}\ \bibnamefont {Mahato}}, \bibinfo {author} {\bibfnamefont {R.}~\bibnamefont {Bracher}}, \bibinfo {author} {\bibfnamefont {J.}~\bibnamefont {McLauchlan}}, \bibinfo {author} {\bibfnamefont {R.~L.}\ \bibnamefont {Harniman}},\ and\ \bibinfo {author} {\bibfnamefont {J.~S.}\ \bibnamefont {Walker}},\ }\bibfield  {title} {\bibinfo {title} {The impact dynamics of picolitre aerosol droplets depositing on surfaces: Effect of wettability, inertia and viscosity},\ }\href@noop {} {\bibfield  {journal} {\bibinfo  {journal} {Aerosol Science and Technology}\ }\textbf {\bibinfo {volume} {0}},\ \bibinfo {pages} {1} (\bibinfo {year} {2025})}\BibitemShut {NoStop}%
\bibitem [{\citenamefont {Tai}\ \emph {et~al.}(2021)\citenamefont {Tai}, \citenamefont {Zhao}, \citenamefont {Guo}, \citenamefont {Li}, \citenamefont {Wang},\ and\ \citenamefont {Xia}}]{Tai2021}%
  \BibitemOpen
  \bibfield  {author} {\bibinfo {author} {\bibfnamefont {Y.}~\bibnamefont {Tai}}, \bibinfo {author} {\bibfnamefont {Y.}~\bibnamefont {Zhao}}, \bibinfo {author} {\bibfnamefont {X.}~\bibnamefont {Guo}}, \bibinfo {author} {\bibfnamefont {L.}~\bibnamefont {Li}}, \bibinfo {author} {\bibfnamefont {S.}~\bibnamefont {Wang}},\ and\ \bibinfo {author} {\bibfnamefont {Z.}~\bibnamefont {Xia}},\ }\bibfield  {title} {\bibinfo {title} {Research on the contact time of a bouncing microdroplet with lattice boltzmann method},\ }\href@noop {} {\bibfield  {journal} {\bibinfo  {journal} {Physics of Fluids}\ }\textbf {\bibinfo {volume} {33}},\ \bibinfo {pages} {042011} (\bibinfo {year} {2021})}\BibitemShut {NoStop}%
\bibitem [{\citenamefont {Zhang}\ \emph {et~al.}(2016{\natexlab{a}})\citenamefont {Zhang}, \citenamefont {Lei}, \citenamefont {Wang},\ and\ \citenamefont {Zhang}}]{Zhang2016}%
  \BibitemOpen
  \bibfield  {author} {\bibinfo {author} {\bibfnamefont {B.}~\bibnamefont {Zhang}}, \bibinfo {author} {\bibfnamefont {Q.}~\bibnamefont {Lei}}, \bibinfo {author} {\bibfnamefont {Z.}~\bibnamefont {Wang}},\ and\ \bibinfo {author} {\bibfnamefont {X.}~\bibnamefont {Zhang}},\ }\bibfield  {title} {\bibinfo {title} {Droplets can rebound toward both directions on textured surfaces with a wettability gradient},\ }\href@noop {} {\bibfield  {journal} {\bibinfo  {journal} {Langmuir}\ }\textbf {\bibinfo {volume} {32}},\ \bibinfo {pages} {346} (\bibinfo {year} {2016}{\natexlab{a}})}\BibitemShut {NoStop}%
\bibitem [{\citenamefont {COMSOL}(2025)}]{comsol}%
  \BibitemOpen
  \bibfield  {author} {\bibinfo {author} {\bibnamefont {COMSOL}},\ }\href@noop {} {\bibinfo {title} {Comsol multiphysics}} (\bibinfo {year} {2025}),\ \bibinfo {note} {version 6.1}\BibitemShut {NoStop}%
\bibitem [{\citenamefont {Xia}\ \emph {et~al.}(2023)\citenamefont {Xia}, \citenamefont {Chen}, \citenamefont {Liu}, \citenamefont {Zhang}, \citenamefont {Tian},\ and\ \citenamefont {Zhang}}]{Phase1}%
  \BibitemOpen
  \bibfield  {author} {\bibinfo {author} {\bibfnamefont {L.}~\bibnamefont {Xia}}, \bibinfo {author} {\bibfnamefont {F.}~\bibnamefont {Chen}}, \bibinfo {author} {\bibfnamefont {T.}~\bibnamefont {Liu}}, \bibinfo {author} {\bibfnamefont {D.}~\bibnamefont {Zhang}}, \bibinfo {author} {\bibfnamefont {Y.}~\bibnamefont {Tian}},\ and\ \bibinfo {author} {\bibfnamefont {D.}~\bibnamefont {Zhang}},\ }\bibfield  {title} {\bibinfo {title} {Phase-field simulations of droplet impact on superhydrophobic surfaces},\ }\href@noop {} {\bibfield  {journal} {\bibinfo  {journal} {International Journal of Mechanical Sciences}\ }\textbf {\bibinfo {volume} {240}},\ \bibinfo {pages} {107957} (\bibinfo {year} {2023})}\BibitemShut {NoStop}%
\bibitem [{\citenamefont {Zhang}\ \emph {et~al.}(2016{\natexlab{b}})\citenamefont {Zhang}, \citenamefont {Qian},\ and\ \citenamefont {Wang}}]{Phase2}%
  \BibitemOpen
  \bibfield  {author} {\bibinfo {author} {\bibfnamefont {Q.}~\bibnamefont {Zhang}}, \bibinfo {author} {\bibfnamefont {T.-Z.}\ \bibnamefont {Qian}},\ and\ \bibinfo {author} {\bibfnamefont {X.-P.}\ \bibnamefont {Wang}},\ }\bibfield  {title} {\bibinfo {title} {Phase field simulation of a droplet impacting a solid surface},\ }\href@noop {} {\bibfield  {journal} {\bibinfo  {journal} {Physics of Fluids}\ }\textbf {\bibinfo {volume} {28}},\ \bibinfo {pages} {022103} (\bibinfo {year} {2016}{\natexlab{b}})}\BibitemShut {NoStop}%
\bibitem [{\citenamefont {{Akhlaghi Amiri}}\ and\ \citenamefont {Hamouda}(2013)}]{Phase3}%
  \BibitemOpen
  \bibfield  {author} {\bibinfo {author} {\bibfnamefont {H.}~\bibnamefont {{Akhlaghi Amiri}}}\ and\ \bibinfo {author} {\bibfnamefont {A.}~\bibnamefont {Hamouda}},\ }\bibfield  {title} {\bibinfo {title} {Evaluation of level set and phase field methods in modeling two phase flow with viscosity contrast through dual-permeability porous medium},\ }\href@noop {} {\bibfield  {journal} {\bibinfo  {journal} {International Journal of Multiphase Flow}\ }\textbf {\bibinfo {volume} {52}},\ \bibinfo {pages} {22} (\bibinfo {year} {2013})}\BibitemShut {NoStop}%
\bibitem [{\citenamefont {Wang}\ \emph {et~al.}(2024)\citenamefont {Wang}, \citenamefont {Cai}, \citenamefont {Wang}, \citenamefont {Gao}, \citenamefont {Yang}, \citenamefont {Zheng}, \citenamefont {Lee},\ and\ \citenamefont {Wang}}]{Bound2}%
  \BibitemOpen
  \bibfield  {author} {\bibinfo {author} {\bibfnamefont {Y.-F.}\ \bibnamefont {Wang}}, \bibinfo {author} {\bibfnamefont {Z.-H.}\ \bibnamefont {Cai}}, \bibinfo {author} {\bibfnamefont {Y.-B.}\ \bibnamefont {Wang}}, \bibinfo {author} {\bibfnamefont {S.-R.}\ \bibnamefont {Gao}}, \bibinfo {author} {\bibfnamefont {Y.-R.}\ \bibnamefont {Yang}}, \bibinfo {author} {\bibfnamefont {S.-F.}\ \bibnamefont {Zheng}}, \bibinfo {author} {\bibfnamefont {D.-J.}\ \bibnamefont {Lee}},\ and\ \bibinfo {author} {\bibfnamefont {X.-D.}\ \bibnamefont {Wang}},\ }\bibfield  {title} {\bibinfo {title} {Progressive evolution of the viscous dissipation mechanism from the macroscale to the nanoscale},\ }\href@noop {} {\bibfield  {journal} {\bibinfo  {journal} {Journal of Fluid Mechanics}\ }\textbf {\bibinfo {volume} {998}},\ \bibinfo {pages} {A21} (\bibinfo {year} {2024})}\BibitemShut {NoStop}%
\bibitem [{\citenamefont {Okumura}\ \emph {et~al.}(2003)\citenamefont {Okumura}, \citenamefont {Chevy}, \citenamefont {Richard}, \citenamefont {Quéré},\ and\ \citenamefont {Clanet}}]{Okumura2003}%
  \BibitemOpen
  \bibfield  {author} {\bibinfo {author} {\bibfnamefont {K.}~\bibnamefont {Okumura}}, \bibinfo {author} {\bibfnamefont {F.}~\bibnamefont {Chevy}}, \bibinfo {author} {\bibfnamefont {D.}~\bibnamefont {Richard}}, \bibinfo {author} {\bibfnamefont {D.}~\bibnamefont {Quéré}},\ and\ \bibinfo {author} {\bibfnamefont {C.}~\bibnamefont {Clanet}},\ }\bibfield  {title} {\bibinfo {title} {Water spring: A model for bouncing drops},\ }\href@noop {} {\bibfield  {journal} {\bibinfo  {journal} {Europhysics Letters}\ }\textbf {\bibinfo {volume} {62}},\ \bibinfo {pages} {237} (\bibinfo {year} {2003})}\BibitemShut {NoStop}%
\bibitem [{\citenamefont {Thenarianto}\ \emph {et~al.}(2023)\citenamefont {Thenarianto}, \citenamefont {Koh}, \citenamefont {Lin}, \citenamefont {Jokinen},\ and\ \citenamefont {Daniel}}]{Bound1}%
  \BibitemOpen
  \bibfield  {author} {\bibinfo {author} {\bibfnamefont {C.}~\bibnamefont {Thenarianto}}, \bibinfo {author} {\bibfnamefont {X.~Q.}\ \bibnamefont {Koh}}, \bibinfo {author} {\bibfnamefont {M.}~\bibnamefont {Lin}}, \bibinfo {author} {\bibfnamefont {V.}~\bibnamefont {Jokinen}},\ and\ \bibinfo {author} {\bibfnamefont {D.}~\bibnamefont {Daniel}},\ }\bibfield  {title} {\bibinfo {title} {Energy loss for droplets bouncing off superhydrophobic surfaces},\ }\href@noop {} {\bibfield  {journal} {\bibinfo  {journal} {Langmuir}\ }\textbf {\bibinfo {volume} {39}},\ \bibinfo {pages} {3162} (\bibinfo {year} {2023})}\BibitemShut {NoStop}%
\bibitem [{\citenamefont {Quéré}(2008)}]{Quere}%
  \BibitemOpen
  \bibfield  {author} {\bibinfo {author} {\bibfnamefont {D.}~\bibnamefont {Quéré}},\ }\bibfield  {title} {\bibinfo {title} {Wetting and roughness},\ }\href@noop {} {\bibfield  {journal} {\bibinfo  {journal} {Annual Review of Materials Research}\ }\textbf {\bibinfo {volume} {38}},\ \bibinfo {pages} {71} (\bibinfo {year} {2008})}\BibitemShut {NoStop}%
\bibitem [{\citenamefont {Clanet}\ \emph {et~al.}(2004)\citenamefont {Clanet}, \citenamefont {Béguin}, \citenamefont {Richard},\ and\ \citenamefont {Quéré}}]{Clanet2004}%
  \BibitemOpen
  \bibfield  {author} {\bibinfo {author} {\bibfnamefont {C.}~\bibnamefont {Clanet}}, \bibinfo {author} {\bibfnamefont {C.}~\bibnamefont {Béguin}}, \bibinfo {author} {\bibfnamefont {D.}~\bibnamefont {Richard}},\ and\ \bibinfo {author} {\bibfnamefont {D.}~\bibnamefont {Quéré}},\ }\bibfield  {title} {\bibinfo {title} {Maximal deformation of an impacting drop},\ }\href@noop {} {\bibfield  {journal} {\bibinfo  {journal} {Journal of Fluid Mechanics}\ }\textbf {\bibinfo {volume} {517}},\ \bibinfo {pages} {199} (\bibinfo {year} {2004})}\BibitemShut {NoStop}%
\bibitem [{\citenamefont {Eggers}\ \emph {et~al.}(2010)\citenamefont {Eggers}, \citenamefont {Fontelos}, \citenamefont {Josserand},\ and\ \citenamefont {Zaleski}}]{Eggers2010}%
  \BibitemOpen
  \bibfield  {author} {\bibinfo {author} {\bibfnamefont {J.}~\bibnamefont {Eggers}}, \bibinfo {author} {\bibfnamefont {M.~A.}\ \bibnamefont {Fontelos}}, \bibinfo {author} {\bibfnamefont {C.}~\bibnamefont {Josserand}},\ and\ \bibinfo {author} {\bibfnamefont {S.}~\bibnamefont {Zaleski}},\ }\bibfield  {title} {\bibinfo {title} {Drop dynamics after impact on a solid wall: Theory and simulations},\ }\href@noop {} {\bibfield  {journal} {\bibinfo  {journal} {Physics of Fluids}\ }\textbf {\bibinfo {volume} {22}},\ \bibinfo {pages} {062101} (\bibinfo {year} {2010})}\BibitemShut {NoStop}%
\bibitem [{\citenamefont {Villermaux}\ and\ \citenamefont {Bossa}(2011)}]{BOSSA_2011}%
  \BibitemOpen
  \bibfield  {author} {\bibinfo {author} {\bibfnamefont {E.}~\bibnamefont {Villermaux}}\ and\ \bibinfo {author} {\bibfnamefont {B.}~\bibnamefont {Bossa}},\ }\bibfield  {title} {\bibinfo {title} {Drop fragmentation on impact},\ }\href@noop {} {\bibfield  {journal} {\bibinfo  {journal} {Journal of Fluid Mechanics}\ }\textbf {\bibinfo {volume} {668}},\ \bibinfo {pages} {412–435} (\bibinfo {year} {2011})}\BibitemShut {NoStop}%
\bibitem [{\citenamefont {Laan}\ \emph {et~al.}(2014)\citenamefont {Laan}, \citenamefont {de~Bruin}, \citenamefont {Bartolo}, \citenamefont {Josserand},\ and\ \citenamefont {Bonn}}]{Bruin2014}%
  \BibitemOpen
  \bibfield  {author} {\bibinfo {author} {\bibfnamefont {N.}~\bibnamefont {Laan}}, \bibinfo {author} {\bibfnamefont {K.~G.}\ \bibnamefont {de~Bruin}}, \bibinfo {author} {\bibfnamefont {D.}~\bibnamefont {Bartolo}}, \bibinfo {author} {\bibfnamefont {C.}~\bibnamefont {Josserand}},\ and\ \bibinfo {author} {\bibfnamefont {D.}~\bibnamefont {Bonn}},\ }\bibfield  {title} {\bibinfo {title} {Maximum diameter of impacting liquid droplets},\ }\href@noop {} {\bibfield  {journal} {\bibinfo  {journal} {Phys. Rev. Appl.}\ }\textbf {\bibinfo {volume} {2}},\ \bibinfo {pages} {044018} (\bibinfo {year} {2014})}\BibitemShut {NoStop}%
\bibitem [{\citenamefont {Wildeman}\ \emph {et~al.}(2016)\citenamefont {Wildeman}, \citenamefont {Visser}, \citenamefont {Sun},\ and\ \citenamefont {Lohse}}]{Wildeman2016}%
  \BibitemOpen
  \bibfield  {author} {\bibinfo {author} {\bibfnamefont {S.}~\bibnamefont {Wildeman}}, \bibinfo {author} {\bibfnamefont {C.~W.}\ \bibnamefont {Visser}}, \bibinfo {author} {\bibfnamefont {C.}~\bibnamefont {Sun}},\ and\ \bibinfo {author} {\bibfnamefont {D.}~\bibnamefont {Lohse}},\ }\bibfield  {title} {\bibinfo {title} {On the spreading of impacting drops},\ }\href@noop {} {\bibfield  {journal} {\bibinfo  {journal} {Journal of Fluid Mechanics}\ }\textbf {\bibinfo {volume} {805}},\ \bibinfo {pages} {636–655} (\bibinfo {year} {2016})}\BibitemShut {NoStop}%
\bibitem [{\citenamefont {Sanjay}\ and\ \citenamefont {Lohse}(2025)}]{sanjay2024}%
  \BibitemOpen
  \bibfield  {author} {\bibinfo {author} {\bibfnamefont {V.}~\bibnamefont {Sanjay}}\ and\ \bibinfo {author} {\bibfnamefont {D.}~\bibnamefont {Lohse}},\ }\bibfield  {title} {\bibinfo {title} {Unifying theory of scaling in drop impact: Forces and maximum spreading diameter},\ }\href@noop {} {\bibfield  {journal} {\bibinfo  {journal} {Phys. Rev. Lett.}\ }\textbf {\bibinfo {volume} {134}},\ \bibinfo {pages} {104003} (\bibinfo {year} {2025})}\BibitemShut {NoStop}%
\bibitem [{\citenamefont {Kamusewitz}\ and\ \citenamefont {Possart}(1985)}]{Kamuse85}%
  \BibitemOpen
  \bibfield  {author} {\bibinfo {author} {\bibfnamefont {H.}~\bibnamefont {Kamusewitz}}\ and\ \bibinfo {author} {\bibfnamefont {W.}~\bibnamefont {Possart}},\ }\bibfield  {title} {\bibinfo {title} {The static contact angle hysteresis obtained by different experiments for the system ptfe/water},\ }\href@noop {} {\bibfield  {journal} {\bibinfo  {journal} {International Journal of Adhesion and Adhesives}\ }\textbf {\bibinfo {volume} {5}},\ \bibinfo {pages} {211} (\bibinfo {year} {1985})}\BibitemShut {NoStop}%
\end{thebibliography}
